\documentclass[useAMS,usenatbib]{mn2e}

\usepackage{latexsym,graphicx}
\usepackage{color}
\usepackage{fixltx2e}
\usepackage{verbatim}
\usepackage{float}
\usepackage{amsmath,amssymb}
\usepackage{times}
\usepackage{soul}

\usepackage{setspace}

\def\apgt{\ {\raise-.5ex\hbox{$\buildrel>\over\sim$}}\ }
\def\aplt{\ {\raise-.5ex\hbox{$\buildrel<\over\sim$}}\ }

\newcommand{\degree}{\ensuremath{^\circ}}

\def\HII{\rm{H}\,{\textsc {ii}}}

\title[Models of the circumstellar medium of evolving, massive runaway stars]{Models of the circumstellar medium of evolving, massive runaway stars moving through the Galactic plane}

\author[D. M.-A.~Meyer et al.]
       {D. M.-A.~Meyer,$^{1}$\thanks{E-mail: dmeyer@astro.uni-bonn.de}
       J.~Mackey,$^{1}$ N.~Langer,$^{1}$\thanks{Alexander von Humboldt Professor} V. V.~Gvaramadze,$^{2,3}$ A.~Mignone,$^4$
     	\newauthor R. G. Izzard$^{1}$ and L. Kaper$^{5}$\\
        $^{1}$Argelander-Institut f\"ur Astronomie der Universit\"at Bonn, Auf dem H\"ugel 71, 53121, Bonn, Germany \\
        $^{2}$Sternberg Astronomical Institute, Lomonosov Moscow State University, Universitetskij Pr. 13, Moscow 119992, Russia\\
        $^{3}$Isaac Newton Institute of Chile, Moscow Branch, Universitetskij Pr. 13, Moscow 119992, Russia\\
        $^{4}$Dipartimento di Fisica Generale Facolt\`a di Scienze M.F.N., Universit\`a degli Studi di Torino, Via Pietro Giuria 1, 10125 Torino, Italy\\
        $^{5}$Astronomical Institute Anton Pannekoek, University of Amsterdam, Science Park 904, 1098 XH Amsterdam, The Netherlands\\ 
        }

\begin{document}

\date{Received May 6, 2014; accepted August 8, 2014}

\maketitle
   
\label{firstpage}

\begin{abstract}

\textcolor{black}{
  At least 5 per cent of the massive stars are moving supersonically through the
interstellar medium (ISM) and are expected to produce a stellar wind bow shock.
We explore how the mass loss and space velocity of massive runaway stars affect
the morphology of their bow shocks. We run two-dimensional axisymmetric hydrodynamical simulations following the
evolution of the circumstellar medium of these stars in the Galactic plane from
the main sequence to the red supergiant phase. We find that thermal conduction
is an important process governing the shape, size and structure of the bow shocks around hot
stars, and that they have an optical luminosity mainly produced by forbidden
lines, e.g. [O\,{\sc iii}]. The H$\alpha$ emission of the bow shocks around hot
stars originates from near their contact discontinuity. The H$\alpha$ emission
of bow shocks around cool stars originates from their forward shock, and is too
faint to be observed for the bow shocks that we simulate. The emission of
optically-thin radiation mainly comes from the shocked ISM material. All bow
shock models are brighter in the infrared, i.e. the infrared is the most
appropriate waveband to search for bow shocks. Our study suggests that the
infrared emission comes from near the contact discontinuity for bow shocks of
hot stars and from the inner region of shocked wind for bow shocks around cool
stars. We predict that, in the Galactic plane, the brightest, i.e. the most
easily detectable bow shocks are produced by high-mass stars moving
with small space velocities. 
}

\end{abstract}

\begin{keywords}
methods: numerical -- shock waves - circumstellar matter -- stars: massive.
\end{keywords}


\section{Introduction}
\label{sect:introduction}

Massive stars have strong winds and evolve through distinct stellar evolutionary
phases which shape their surroundings. Releasing material and radiation, they
give rise to ISM structures whose geometries strongly depend on the
properties of their driving star, e.g. rotation~\citep{langer_ApJ_520_1999,
vanmarle_aa_478_2008, chita_aa_488_2008},
motion~\citep{brighenti_mnras_277_1995, brighenti_mnras_273_1995}, internal
pulsation\textcolor{black}{~\citep[see chapter 5 in][]{vanveelen_phd}, duplicity~\citep{stevens_apj_386_1992}
or stellar evolution~\citep[e.g. the Napoleon's hat generated by the progenitor of 
the supernova SN1987A and overhanging its remnant, see][]{wang_MNRAS_261_1993}}.
At the end of their lives, most massive stars explode as a
supernova or generate a gamma-ray burst event~\citep{woosley_rvmp_74_2002} and their ejecta interact with their
circumstellar medium~\citep{borkowski_apj_400_1992, vanveelen_aa_50._2009,
chiotellis_aa_537_2012}. Additionally, massive stars are important engines for
chemically enriching the interstellar medium (ISM) of galaxies, e.g. via their
metal-rich winds and supernova ejecta, and returning kinetic energy and
momentum to the ISM~\citep{vink_asp_353_2006}.

Between 10 and 25 per cent of the O stars
are runaway stars~\citep{gies_apjs_64_1987,blau1993ASPC...35..207B} and about 40 per cent of these, i.e.
about between 4 and 10 per cent of all O stars~\citep[see][]{huthoff_aa_383_2002}, have
identified bow shocks. The bow shocks can be detected at
X-ray~\citep{lopez_apj_757_2012}, ultraviolet~\citep{lebertre_mnras_422_2012},
optical~\citep{gull_apj_230_1979}, infrared~\citep{buren_apj_329_1988} and
radio~\citep{benaglia_aa_517_2010} wavelengths. The bow-shock-producing stars
are mainly on the main sequence or blue
supergiants~\citep{vanburen_aj_110_1995,peri_aa_538_2012}. There are also known
bow shocks around red supergiants,
Betelgeuse~\citep{noriegacrespo_aj_114_1997, decin_aa_548_2012}, $\mu$
Cep~\citep{cox_aa_537_2012} and IRC$-$10414~\citep{Gvaramadze_2014} or
asymptotic giant branch stars~\citep{cox_aa_537_2012,jorissen_532_aa_2011}. Bow
shocks are used to find new runaway stars~\citep*{gvaramadze_519_aa_2010}, to
identify star clusters from which these stars have been
ejected~\citep{gvaramadze_aa_490_2008} and to constrain the properties of their
central stars, e.g. mass-loss rate~\citep{gull_apj_230_1979,
gvaramadze_mnras_427_2012}, or the density of the local
ISM~\citep{kaper_apj_475_1997, Gvaramadze_2014}.

\begin{figure}
          \centering 
	  \includegraphics[width=0.4\textwidth]{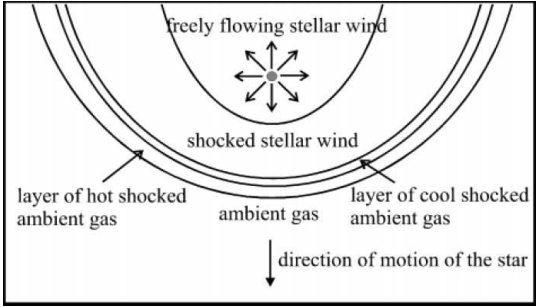}    
	  \caption{ Typical structure of a bow shock generated by a hot runaway star. 
	            The figure is taken from~\citet{comeron_aa_338_1998}. }
          \label{fig:sketch}
\end{figure}

The structure of such bow shocks is sketched in Fig.~\ref{fig:sketch}.
However the layers of shocked ISM develop
differently as a function of the wind power and ISM properties. The wind and ISM
pressure balance at the contact discontinuity. It separates the regions of
shocked material bordered by the forward and reverse shocks. The distance from
the star to the contact discontinuity in the direction of the relative motion
between wind and ISM defines the stand-off distance of the bow
shock~\citep*{baranov_sphd_15_1971}. The shape of isothermal bow shocks, in
which the shocked regions are thin, is analytically
approximated in~\citet{wilkin_459_apj_1996}. 

A numerical study by~\citet{comeron_aa_338_1998} compares wind-ISM interactions
with (semi-)analytical models and concludes that the thin-shell approximation
has partial validity. This work describes the variety of shapes which could be
produced in bow shocks of OB stars. It details how the action of the wind on the
ISM, together with the cooling in the shocked gas, shapes the circumstellar
medium, determines the relative thickness of the layers composing a bow shock,
and determines its (in)stability. It shows the importance of heat
conduction~\citep{spitzer_1962, cowie_apj_211_1977} to  the size of these bow
shocks, and that rapid cooling distorts them. The shocked regions are thick if
the shock is weak, but they cool rapidly and become denser and thinner for the
regime involving either high space velocities or strong winds and/or high
ambient medium densities. This leads to distorting instabilities such as the
transverse acceleration instability~\citep{blondin_na_57_1998} or the non-linear
thin shell instability~\citep{dgani_apj_461_1996, dgani_apr_461_1996}.
\citet{maclow_apj_369_1991} models bow shocks around main sequence stars in
dense molecular clouds. The bow shock models in~\citet{comeron_aa_338_1998} are
set in low-density ambient medium.

Models for bow shocks around evolved, cool runaway stars exist for several
stellar evolutionary phases, such as red supergiants
\citep{brighenti_mnras_277_1995, mohamed_aa_541_2012, decin_aa_548_2012} or
asymptotic giant branch (AGB)
phases~\citep{WareingZijlstraOBrien_2007_MNRAS_382_1233_AGB_bowshocks,
villaver_apj_748_2012}. When a bow shock around a red supergiant forms, the
new-born shell swept up by the cool wind succeeds the former bow shock from the
main sequence. A collision between the old and new shells of different densities
precedes the creation of a second bow shock~\citep{mackey_apjlett_751_2012}.
Bow shocks around cool stars are more
likely to generate vortices~\citep{wareing_apj_660_2007} and their substructures
are Rayleigh-Taylor and Kelvin-Helmholtz unstable~\citep{decin_aa_548_2012}. The
dynamics of ISM dust grains penetrating into the bow shocks of red supergiants
is numerically investigated in~\citet{vanmarle_apjl_734_2011}. The effect
of the space velocity and the ISM density on the morphology of Betelgeuse's bow
shock is explored in~\citet{mohamed_aa_541_2012}, however this study
considers a single mass-loss rate and does not allow to appreciate how the wind
properties modify the bow shock's shape or luminosity. In addition,
\citet{vanmarle_aa_561_2014} show the stabilizing effect of a weak ISM magnetic
field on the bow shock of Betelgeuse.

In this study, we explore in a grid of 2D models the combined role of the star's
mass-loss and its space velocity on the dynamics and morphology of bow shocks of
various massive stars moving within the Galactic plane. We use representative
initial masses and space velocities of massive
stars~\citep*{eldridge_mnras_414_2011}. Stellar evolution is followed from the
main sequence to the red supergiant phase.  The treatment of the dissipative
processes and the discrimination between wind and ISM material allows us to
calculate the bow shock luminosities and to discuss the origin of their
emission. We also estimate the luminosity of the bow shocks to predict the best
way to observe them. The project differs from previous studies~\citep[e.g.][]{
comeron_aa_338_1998, mohamed_aa_541_2012} in that we use more realistic cooling
curves, we include stellar evolution in the models \textcolor{black}{and because 
we focus on the emitting properties and observability of our bow shocks}. We 
do not take into account the inhomogeneity and the magnetic field of the ISM.

This paper is organised as follows. We first begin our Section 2 by presenting
our method, stellar evolution models, included physics and the numerical code.
Models for the main sequence, \textcolor{black}{the stellar phase transition} and
red supergiant phases are presented in Sections 3, \textcolor{black}{4} and
\textcolor{black}{5}, respectively. We describe the grid of 2D simulations of bow
shocks around massive stars, discuss their morphology, compare their
substructures to an analytical solution for infinitely thin bow shock and
present their luminosities and H$\alpha$ surface brightnesses. Section
\textcolor{black}{6} discusses our results. We conclude in Section
\textcolor{black}{7}.


\section{Numerical scheme and initial parameters}
\label{sect:method}


\subsection{Hydrodynamics, boundary conditions and numerical scheme}
\label{subsect:hydro}

The governing equations are the Euler equations of classical hydrodynamics,
including radiative cooling and heating for an optically-thin plasma and taking
into account electronic thermal conduction, which are,
\begin{equation}
	   \frac{\partial \rho}{\partial t}  + 
	   \bmath{\nabla}  \cdot (\rho\bmath{v}) =   0,
\label{eq:euler1}
\end{equation}
\begin{equation}
	   \frac{\partial \rho \bmath{v} }{\partial t}  + 
           \bmath{\nabla} \cdot ( \bmath{v} \otimes \rho \bmath{v}) 	      + 
           \bmath{\nabla}p 			      =   \bmath{0},
\label{eq:euler2}
\end{equation}
and,
\begin{equation}
	  \frac{\partial E }{\partial t}   + 
	  \bmath{\nabla} \cdot(E\bmath{v})   +
	  \bmath{\nabla} \cdot (p \bmath{v})   =	   
	  \itl{\Phi}(T,\rho) +
	  \bmath{\nabla} \cdot \bmath{{F}_{\rm c}}.
\label{eq:euler3}
\end{equation}
In the system of equations~(\ref{eq:euler1})$-$(\ref{eq:euler3}), $\bmath{v}$ is
the gas velocity in the frame of reference of the star, $\rho$ is the gas mass
density and $p$ is its thermal pressure. The total number density $n$ is defined
by $\rho=\mu n m_{\rm H}$, where $\mu$ is the mean molecular weight in units of 
the mass of the hydrogen atom $m_{\rm H}$. The total energy density is the sum of 
its thermal and kinetic parts,
\begin{equation}
	E = \frac{p}{(\gamma - 1)} + \frac{\rho v^{2}}{2},
\label{eq:energy}
\end{equation}
where $\gamma$ is the ratio of specific heats for an ideal gas, i.e. $\gamma=5/3$.
The temperature inside a given layer of the bow shock is given by, 
\begin{equation}
	T =  \mu \frac{ m_{\mathrm{H}} }{ k_{\rm{B}} } \frac{p}{\rho},
\label{eq:temperature}
\end{equation}
where $k_{\rm B}$ is the Boltzmann constant. The quantity $\it \Phi$ in the energy
equation~(\ref{eq:euler3}) gathers the rates $\it \Lambda$ for optically-thin
radiative cooling and $\it \Gamma_{\alpha}$ for heating, 
\begin{equation}  
	 \itl \Phi(T,\rho)  =  n^{\alpha}_{\mathrm{H}}\itl{\Gamma}_{\alpha}(T)   
		   		 -  n^{2}_{\mathrm{H}}\itl{\Lambda}(T),
\label{eq:dissipation}
\end{equation}
where the exponent $\alpha$ depends on the ionization of the medium (see
Section~\ref{subsect:radlosses}), and $n_{\rm H}$ is the hydrogen number
density. The heat flux is symbolised by the vector $\bmath{F_{\rm c}}$. The
relation $c_{\rm s} = \sqrt{\gamma p/\rho}$ closes the system of partial
differential equations (\ref{eq:euler1})$-$(\ref{eq:euler3}), where $c_{\rm s}$
is the adiabatic speed of sound.

We perform calculations on a 2D rectangular computational domain in a
cylindrical frame of reference $(O; R, z)$ of origin $O$, imposing rotational
symmetry about $R=0$. We use a uniform grid divided into $N_{\rm R}\times
N_{\rm z}$ cells, and we pay attention to the number of cells resolving the
layers of the bow shocks~\citep{comeron_aa_338_1998}. We choose the size of the
computational domain such that the tail of the bow shock only crosses the
downstream boundary $z=z_{\rm min}$. Following the methods of~\citet{comeron_aa_338_1998}
and~\citet{vanmarle_aa_460_2006}, a stellar wind is released into the domain by
a half circle of radius $20 $ cells centred on the origin. We impose at every
timestep a wind density $\rho_{\rm w} \propto r^{-2}$ onto this circle, where
$r$ is the distance to $O$. We work in the frame of reference of the runaway
star. Outflow boundary conditions are assigned at the $z=z_{\rm min}$ and
$R=R_{\rm max}$ borders of the domain, whereas ISM material flows into the
domain from the $z=z_{\rm max}$ border. The choice of a 2D cylindrical
coordinate system possessing an intrinsic axisymmetric geometry limits us to the
modelling of symmetric bow shocks only.

We solve the equations with the magneto-hydrodynamics code {\sc pluto}
~\citep{mignone_apj_170_2007,migmone_apjs_198_2012}. We use a finite volume
method with the Harten-Lax-van Leer approximate Riemann solver for the fluid
dynamics, controlled by the standard Courant-Friedrich-Levy (CFL) parameter
initially set to $C_{\rm cfl}=0.1$. The equations are integrated with a second
order, unsplit, time-marching algorithm. This Godunov-type scheme is second
order accurate in space and in time. \textcolor{black}{Optically-thin radiative losses are
linearly interpolated from tabulated cooling curves} and the
corresponding rate of change is subtracted from the pressure. The parabolic term
in the equation (\ref{eq:euler3}), corresponding to the heat conduction is
treated with the Super-Time-Stepping algorithm~\citep{alexiades_cnme_12_1996} in
an operator-split, first order accurate in time algorithm.

We use {\sc pluto} 4.0 where linear interpolation in cylindrical coordinates is
correctly performed by taking into account the geometrical centroids rather than
the cell centre~\textcolor{black}{\citep{mignone_JCoPh_270_2014}}. 
We have found that this leads to better results compared to
{\sc pluto} 3.1, especially in close proximity to the axis. The diffusive solver
chosen to carry out the simulations damps the dramatic numerical instabilities
along the symmetry axis at the apex of the bow shocks~\citep{vieser_472_aa_2007,
kwak_apj_739_2011} and is more robust for hypersonic flows. All the physical
components of the model are included from the first timestep of the simulations.


\subsection{Wind model}
\label{subsect:wind}

\begin{figure}
          \centering 
	  \includegraphics[width=0.45\textwidth]{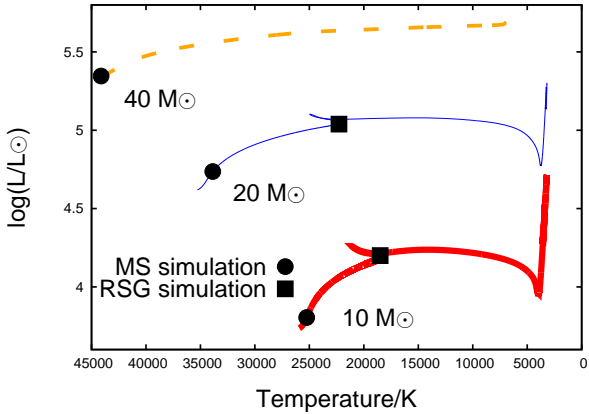}    
	  \caption{ Stellar evolutionary tracks used in the simulations.
		   Thick solid red line, thin solid blue line and dashed orange line 
		   are the evolutionary tracks for our $10$, $20$ and $40$ 
		   $\mathrm{M}_{\odot}$ models, respectively.
		   Circles indicate the time of the beginning of the simulations for the main sequence phase
		   and squares for the red supergiant phase. }
          \label{fig:hrd}
\end{figure}

Stellar evolution models provide us with the wind parameters throughout the
star's life from the main sequence to the red supergiant phase (see evolutionary
tracks in Fig.~\ref{fig:hrd}). We obtain the wind inflow boundary conditions
from a grid of evolutionary models for non-rotating massive stars with solar
metallicity~\citep{brott_aa_530_2011a}. Their initial masses are $M_{\star}=10,
20$ and $40\, \mathrm{M}_{\odot}$ (the masses of the stars quoted hereafter are
the zero-age main sequence masses, unless otherwise stated), and they have been
modelled with the Binary Evolution Code (BEC)~\citep{heger_apj_apj_2000,
yoon_443_aa_2005} including mass-loss but ignoring overshooting. The mass-loss
rate calculation includes the prescriptions for O-type stars
by~\citet{vink_aa_362_2000, vink_aa_369_2001} and for cool stars
by~\citet{jager_aas_72_1988}.

\begin{figure*}
	\begin{minipage}[b]{0.32\textwidth}
		\includegraphics[width=1.07\textwidth,angle=0]{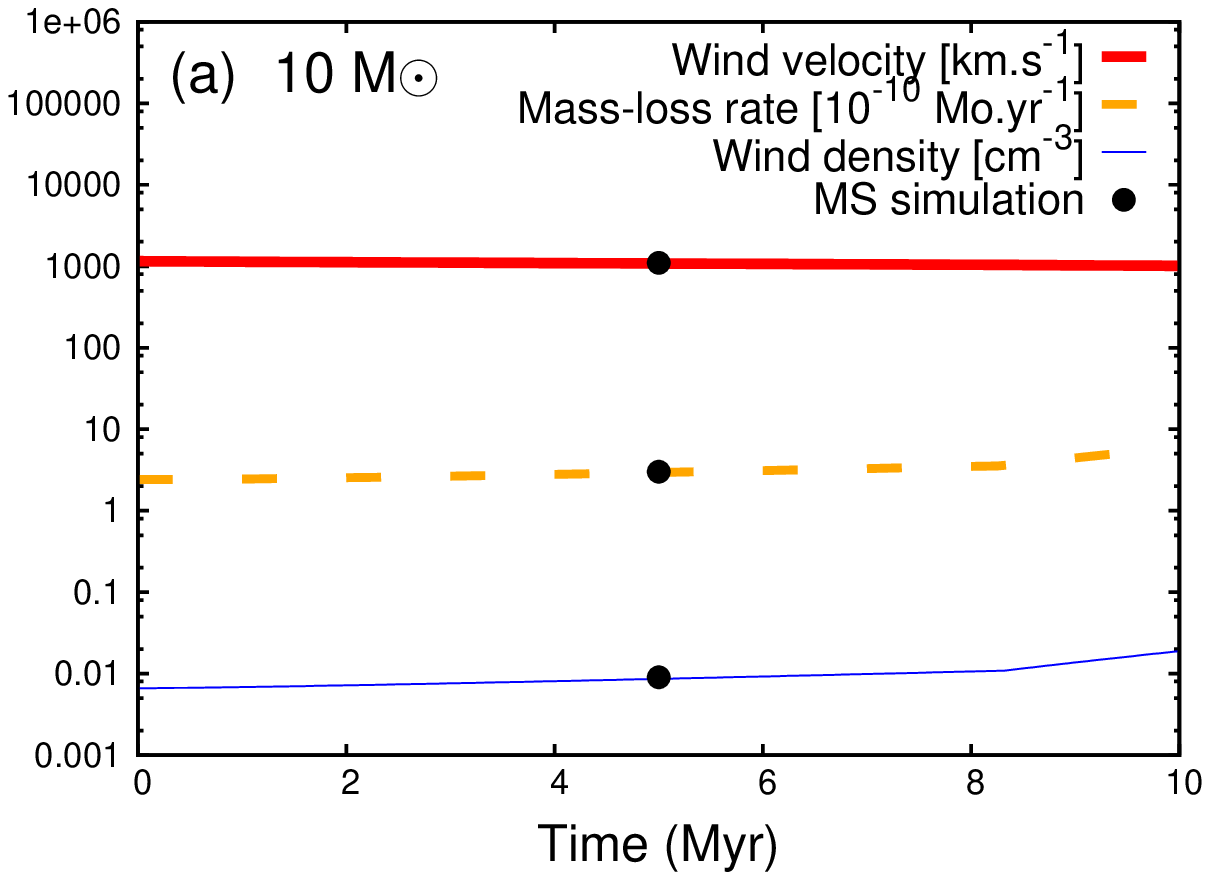}
	\end{minipage}
	\begin{minipage}[b]{0.32\textwidth}
		\includegraphics[width=1.07\textwidth,angle=0]{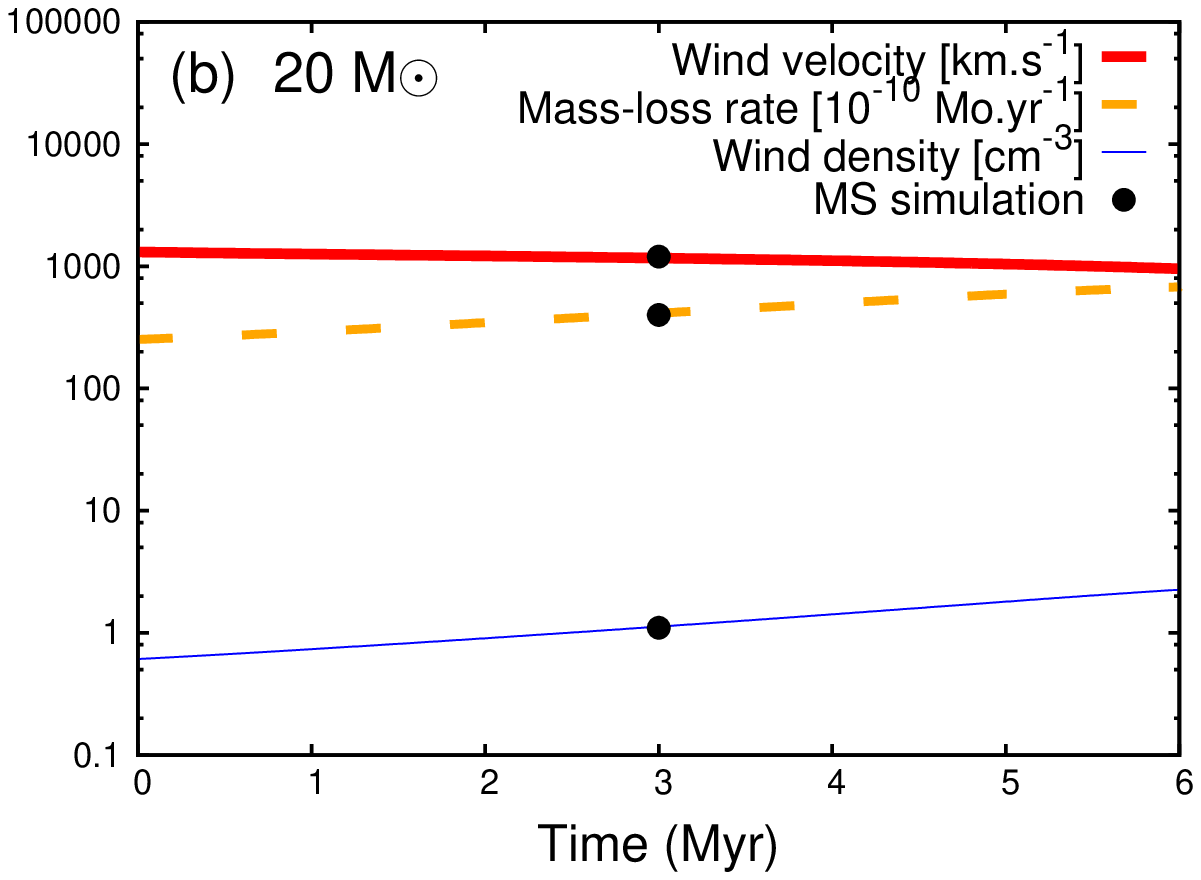}
	\end{minipage}
	\begin{minipage}[b]{0.32\textwidth}
		\includegraphics[width=1.07\textwidth,angle=0]{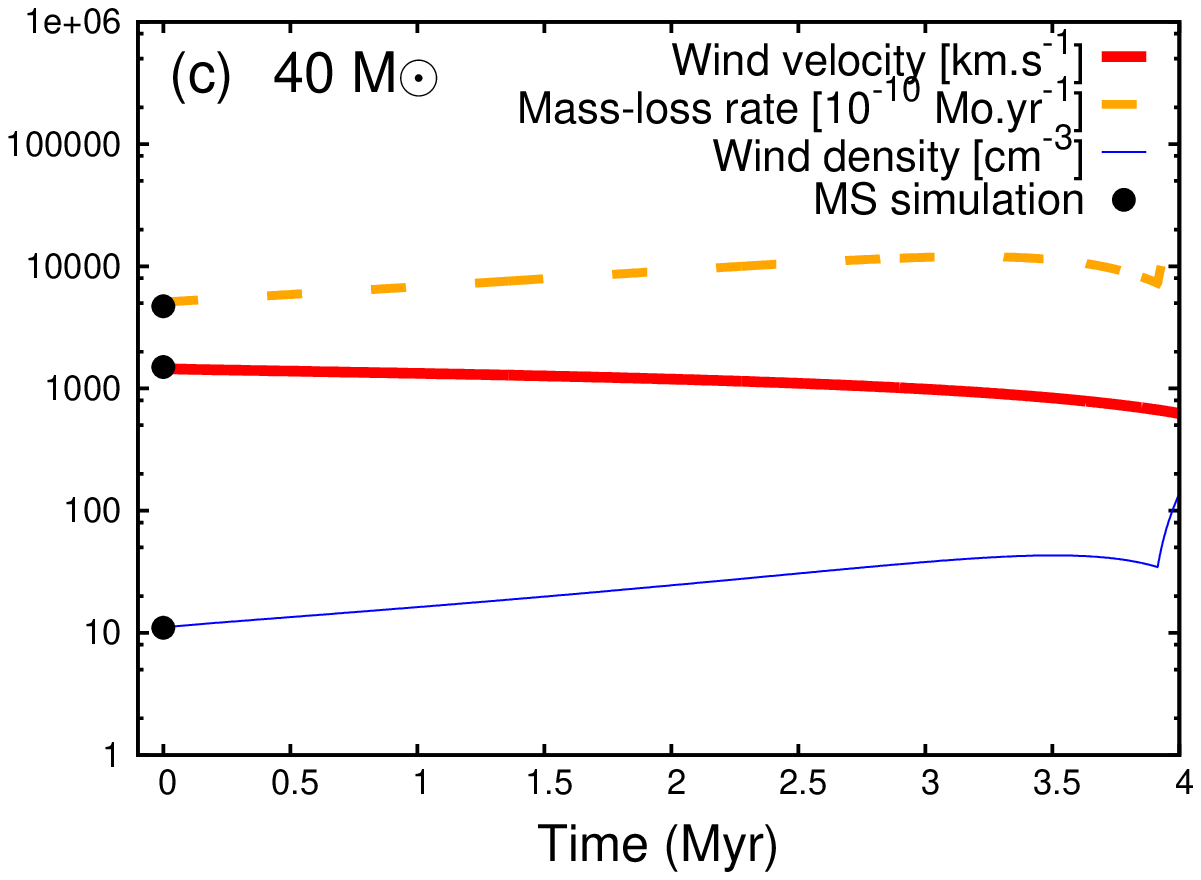}
	\end{minipage}  \\
	\begin{minipage}[b]{0.32\textwidth}
		\includegraphics[width=1.07\textwidth,angle=0]{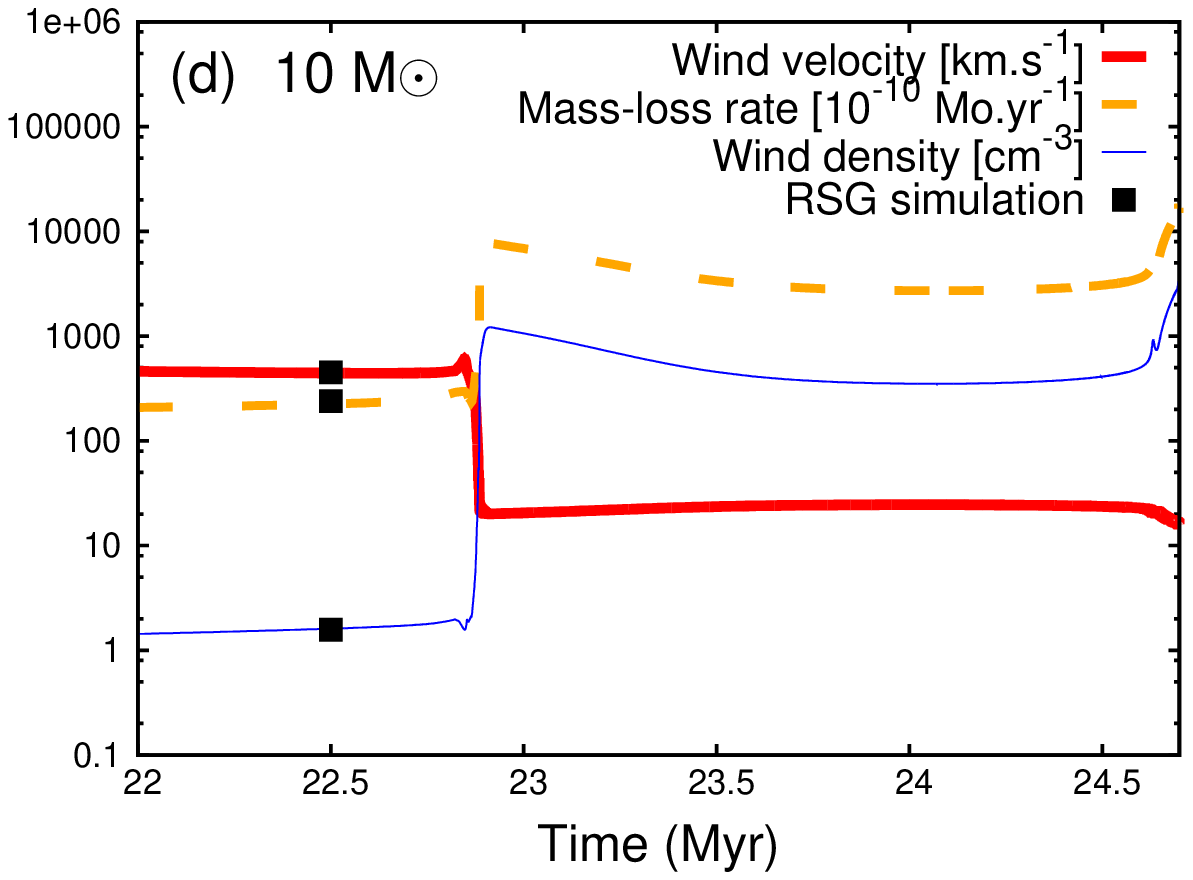}
	\end{minipage}
	\begin{minipage}[b]{0.32\textwidth}
		\includegraphics[width=1.07\textwidth,angle=0]{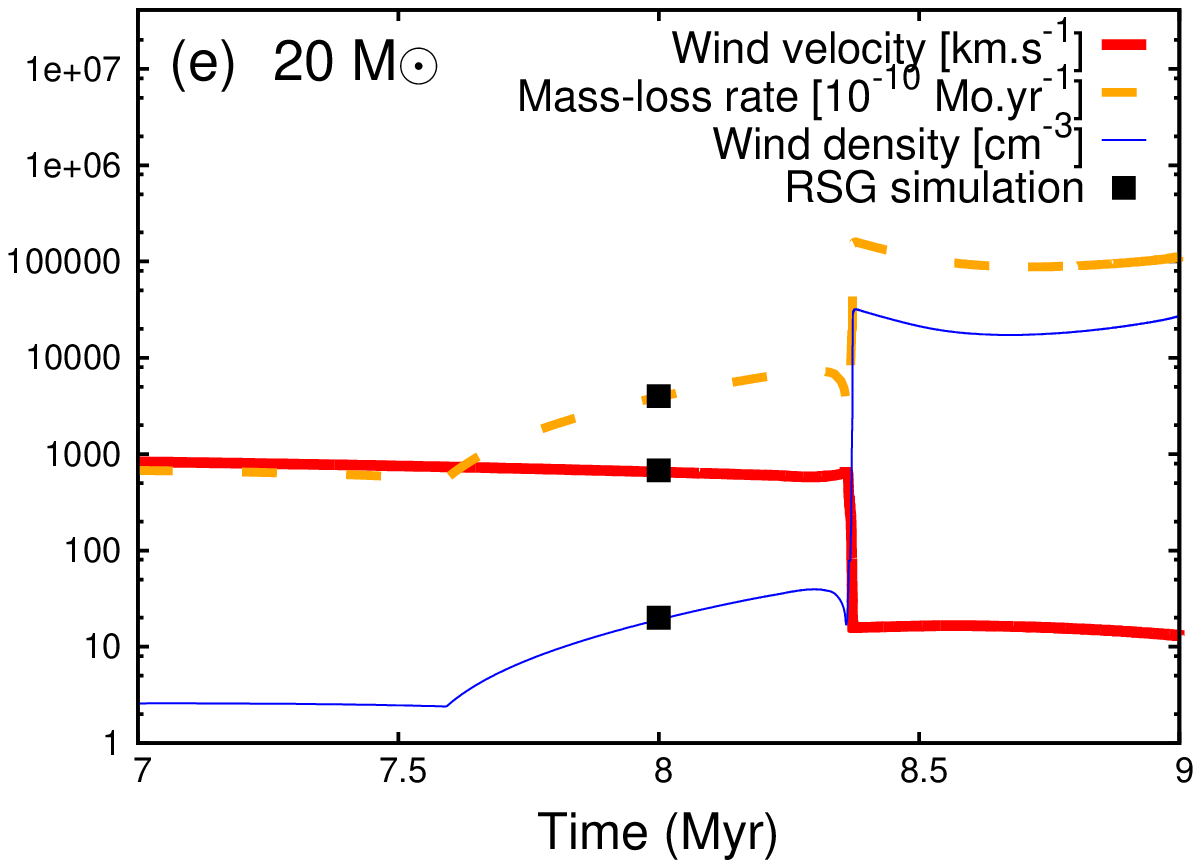}
	\end{minipage}
	\begin{minipage}[b]{0.32\textwidth}
		\includegraphics[width=1.07\textwidth,angle=0]{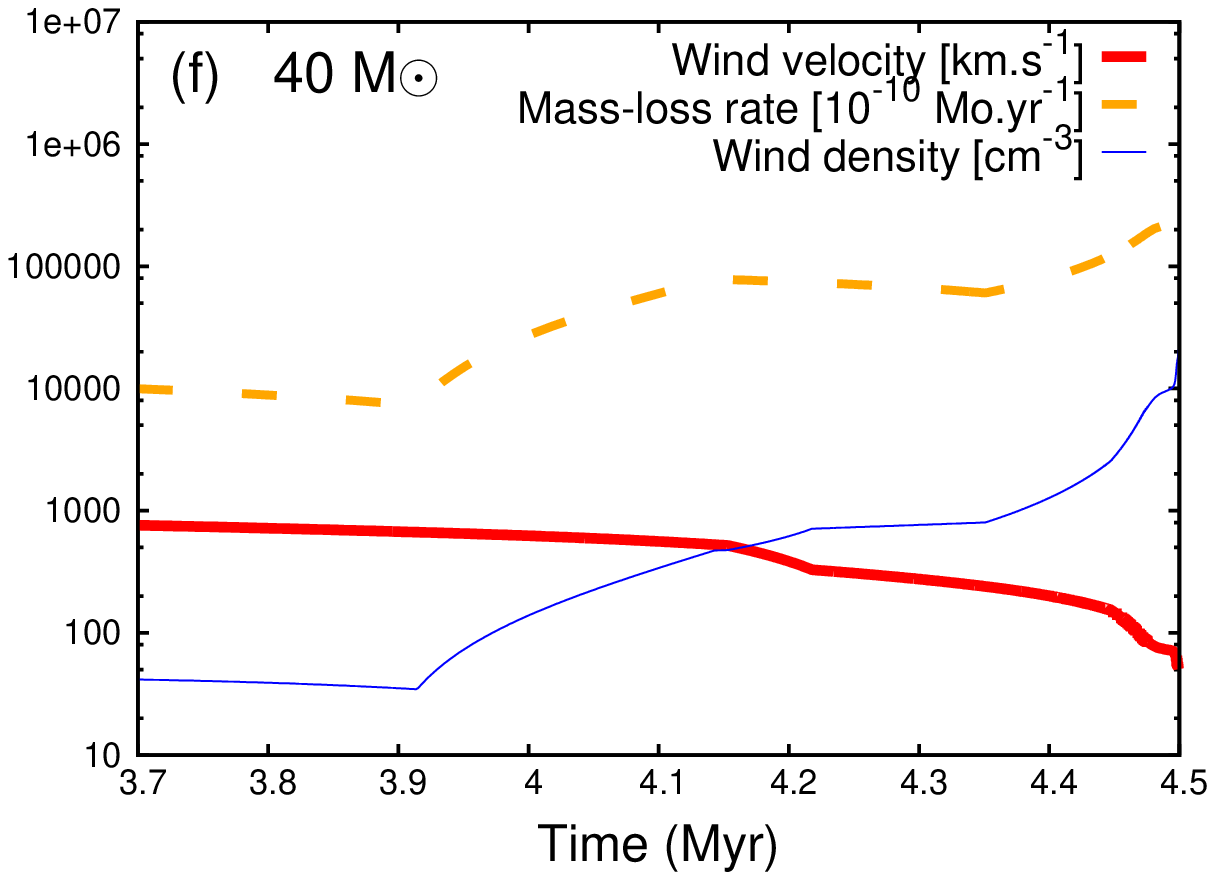}
	\end{minipage} 
	\caption{Physical parameters of the stellar winds used in our simulations. 
		 The top panels represent the wind velocity $v_{\mathrm{w}}$ (thick solid red line),
		 the mass-loss rate $\dot{M}$ (dashed orange line) and the		 
		 wind number density $n_{\mathrm{w}}$ (thin solid blue line) during 
		 the main sequence phase of the $10$, $20\, \mathrm{and}\, 40\, \mathrm{M_{\odot}}$ stars, 
		 whereas the bottom panels show these parameters during the red supergiant phase of the same stars. 
		 Wind number density is calculated at $0.01\, \mathrm{pc}$ from the star
		 and are proportional to the mass-loss rate $\dot{M}$ (see Eq.~\ref{eq:wind}).
		 Black dots show the beginning of the simulations for the main sequence phase and black squares 
		 for the red supergiant phase (Table~\ref{tab:ms}). 
		 }
	\label{fig:models_ms_rsg}  
\end{figure*}

Fig.~\ref{fig:models_ms_rsg} shows the stellar wind properties of the different
models at a radius of $r=0.01\, \mathrm{pc}$ from the star. Mass-loss rate
$\dot{M}$, wind density $\rho_w$ and velocity $v_{\rm w}$ are linked by,
\begin{equation}
	\rho_{w} = \frac{ \dot{M} }{ 4\pi r^{2} v_{\rm w} }.
\label{eq:wind}
\end{equation}
The wind terminal velocity is calculated from the escape velocity
$v_{\mathrm{esc}}$ using $v_{\rm w}^{2}=\beta_{w}(T)
v_{\mathrm{esc}}^{2}$~\citep{eldridge_mnras_367_2006}, with $\beta_{w}$ a
parameter given in their table 1.

The mass-loss rate of the star has a constant value of around $10^{-9.5}$,
$10^{-7.3}$ and  $10^{-6.2}\, \mathrm{M}_{\odot}\,
\mathrm{yr}^{-1}$ during the main sequence phase of the $10,\, 20$ and $40\,
\mathrm{M}_{\odot}$ stars, respectively. After the transition to a red
supergiant, the mass-loss increases to around $10^{-6.2}$ and around
$10^{-5}\, \mathrm{M}_{\odot}\, \mathrm{yr}^{-1}$ for the $10\,
\mathrm{M}_{\odot}$ and $20\, \mathrm{M}_{\odot}$ stars, respectively. 
\textcolor{black}{The evolutionary model of our $40\,\mathrm{M}_{\odot}$ star 
ends at the beginning of the helium ignition, i.e. it does not have a red supergiant phase 
(Brott, private communication). Such a star may evolve through the 
red supergiant phase but this is not included in our model (see panel (f) of 
Fig.~\ref{fig:models_ms_rsg}). } The wind velocity decreases by two orders of
magnitude from $\sim\, 1000\, \mathrm{km}\, \mathrm{s}^{-1}$ during the main
sequence phase to $\sim\, 10\, \mathrm{km}\, \mathrm{s}^{-1}$ for the red
supergiant phase. The effective temperature of the star decreases from
$T_{\mathrm{eff}} \sim 10^{4}\, \mathrm{K}$ during the main sequence phase to 
$T_{\mathrm{eff}}\sim 2.5$$-$$4.5\times10^{3}\, \mathrm{K}$ when the star
becomes a red supergiant. The thermal pressure of the wind is proportional to
$T_{\mathrm{eff}}$, according to the ideal gas equation of state. It scales as
$r^{-2\gamma}$ and is negligible during all evolutionary phases compared to the
ram pressure of the wind in the free expanding region.

We run two simulations for each $M_{\star}$ and for each considered space velocity $v_{\star}$:
one for the main sequence and one for the red supergiant phase. Simulations are
launched at $5$ and $3\, \rm Myr$ of the main sequence phase for the $10$ and $20\,
\mathrm{M}_{\odot}$ models, and at the zero-age main-sequence for the
$40\, \mathrm{M}_{\odot}$ star, given its short lifetime (see black circles in
Figs.~\ref{fig:hrd} and~\ref{fig:models_ms_rsg}). Red supergiant simulations are
started before the main sequence to red supergiant transition such that a steady
state has been reached when the red supergiant wind begins to expand (see black squares in
Figs.~\ref{fig:hrd} and~\ref{fig:models_ms_rsg}).

The wind material is traced using a scalar marker whose value $Q$ obeys the linear advection equation,
\begin{equation}
	\frac{\partial (\rho Q) }{\partial t } +  \bmath{ \nabla } \cdot  ( \bmath{v} \rho Q) = 0.
\label{eq:tracer}
\end{equation}
This tracer is passively advected with the fluid, allowing us to distinguish between the wind and ISM material.
Its value is set to $Q(\bmath{r})=1$ for the inflowing wind material and to $Q(\bmath{r})=0$ for the ISM material,
where $\bmath{r}$ is the vector position of a given cell of the simulation domain.


\subsection{Interstellar medium}
\label{subsect:ism}

We consider homogeneous and laminar ISM with $n_{\rm H}=0.57\,
\mathrm{cm}^{-3}$, which is typical of the warm neutral medium in the Galactic
plane~\citep{wolfire_apj_587_2003} from where most of runaway massive stars are
ejected. The initial ISM gas velocity is set to $v_{\mathrm{ISM}}=-v_{\star}$.

The photosphere of a main sequence star releases a large flux of hydrogen ionizing photons
$S_{\star}$, that depends on $R_{\star}$ and $T_{\rm eff}$, which allows
us to estimate $S_{\star}=10^{45}\,\rm photon\, \rm s^{-1}$ ($T_{\rm eff}\approx
2.52\times 10^{4}\, \rm K$), $S_{\star}=10^{48}\,\rm photon\, \rm s^{-1}$
($T_{\rm eff}\approx 3.39\times 10^{4}\, \rm K$) and $S_{\star}=10^{49}\,\rm
photon\, \rm s^{-1}$ ($T_{\rm eff}\approx 4.25\times 10^{4}\, \rm K$) for the
$10$, $20$ and $40\, \mathrm{M}_{\odot}$ stars, respectively
~\citep{diazmiller_apj_501_1998}.  These fluxes produce a Str\"omgren sphere
of radius,
\begin{equation}
R_{\rm S} = \Big( \frac{ 3 S_{\star} }{ 4\pi n^{2}\alpha_{\rm rr}^{\rm B} } \Big)^{1/3},
\end{equation}
where $\alpha_{\rm rr}^{\rm B}$ is the case B recombination rate of $\rm H^{+}$,
fitted from~\citet{hummer_mnras_268_1994}. The Str\"omgren sphere is distorted
by the bulk motion of the star in an egg-shaped $\HII$
region~\citep{raga_apj_300_1086, raga_rmxaa_33_1997, mackey_mnras_2013}. 
$R_{\rm S} \approx4.3$, $43$ and $94\, \rm pc$ for the $10$, $20$ and $40\,
\mathrm{M}_{\odot}$ main sequence stars, respectively. $R_{\rm S}$ is larger
than the typical scale of a stellar bow shock (i.e. larger than the full size
of the computational domain of $\sim \rm pc$). Because of this, we treated the
plasma on the full simulation domain as photoionized with the corresponding
dissipative processes (see panel (a) of Fig.~\ref{fig:coolingcurves}), \textcolor{black}{i.e., we neglect 
the possiblity that a dense circumstellar structure could trap the 
stellar radiation field~\citep{weaver_apj_218_1977}}. We
consider that both the wind and the ISM are fully ionized until the end of the
main sequence, and we use an initial $T_{\mathrm{ISM}}\approx 8000\, \mathrm{K}$
which is the equilibrium temperature of the photoionized cooling curve (see
panel (a) of Fig.~\ref{fig:coolingcurves}).

In the case of models without an ionizing radiation field, involving a phase
transition or a red supergiant, the plasma is assumed to be in collisional
ionization equilibrium (CIE). We adopt $T_{\mathrm{ISM}} \approx 3300\,
\mathrm{K}$, which corresponds to the equilibrium temperature of the CIE cooling
curve for the adopted ISM density (see panel (b) of Fig.~\ref{fig:coolingcurves}).


\subsection{Radiative losses and heating}
\label{subsect:radlosses}

\begin{figure*}
	\begin{minipage}[b]{0.48\textwidth}
                \centering 
		\includegraphics[width=0.7\textwidth,angle=270]{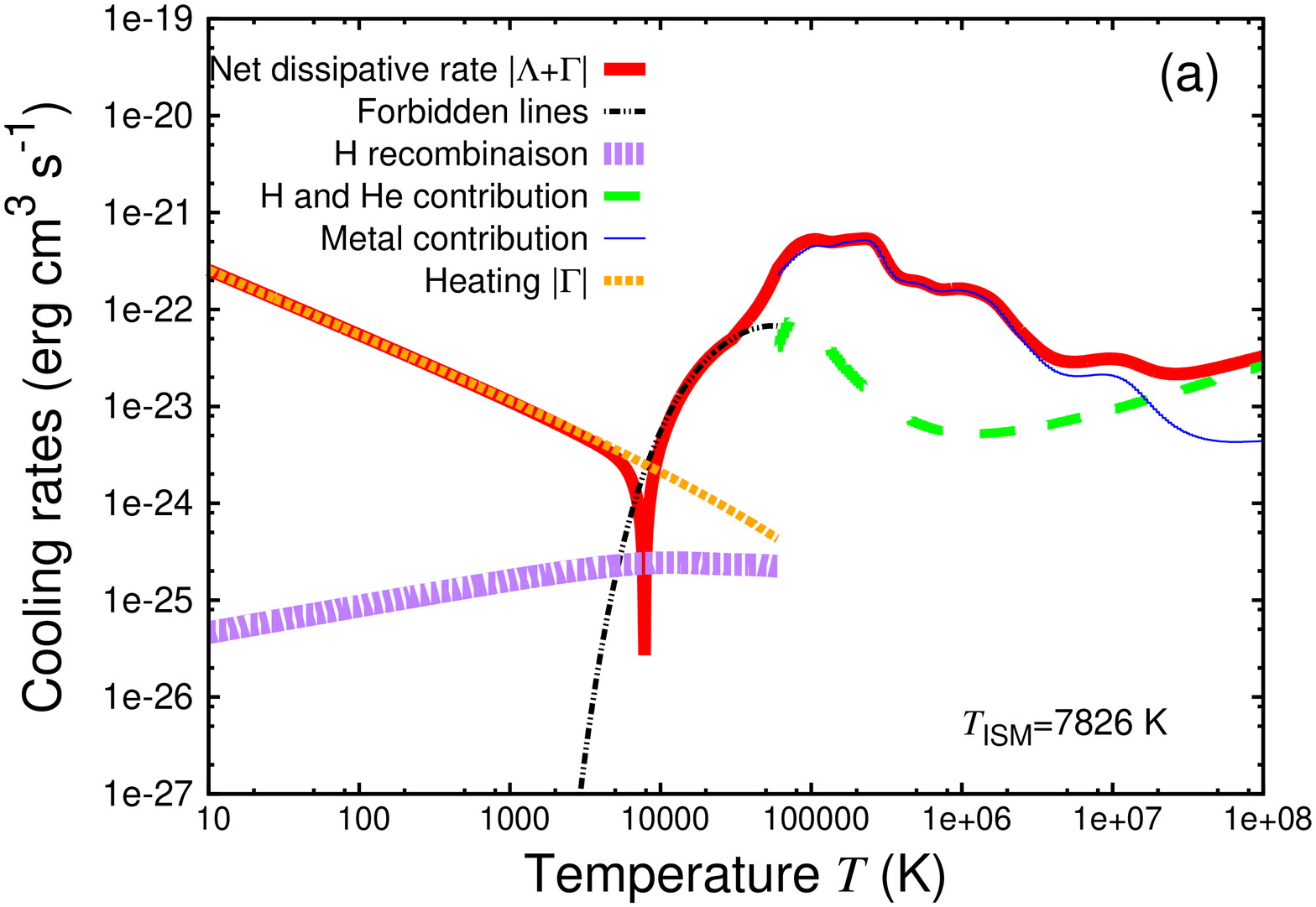} 
   	\end{minipage}
	\begin{minipage}[b]{0.48\textwidth}
                \centering 
		\includegraphics[width=0.7\textwidth,angle=270]{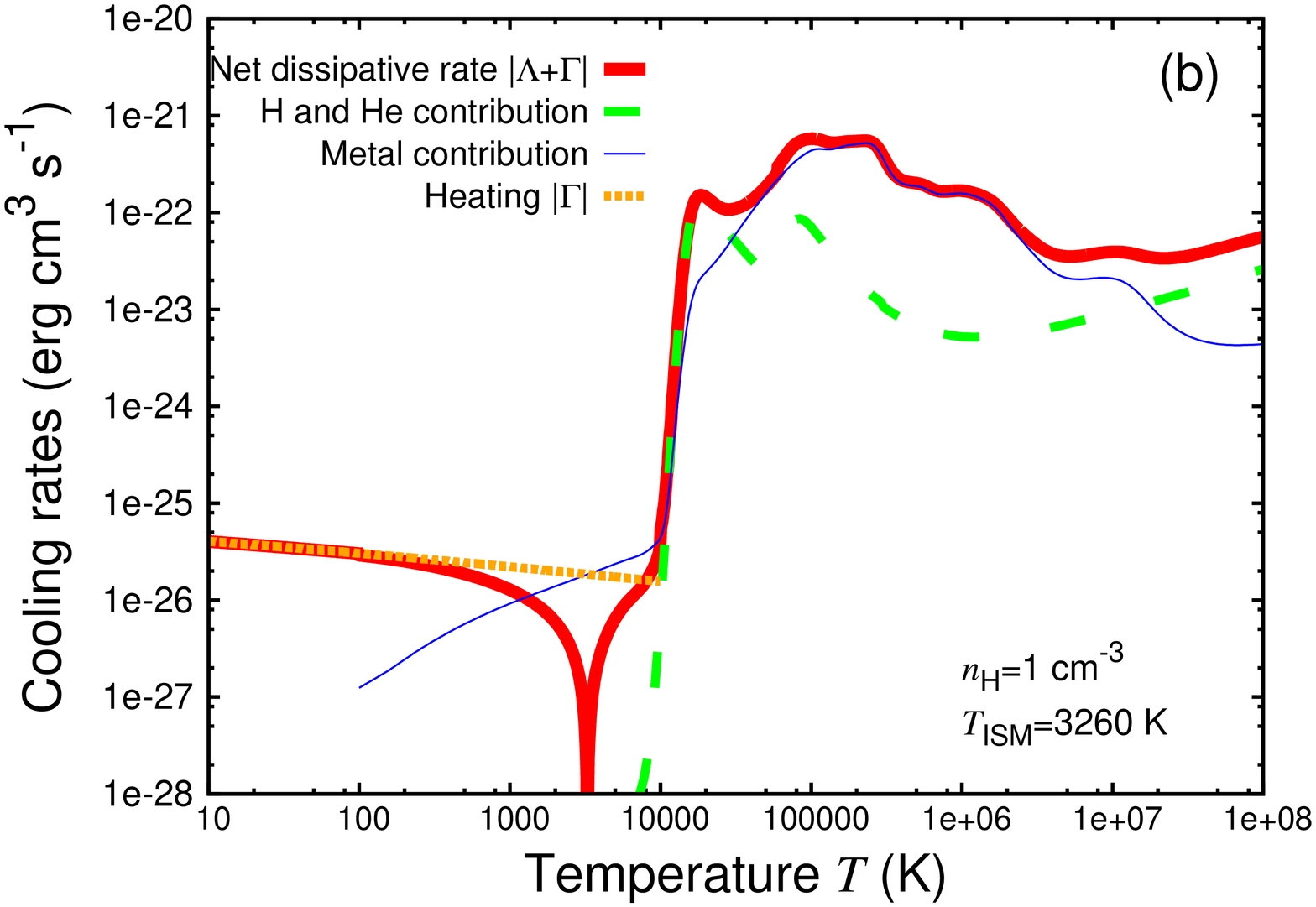}  
	\end{minipage}
	\caption{Cooling and heating rates as a function of temperature for photoionized (a) and collisional ionization equilibrium (b) medium. 
		 The solid thick red line is the curve representing the net rate of emitted energy, i.e. the
		 absolute value of the sum of the luminosity due to cooling $\it \Lambda$ and heating $\it \Gamma$. 
		 Dotted and thin lines correspond to the different processes the model takes 
		 into account: emission from forbidden lines (dotted dashed thin black), H recombination lines (dotted thick purple), 
		 hydrogen and helium (dashed green) and metals (solid thin blue) as well as the heating rate $\it \Gamma$ (dotted orange). 
		 All luminosity from the different coolants and heating rate of processes are presented for $n_{\rm H}=1\, \rm cm^{-3}$, 
		 within their range of interest. 
	         The $x$-axis represents temperature (in $\mathrm{K}$) and the $y$-axis the emitted energy (in $\rm erg\, s^{-1}\, \rm cm^{3}$). }
        \label{fig:coolingcurves}
\end{figure*}

A cooling curve for photoionized material has been implemented, whereas another
assuming CIE is used for the gas that is not exposed to ionizing
radiation. In terms of Eq.~(\ref{eq:dissipation}), we set $\alpha=2$ for photoionized gases and
$\alpha=1$ for the CIE medium. 
The cooling component $\it \Lambda$ of Eq.~(\ref{eq:dissipation}) is,
\begin{equation}
      \itl{ \Lambda   = \Lambda_{\mathrm{H+He}} + \Lambda_{\mathrm{Z}} 
		+ \Lambda_{\mathrm{RR}}   + \Lambda_{\mathrm{FL}} },
      \label{eq:coolants}
\end{equation}
where $\it \Lambda_{\mathrm{H+He}}$ and $\it \Lambda_{\mathrm{Z}}$ represent the
cooling from hydrogen plus helium, and metals $Z$ respectively~\citep{wiersma_mnras_393_2009}
for a medium with the solar helium abundance
$\itl{\chi}_{\mathrm{He}}=n_{\mathrm{He}}/n_{\mathrm{H}}=0.097$~\citep{
asplund_araa_47_2009}. $\itl \Lambda_{\mathrm{H+He}}+ \itl  \Lambda_{\mathrm{Z}}$ dominates
the cooling at high $T$ (see panel (a) of Fig.~\ref{fig:coolingcurves}). A
cooling term for hydrogen recombination $\it \Lambda_{\mathrm{RR}}$ is obtained
by fitting the case B energy loss coefficient
$\beta_{B}$~\citep{hummer_mnras_268_1994}. The rate of change of $E$ is also
affected by collisionally excited forbidden lines from elements heavier than
helium, e.g. oxygen and carbon~\citep{raga_apjs_109_1997}. The corresponding
cooling term $\it \Lambda_{\mathrm{FL}}$ is adapted from a fit of [O\,{\sc ii}]
and [O\,{\sc iii}] lines (see Eq. A9 of Henney et al., 2009) with the abundance
of $n_{\mathrm{O}}/n_{\mathrm{H}}=4.89\times
10^{-4}$~\citep{asplund_araa_47_2009}.

The heating rate $\itl{ \Gamma}_{2}$ in Eq.~(\ref{eq:dissipation}) represents
the effect of photons emitted by the hot stars ionizing the recombining $\rm
H^{+}$ ions and liberating energetic electrons. It is calculated as the energy
of an ionizing photon after subtracting the reionization potential of an
hydrogen atom, i.e. $5\, \mathrm{eV}$ for a typical main sequence
star~\citep{osterbrock_1989}, weighted by $\alpha_{\mathrm{rr}}^{\mathrm{B}}$. 

At low temperatures ($T< 6\times 10^{4}\, \mathrm{K}$), the cooling rate is the
sum of all terms $\it \Lambda_{\mathrm{H+He}}$, $\it \Lambda_{\mathrm{Z}}$, $\it
\Lambda_{\mathrm{RR}}$ and $\it \Lambda_{\mathrm{FL}}$, whereas for higher
temperatures ($T> 6\times 10^{4}\, \mathrm{K}$) only the ones for hydrogen,
helium and $Z$ are used. The two parts of the curve are linearly interpolated in
the range of $4.5\times 10^{4} < T < 6.0\times 10^{4}\, \mathrm{K}$. \\

The CIE cooling curve (see panel (b) of Fig.~\ref{fig:coolingcurves}) also
assumes solar abundances~\citep{wiersma_mnras_393_2009} for hydrogen, helium
and $Z$. The heating term $\itl{ \Gamma}_{1}$ represents the photoelectric
heating of dust grains by the Galactic far-UV background. For $T \le 1000\,
\mathrm{K}$, we used equation \,C5 of~\citet{wolfire_apj_587_2003}. We impose a
low temperature ($T<1000\, \mathrm{K}$) electron number density profile $n_{\rm
e}$ using eq.\,C3 of~\citet{wolfire_apj_587_2003}. For $T>1000\, \mathrm{K}$ we
take the value of $n_{\rm e}$ interpolated from the CIE curve
by~\citet{wiersma_mnras_393_2009}.

\begin{figure}
          \centering 
	  \includegraphics[width=0.34\textwidth,angle=270]{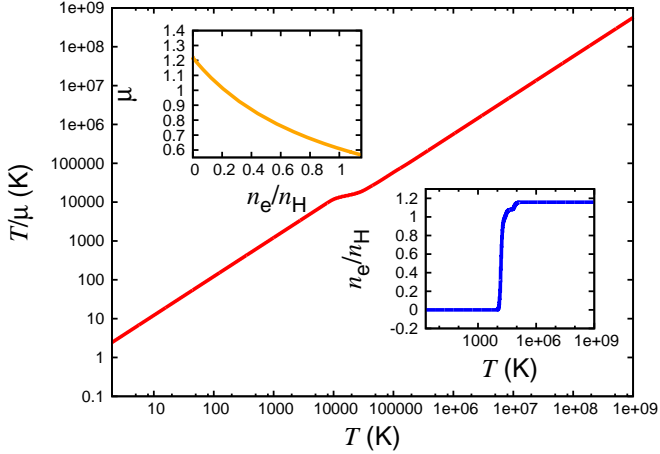}    
	  \caption{ Temperature divided by the mean mass per particle $T/\mu$ (in $\mathrm{K}$) 
		    as a function of temperature $T$ (in $\mathrm{K}$). Data are derived
                    from the collisional ionization equilibrium cooling curves~\citep{wiersma_mnras_393_2009}. 
		    Top inset: $\mu$ as a function of electron fraction $n_\mathrm{e}/n_\mathrm{H}$. 
		    Bottom inset: $n_\mathrm{e}/n_\mathrm{H}$ as a function of temperature $T$. }
          \label{fig:ionfrac}
\end{figure}

A transition between the main sequence and the red supergiant phases requires a
transition between photoionized and CIE medium. At the beginning of the red
supergiant phase, our model ceases to consider the dissipation and heating for
photoionized medium and adopts the ones assuming CIE medium. The assumption
of CIE specifies $n_{\mathrm{e}}/n_{\mathrm{H}}$ as a function of
$T$~\citep{wiersma_mnras_393_2009}. The mean mass per particle is calculated as, 

\begin{equation}
	\mu(T) = \frac{ 1 + 4 \itl{\chi}_{\mathrm{He} } }{ ( 1 + \itl{\chi}_{ \mathrm{He} } )[ 1+x(T) ] },
        \label{eq:mu}
\end{equation} 
where, 
\begin{equation}
	x(T) = \Big( \frac{ n_{\mathrm{e}} }{ n_{\mathrm{H}} } \Big)_{T} / \Big( \frac{ n_{\mathrm{e}} }{ n_{\mathrm{H}} } \Big)_{T_\mathrm{max}} ,
        \label{eq:x}
\end{equation}
and $T_{\rm max}$ is the upper limit of the cooling curve temperature
range~\citep{wiersma_mnras_393_2009}, and $x(T)$ is a quantity monotonically
increasing with $T$, that gives the degree of ionization of the medium (see top
inset in Fig.~\ref{fig:ionfrac}). We then have an expression for $\mu$ with low and
high $T$ limits of $\mu=1.27$ and $\mu=0.61$ for neutral and fully ionized
medium, respectively~\citep[e.g.][]{lequeux_edp}. For simulations assuming CIE
we then obtain through $\mu(T)$ a one-to-one correspondence between
$T/\mu\propto p/\rho$ (known) and $T$ (required) for each cell of the
computational domain.     


\subsection{Thermal conduction}
\label{subsect:tc}

The circumstellar medium around runaway main sequence stars presents large
temperature gradients across its shocks and discontinuities (e.g. $\itl{
\Delta} T\approx 10^{7}\, \rm K$ at the reverse shock of the models for the $20$
and $40\, M_{\odot}$ stars), which drive the heat flux~\citep{spitzer_1962,
cowie_apj_211_1977}. Electrons move quickly enough to transfer energy to the
adjacent low temperature gas. The consequent equilibration of the pressure smooths the
density profiles at the discontinuity between the wind and ISM
material~\citep{weaver_apj_218_1977}.

Heat conduction is included in our models over the whole computational domain.
For the models with partially neutral gas, e.g. during a phase transition or for
models involving a red supergiant, $\bmath{ F}_{\rm c}$ is calculated at
$T<1000\, \rm K$ with $n_{\rm e}$  from eq.\,C3 in~\citet{wolfire_apj_587_2003}.
Our study does not consider either the stellar or interstellar magnetic field,
which make the heat conduction anisotropic~\citep{orlando_apj_678_2008}.


\subsection{Relevant characteristic quantities of a stellar wind bow shock}
\label{subsect:shockproperties}

A stellar wind bow shock generally has four distinct regions: the
unperturbed ISM, the shocked ISM, the shocked wind material and the
freely-expanding wind. The shocked materials are separated by a contact
discontinuity, the expanding wind from the star is separated from shocked wind by the reverse shock and the
structure's outermost border is marked by the forward
shock~\citep[e.g.][]{vanburen_aspc_35_1993}.

The stand-off distance of the bow shock is,
\begin{equation}
	R(0) = \sqrt{ \frac{\dot{M}v_{\mathrm{w}}}{4\pi\rho_{\mathrm{ISM}}v_{\star}^{2} } }
\label{eq:Ro}
\end{equation}
~\citep{baranov_sphd_15_1971}.
The analytical approximation for the shape of an infinitely thin bow shock is,
\begin{equation}
	R({\theta}) = R(0) \mathrm{cosec}(\theta \frac{\pi}{180}) \sqrt{ 3(1-\theta \frac{\pi}{180} )\mathrm{cotan}(\theta \frac{\pi}{180} ) },
\label{eq:wilkin}
\end{equation}
where $\theta$ is the angle from the direction of motion in degrees and $R(0)$ is given by Eq.~(\ref{eq:Ro}).

\textcolor{black}{ The dynamical timescale of a layer constituting a stellar wind bow shock 
is equal to the time a fluid element spends in it before it is advected 
downstream, } 
\begin{equation}
t_{\mathrm{dyn}}=\frac{ \Delta z }{ v }, 
\label{eq:time_dyn}
\end{equation}
\textcolor{black}{ where $\Delta z$ is the thickness of the layer along the $Oz$ direction and 
$v$ is a characteristic velocity of the gas in the considered region, i.e. the post-shock velocity
$v \simeq v_{\rm w}/4$ in the shocked wind or $v \simeq v_{\star}/4$ in the shocked ISM.} 
The gas density and pressure govern the cooling timescale,
\begin{equation}
t_{\mathrm{cool}} =  \frac{E}{\dot{E}}  = \frac{p}{(\gamma-1)\itl{\Lambda}(T) n_{\rm H}^{2}}. 
\label{eq:time_cool}
\end{equation}
These two timescales determine whether a shock is adiabatic ($ t_{\mathrm{dyn}} \ll t_{\mathrm{cool}} $) 
or radiative ($ t_{\mathrm{dyn}} \gg t_{\mathrm{cool}} $).


\subsection{Presentation of the simulations}
\label{subsect:models}

The parameters used in our simulations are gathered together with information
concerning the size of the computational domain in Table~\ref{tab:ms}. The size
of the computational domain is inspired by~\citet{comeron_aa_338_1998}, i.e. we
use a sufficient number of cells $N_{\rm R}$ to adequately resolve the
substructures of each bow shock in the direction of the stellar motion. As
$v_{\star}$ increases, the bow shock and the domain size decreases, so the
spatial resolution $\it \Delta=R_{\rm max}/N_{\rm R}$ also decreases. The
dimensions of the domain are chosen such that the tail of the bow shock only
crosses the $\it z=\it z_{\rm min}$ boundary, but never intercepts the outer
radial border at $\it R=\it R_{\rm max}$ to avoid numerical boundary effects.

We model bow shocks for a space velocity $20\, \le v_{\star} \le 70\, \rm km\,
\rm s^{-1}$, since these include the most probable space velocities of runaway
stars and ranges from supersonic to hypersonic~\citep{eldridge_mnras_414_2011}.
For the bow shocks of main sequence stars the label is MS, and the models for
the red supergiant phase are labelled with the prefix RSG. In our nomenclature,
the four digits following the prefix of a model indicate the zero age main
sequence mass (first two digits) and the space velocity (next two digits).

Simulations of bow shocks involving a main sequence star are started at a time
$t_{\rm start}$ in the middle of their stellar evolutionary phase in order to
model bow shocks with roughly constant wind properties. The distortion of the
initially spherically expanding bubble into a steady bow shock takes up to
$\approx 16\, t_{\rm cross}$, where $t_{\rm cross} = R(0)/v_{\star}$ is the bow
shock crossing-time. We stop the simulations at least $32\, t_{\rm cross}$ after
the beginning of the integration, except for model MS4020 for which such a time
is larger than the main sequence time.

\begin{table*}
	\centering
	\caption{Nomenclature and grid parameters used in our hydrodynamical simulations. 
	 Parameters $\it \Delta$, $R_{\mathrm{max}}$ and $z_{\mathrm{min}}$ are 
	 the resolution of the uniform grid (in $\mathrm{pc}\, \mathrm{{cell}^{-1}}$) and respectively
	 the upper and lower limits of the domain along the $x$-axis and $y$-axis (in $\mathrm{pc}$). 
	 $N_{\mathrm{R}}$ and $N_{\mathrm{z}}$ are the number of cells discretising the corresponding directions. 
	 The two last columns contain the starting time $t_\mathrm{start}$ of the simulations relative to the 
	 zero-age main-sequence and the crossing time $\, t_{\rm cross}$ 
	 of the gas because of the stellar motion for each associated bow shock (in $\mathrm{Myr}$).}
	\begin{tabular}{cccccccccc}
	\hline
	\hline
	${\rm {Model}}$ &   $M_{\star}\, (M_{\odot})$                              
 			&   $v_{\star}\, (\mathrm{km}\, \mathrm{s}^{-1})$
			&   $\itl{\Delta}\, (\mathrm{pc}\, \mathrm{cell}^{-1})$ 
			&   $z_{\mathrm{min}}\, (\mathrm{pc})$ 
			&   $R_{\mathrm{max}}\, (\mathrm{pc})$
			&   $N_{\rm R}$  
			&   $N_{\rm z}$
			&   $t_{\mathrm{start}}\, (\mathrm{Myr})$
			&   $t_{\mathrm{cross}}\, (\mathrm{Myr})$ 
			\\ \hline   
	MS1020   &  $10$  &  $20$  &  $1.7\times 10^{-3}$   &  $-0.5$    &  $1.0$  &  $600$   & $600$   & $5.0$   &  $6.3\times 10^{-2}$     \\             
	MS1040   &  $10$  &  $40$  &  $5.7\times 10^{-4}$   &  $-0.2$    &  $0.4$  &  $700$   & $700$   & $5.0$   &  $1.6\times 10^{-3}$     \\        
	MS1070   &  $10$  &  $70$  &  $3.3\times 10^{-4}$   &  $-0.1$    &  $0.2$  &  $600$   & $600$   & $5.0$   &  $4.9\times 10^{-4}$     \\ 
	\hline 
	RSG1020  &  $10$  &  $20$  &  $8.0\times 10^{-3}$   &  $-2.00$   &  $4.0$  &  $500$  & $500$  & $22.62$   &  $1.5 \times 10^{-2}$      \\             
	RSG1040  &  $10$  &  $40$  &  $2.3\times 10^{-3}$   &  $-0.7$    &  $1.4$  &  $600$  & $600$  & $22.78$   &  $5.5 \times 10^{-3}$      \\        
	RSG1070  &  $10$  &  $70$  &  $7.5\times 10^{-4}$   &  $-0.3$    &  $0.6$  &  $800$  & $800$  & $22.86$   &  $2.1 \times 10^{-3}$      \\  
	\hline   
	MS2020   &  $20$  &  $20$  &  $2.0\times 10^{-2}$   &  $-5.0$    &  $10.0$   &  $500$  &  $500$  & $3.0$  &  $7.0\times 10^{-2}$     \\         
	MS2040   &  $20$  &  $40$  &  $8.3\times 10^{-3}$   &  $-2.5$    &  $5.0$    &  $600$  &  $600$  & $3.0$  &  $1.7\times 10^{-2}$     \\     
	MS2070   &  $20$  &  $70$  &  $2.1\times 10^{-3}$   &  $-0.75$   &  $1.5$    &  $700$  &  $700$  & $3.0$  &  $5.4\times 10^{-3}$     \\
	\hline 
	RSG2020  &  $20$  &  $20$  &  $1.0\times 10^{-2}$    &  $-5.0$    &  $15.0$  &  $1000$ & $1500$  & $8.0$  &  $6.8 \times 10^{-2}$    \\         
	RSG2040  &  $20$  &  $40$  &  $7.5\times 10^{-3}$    &  $-3.0$    &  $6.0$   &  $800$  &  $800$  & $8.0$  &  $1.6 \times 10^{-2}$    \\     
	RSG2070  &  $20$  &  $70$  &  $6.7\times 10^{-3}$    &  $-2.0$    &  $4.0$   &  $600$  &  $600$  & $8.0$  &  $4.4 \times 10^{-3}$    \\
	\hline 
	MS4020   &  $40$  &  $20$  &  $6.0\times 10^{-2}$   &  $-15.0$   & $30.0$   &  $500$  &  $500$  & $0.0$  &  $2.8 \times 10^{-1}$      \\              
	MS4040   &  $40$  &  $40$  &  $2.7\times 10^{-2}$   &  $-8.0$    & $16.0$   &  $600$  &  $600$  & $0.0$  &  $7.1 \times 10^{-2}$      \\   
	MS4070   &  $40$  &  $70$  &  $1.1\times 10^{-2}$   &  $-4.0$    &  $8.0$   &  $700$  &  $700$  & $0.0$  &  $2.5 \times 10^{-2}$      \\    
	\hline 
	\end{tabular}
\label{tab:ms}
\end{table*}


\section{The main sequence phase}
\label{sect:result_ms}


\subsection{Physical characteristics of the bow shocks}
\label{subsect:introduction_models_ms}

\begin{figure}
	\begin{minipage}[b]{0.48\textwidth}
		\includegraphics[width=1.0\textwidth]{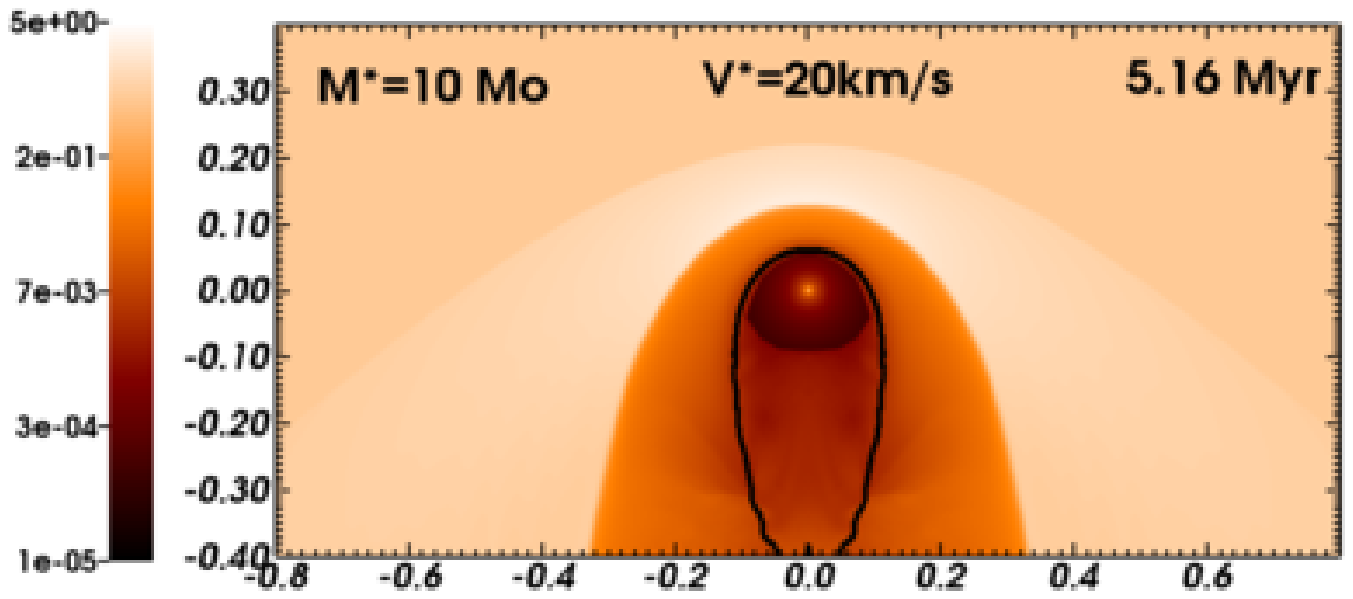}
	\end{minipage} \\
	\begin{minipage}[b]{0.48\textwidth}
		\includegraphics[width=1.0\textwidth]{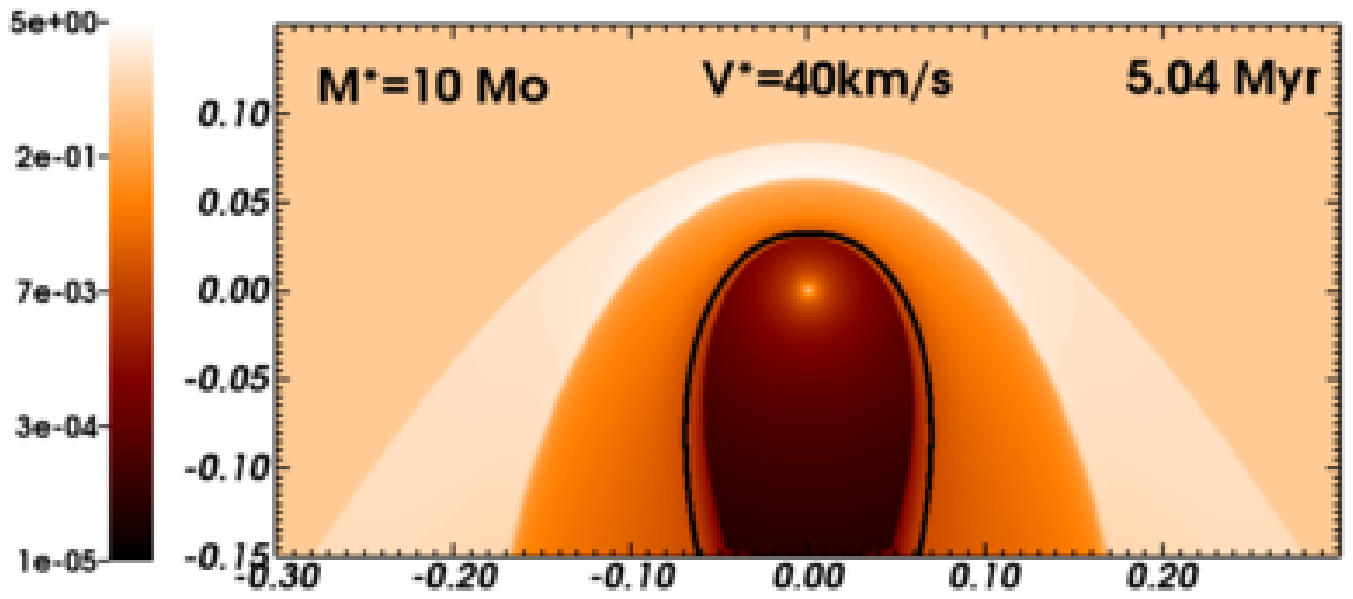}
	\end{minipage}\\
	\begin{minipage}[b]{0.48\textwidth}
		\includegraphics[width=1.0\textwidth]{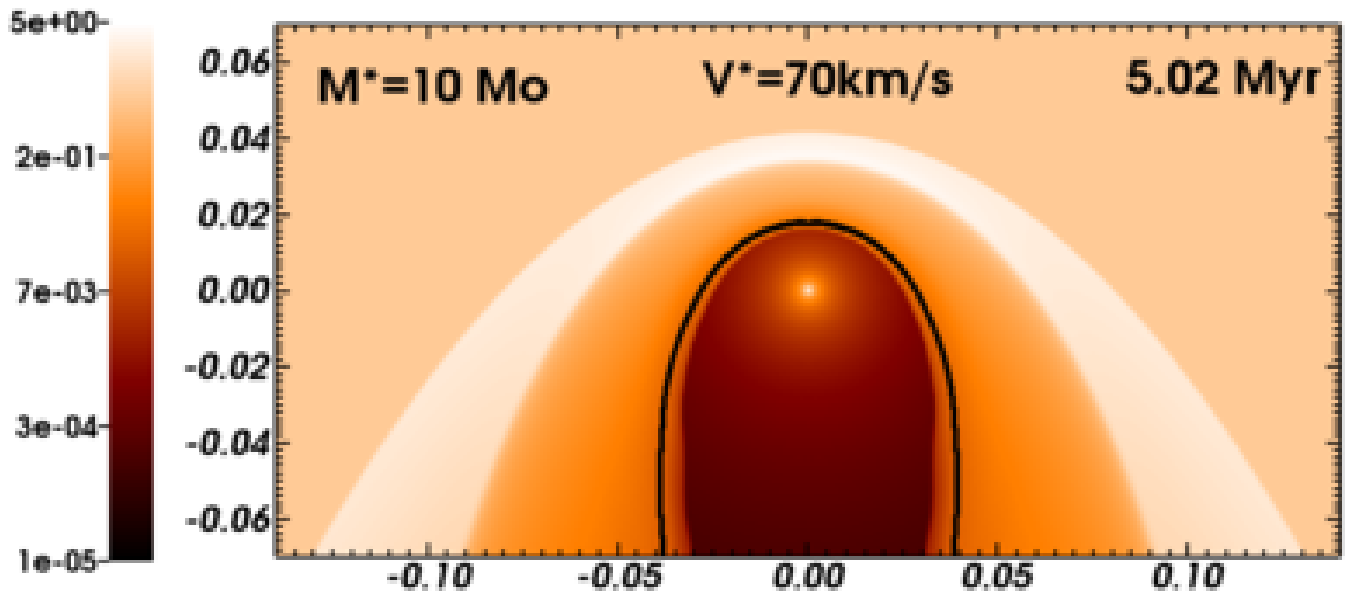}
	\end{minipage}
	\caption{Grid of stellar wind bow shocks from the main sequence phase of the $10\, M_{\odot}$ initial mass star as a function 
		 of the space velocity with respect to the ISM, with $20\, \mathrm{km}\, \mathrm{s}^{-1}$ (top panel), 
		 $40\, \mathrm{km}\, \mathrm{s}^{-1}$ (middle panel) and $70\, \mathrm{km}\, \mathrm{s}^{-1}$ (bottom panel).
		 The nomenclature of the models follows Table~\ref{tab:ms}.  
		 The gas number density is shown with a density range from $10^{-5}$ to $5.0\, \mathrm{cm}^{-3}$ in the 
		 logarithmic scale. 
		 The solid black contour traces the boundary between wind and ISM material $Q(\bmath{r})=1/2$.
		 The $x$-axis represents the radial direction and the $y$-axis the direction of stellar motion (in $\mathrm{pc}$).
		 Only part of the computational domain is shown.  }
	\label{fig:m10ms}  
\end{figure}

\begin{figure}
	\begin{minipage}[b]{0.48\textwidth}
		\includegraphics[width=1.0\textwidth]{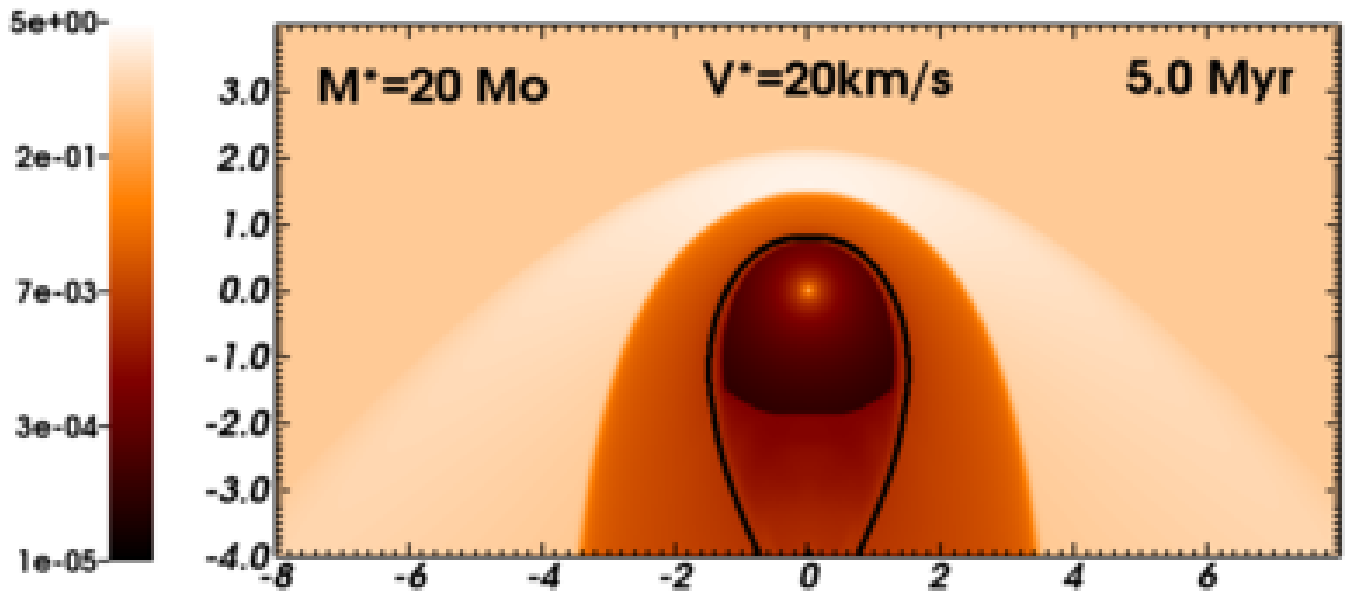}
	\end{minipage}  \\
	\begin{minipage}[b]{0.48\textwidth}
		\includegraphics[width=1.0\textwidth]{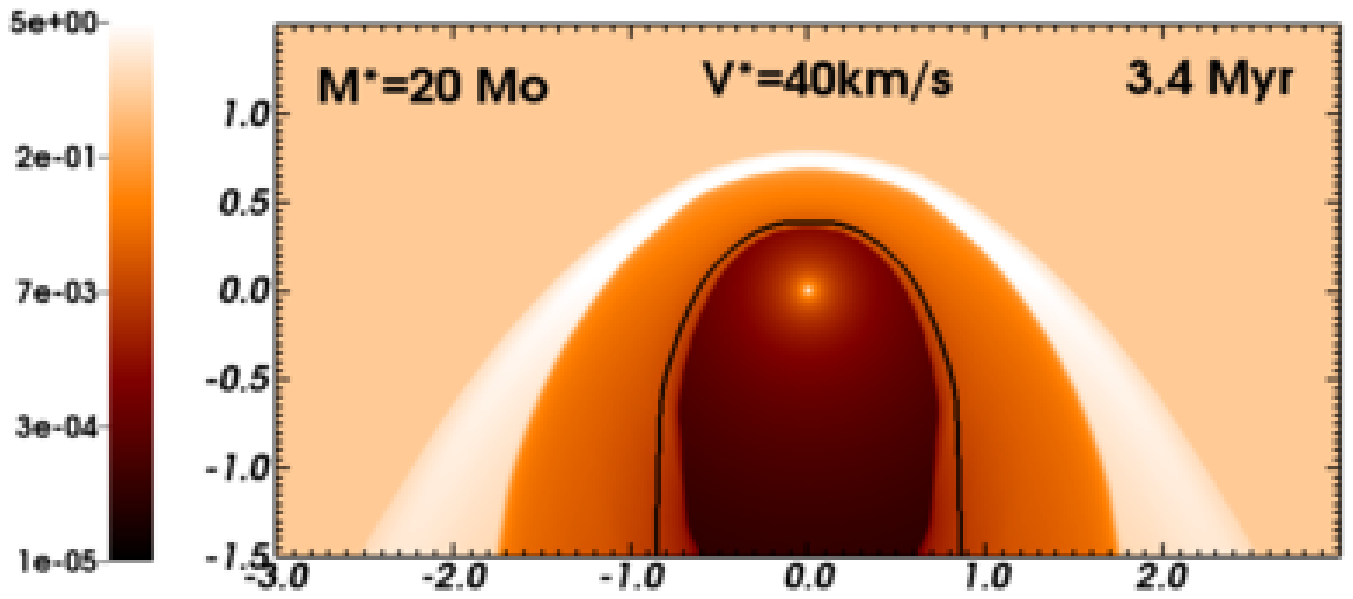}
	\end{minipage} \\
	\begin{minipage}[b]{0.48\textwidth}
		\includegraphics[width=1.0\textwidth]{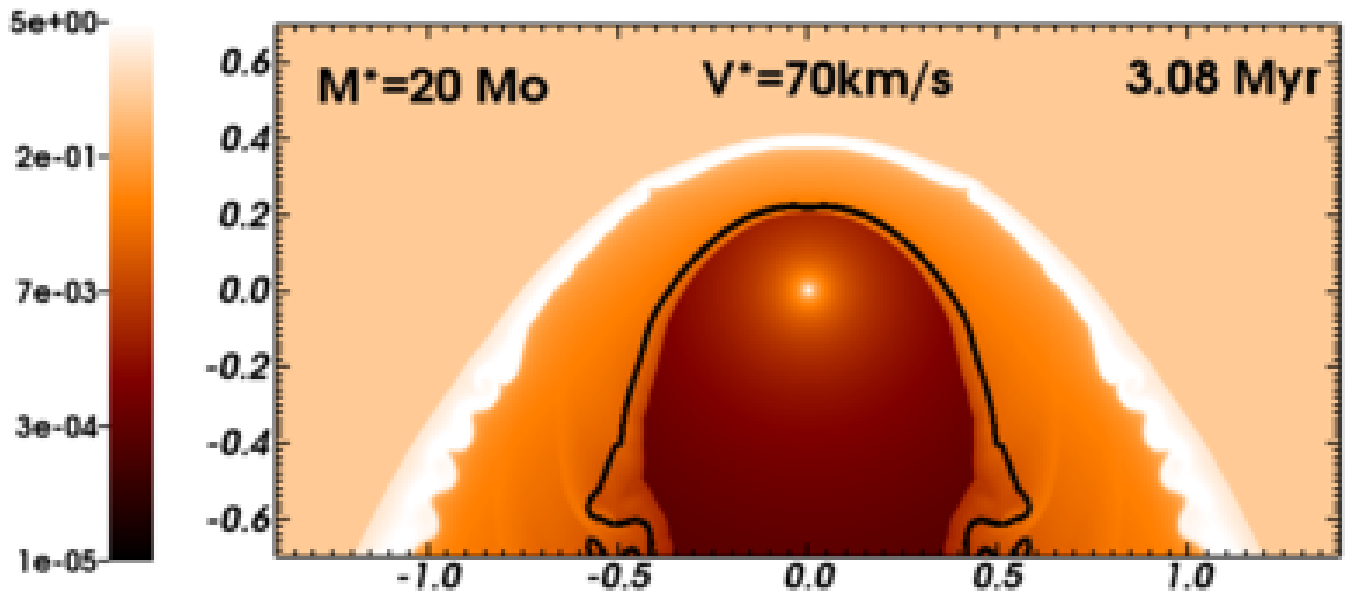}
	\end{minipage}
	\caption{As Fig.~\ref{fig:m10ms}, with an initial stellar mass of $20\, M_{\odot}$.  }
	\label{fig:m20ms}  
\end{figure}

\begin{figure}
	\begin{minipage}[b]{0.48\textwidth}
		\includegraphics[width=1.0\textwidth]{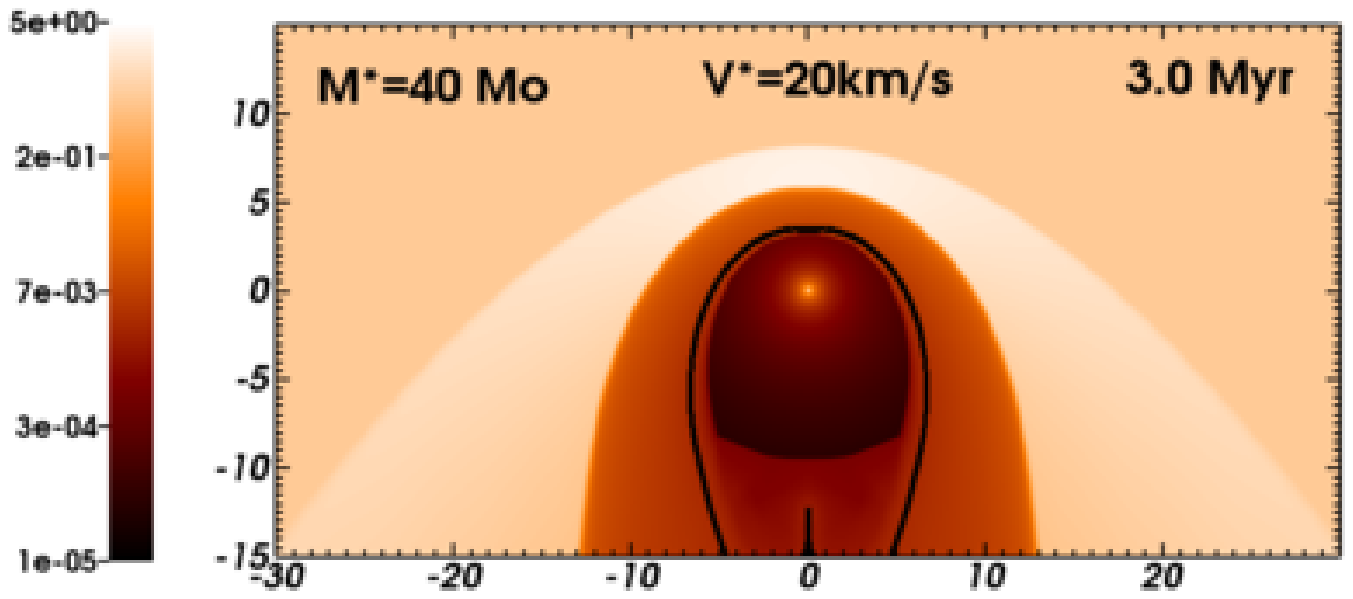}
	\end{minipage}
	\begin{minipage}[b]{0.48\textwidth}
		\includegraphics[width=1.0\textwidth]{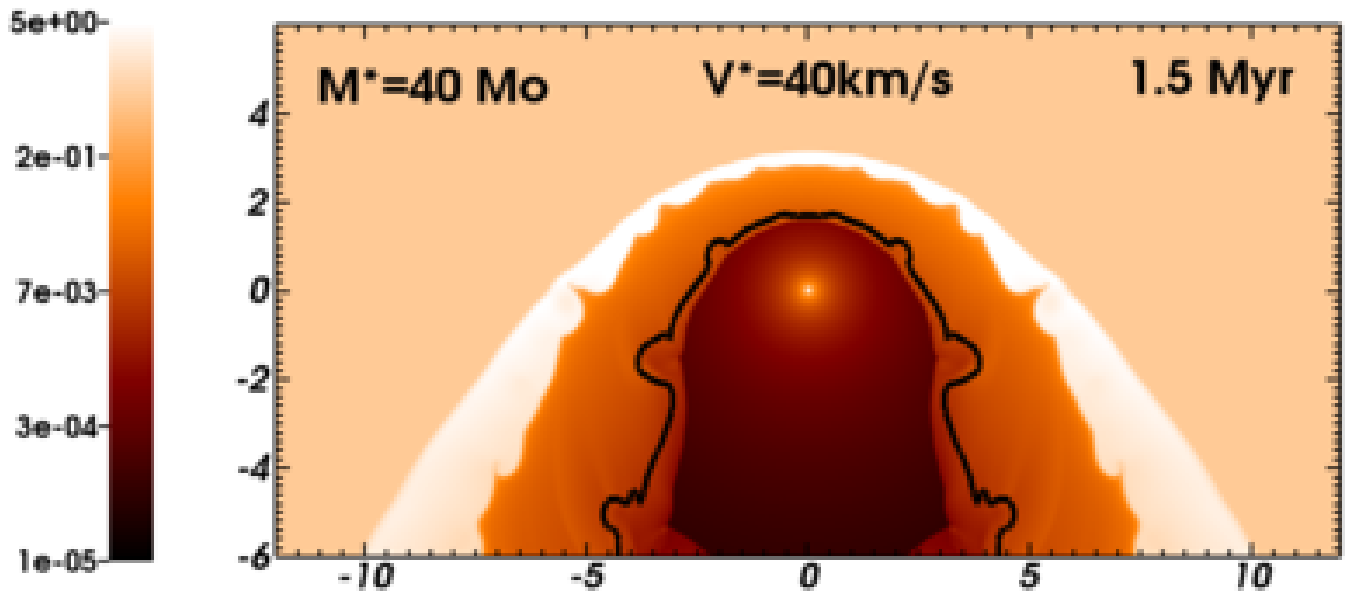}
	\end{minipage}
	\begin{minipage}[b]{0.48\textwidth}
		\includegraphics[width=1.0\textwidth]{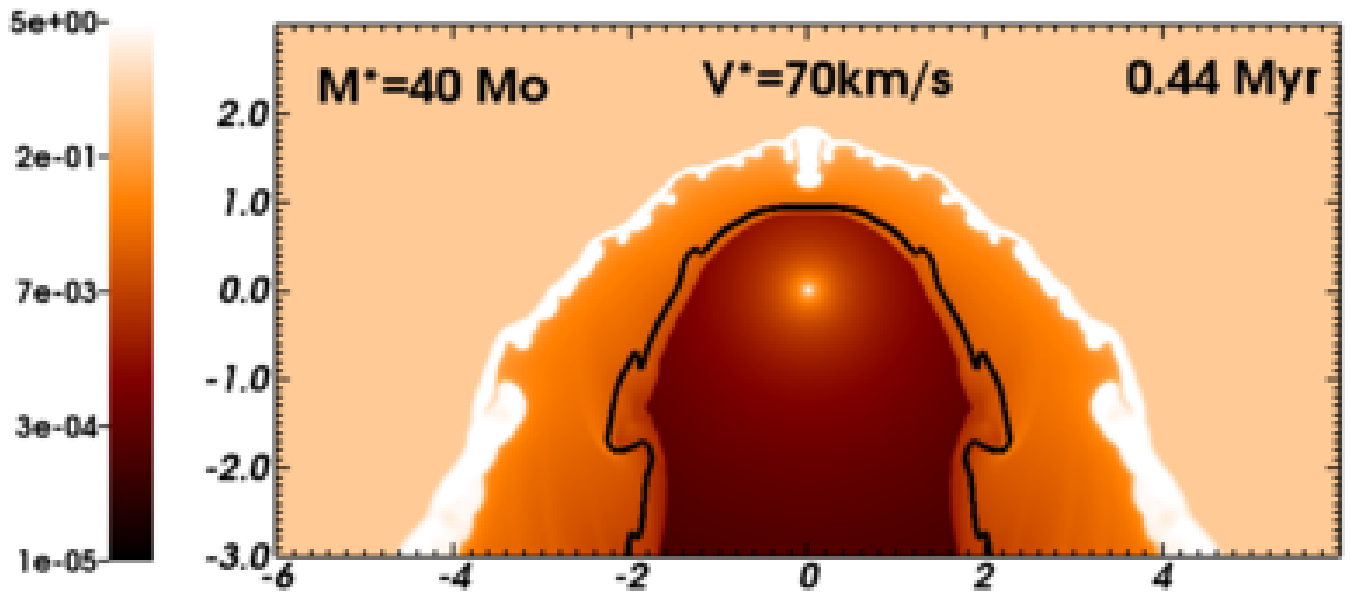}
	\end{minipage}
	\caption{As Fig.~\ref{fig:m10ms}, with an initial stellar mass of $40\, M_{\odot}$.  }
	\label{fig:m40ms}  
\end{figure}

We show the gas density field in our bow shock models of the 
main sequence phase MS1020 ($10\, M_{\odot}$ initial stellar mass, 
$v_{\star}=20\, \mathrm{km}\, \mathrm{s}^{-1}$, upper panel), 
MS1040 ($10\, M_{\odot}$, $40\, \mathrm{km}\, \mathrm{s}^{-1}$, middle panel)
and MS1070 ($10\, M_{\odot}$, $70\, \mathrm{km}\, \mathrm{s}^{-1}$, lower panel)
in Fig.~\ref{fig:m10ms}. Figs.~\ref{fig:m20ms} and~\ref{fig:m40ms} are similar for 
the $20\, M_{\odot}$ and $40\, M_{\odot}$ initial mass stars.
The figures correspond to a time $\approx t_{\rm
start}+32\, t_{\rm cross}$. The model MS4020 has a lifetime $<32\, t_{\rm
cross}$ (see panels (c) and (f) of Fig.~\ref{fig:models_ms_rsg}), and is
therefore shown at a time $\approx 16\, t_{\rm cross}$. In Figs.~\ref{fig:m10ms}
to~\ref{fig:m40ms} the overplotted solid black line is the material
discontinuity, i.e. the border between the wind and ISM gas where the value 
of the material tracer $Q(\bmath{r})=1/2$. The bow shock morphological characteristics such as the
stand-off distance and the axis ratio $R(0)/R(90)$ measured from the simulations
are summarised in Table~\ref{tab:params}.

The theory of~\citet{baranov_sphd_15_1971} predicts that $R(0) \propto
v_{\star}^{-1}$ and $R(0) \propto \dot{M}^{1/2}$ because the stand-off distance
depends on the balance between the wind ram pressure with the ISM ram pressure.
The size of the bow shock decreases as a function of $v_{\star}$: $R(0)$
decreases by a factor of 2 if $v_{\star}$ doubles, e.g. $R(0) \approx 0.13$ in
model MS1020 but $R(0) \approx 0.06$ in model MS1040~\textcolor{black}{(see upper and middle panels 
of Fig.~\ref{fig:m10ms})}. The bow shocks also scale
in size with $\dot{M}$, e.g. at fixed $v_{\star}$ its size for the $10\,
M_{\odot}$ star is smaller by a factor of 10 compared to the size of the bow shock from the $20\,
M_{\odot}$ star, which in turn is smaller by a factor of $\approx 3.5$
compared to one from the $40\, M_{\odot}$ star~\textcolor{black}{(e.g. see middle panels of Figs.~\ref{fig:m10ms} 
to~\ref{fig:m40ms})}. If we look again at $\dot{M}$ in
Fig.~\ref{fig:models_ms_rsg} (a$-$c), we find $\dot{M} \approx 10^{-9.5}$,
$\approx 10^{-7.3}$ and $\approx 10^{-6}\, \rm M_{\odot}\, \rm yr^{-1}$ for the
$10$, $20$ and $40\, M_{\odot}$ star, respectively. We see that these sizes are
in accordance with the theory and arise directly as a result of
Eq.~(\ref{eq:Ro}).

The relative thickness of the substructures varies with the wind and ISM
properties because the gas velocity determines both the post-shock temperature, i.e.
governs the cooling physics at the reverse shock and in the shell, and the
compression of the shocked ISM. Our simulations with $v_{\star}=20\, \mathrm{km}\,
\mathrm{s}^{-1}$ have weak forward shocks, i.e. compression at the forward
shock is not important. The thickness of the layer of shocked ISM gas with
respect to $R(0)$ is roughly independent of $M_{\star}$ for these models (see
upper panels of Figs.~\ref{fig:m10ms}, \ref{fig:m20ms} and~\ref{fig:m40ms}). The
shocked ISM density increases for models with $v_{\star}\ge40\, \mathrm{km}\,
\mathrm{s}^{-1}$ because the high post-shock temperature makes the cooling
efficient. The variations of $\dot{M}$ at a given $v_{\star}$ modify the
morphology of the bow shock because a stronger wind ram pressure enlarges the
size of the bow shock and makes the shell thinner with regard to $R(0)$ (see
models MS1020 and MS4020 \textcolor{black}{in upper panels of Figs.~\ref{fig:m10ms} 
and~\ref{fig:m40ms}}).

Our simulations with $v_{\star}=20\, \mathrm{km}\, \mathrm{s}^{-1}$ all have a
stable density field \textcolor{black}{(see upper panels of Figs.~\ref{fig:m10ms} 
to~\ref{fig:m40ms})}. The simulations with $v_{\star}=40\, \mathrm{km}\,
\mathrm{s}^{-1}$ are bow shocks with radiative forward shocks (i.e. with a
dense and thin layer of shocked ISM). Our simulations for $M_{\star} \ge 20\,
M_{\odot}$ and with $v_{\star}=70\, \rm km\, \rm s^{-1}$ show instabilities at
both the contact and the material discontinuity, \textcolor{black}{see middle 
panel of Fig.~\ref{fig:m20ms} and~\ref{fig:m40ms}. Our models for the $40\,
M_{\odot}$ star with $v_{\star} \ge 40\, \rm km\, \rm s^{-1}$ are similar. Model
MS4040 is slightly more unstable than model MS2070 whereas model MS4070 shows
even stronger instability which develops at its forward shock and dramatically
distorts its dense and thin shell, as shown in the bottom panel 
of Fig.~\ref{fig:m40ms}}. \textcolor{black}{The large density gradient across the material
discontinuity allows Rayleigh-Taylor instabilities to develop}. \textcolor{black}{ The
entire shell of cold ISM gas} has distortions characteristic of the
non-linear thin-shell instability~\citep{vishniac_apj_428_1994,
garciasegura_1996_aa_305}. 

\begin{figure}
          \centering 
	  \includegraphics[width=0.32\textwidth,angle=270]{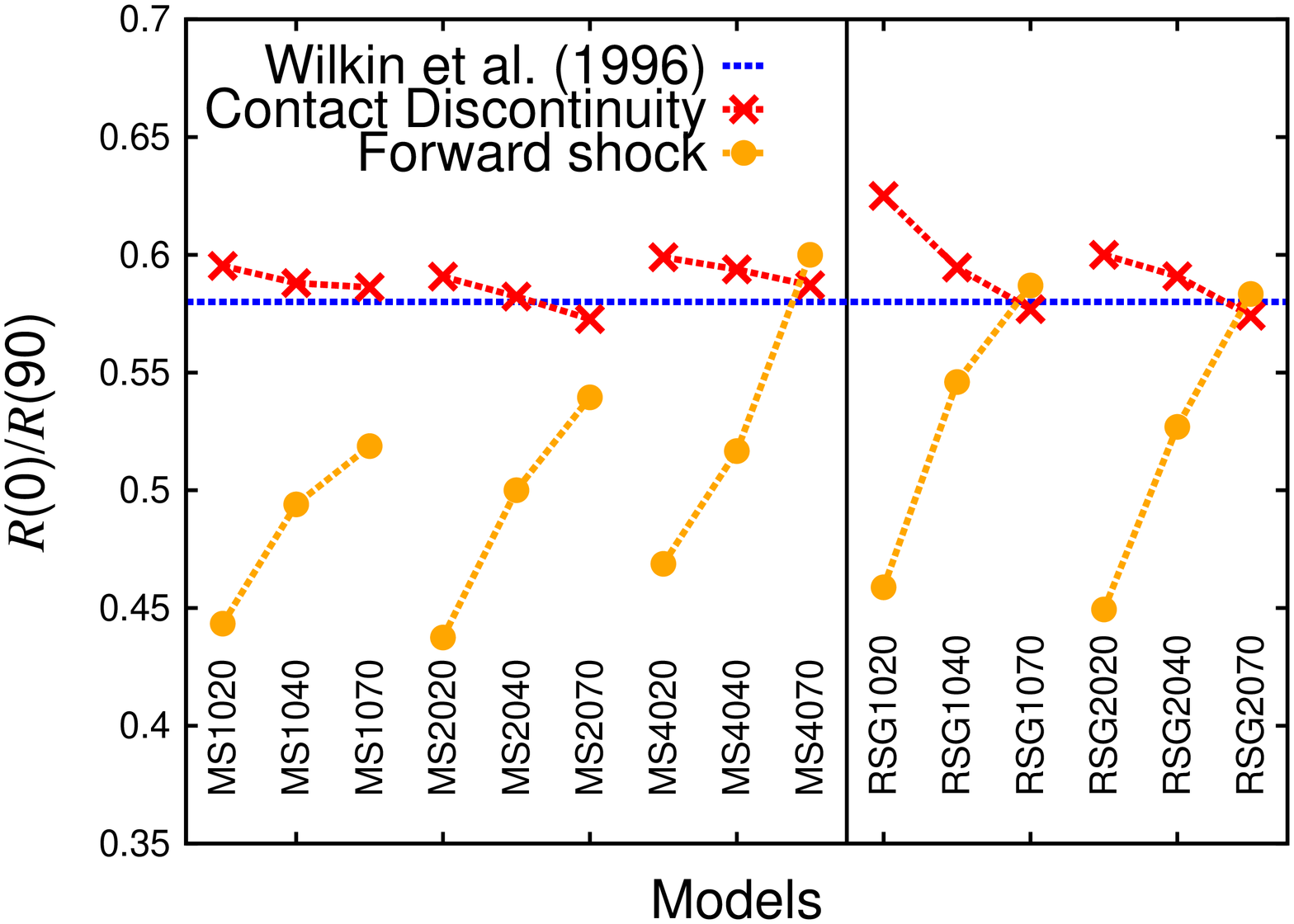}    
	  \caption{Comparison between the ratio $R(0) /\mathrm{R}(90)$ for the main sequence and supergiant models
		   with the theoretical value of $1/\sqrt{3}\approx 0.58$ predicted by~\citet[][horizontal dotted blue line]{wilkin_459_apj_1996}.
		   We distinguish between the contact discontinuity (red crosses)
		   and the forward shock (orange dots) of each model. }
          \label{fig:wilkin}
\end{figure}


\subsection{Comparison of the models with the analytical solution}
\label{subsect:introduction_models_ms}

\begin{figure}
          \centering 
	  \includegraphics[width=0.45\textwidth,angle=0]{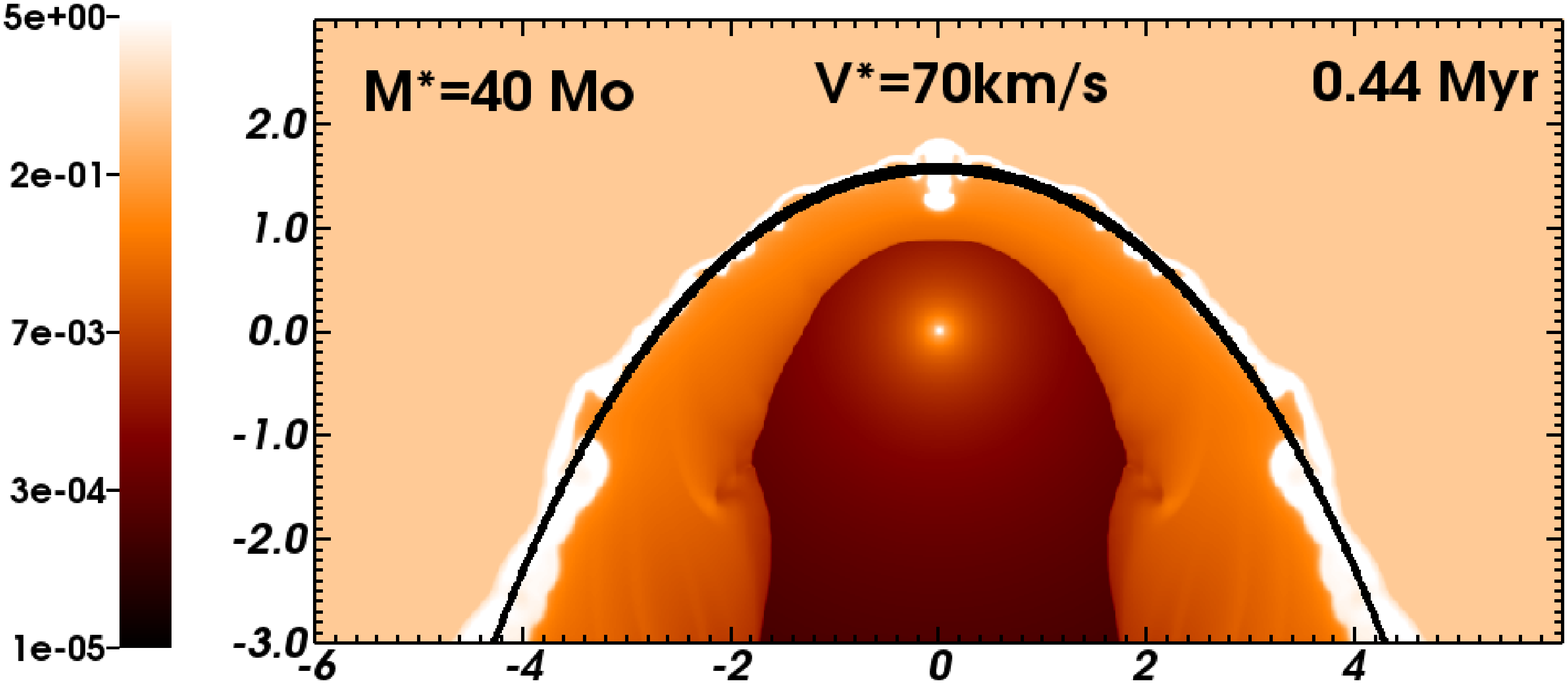}    
	  \caption{Comparison between the density field of model MS4070 presenting a thin shell 
		   and the corresponding analytical solution~\citep[][solid black line]{wilkin_459_apj_1996}.
		   The gas number density is shown with a density range from $10^{-5}$ to $5.0\, \rm cm^{-3}$ 
		   in the logarithmic scale. 
		   The $x$-axis represents the radial direction and the $y$-axis the direction of stellar motion 
		   (in $\mathrm{pc}$). }
          \label{fig:anal}
\end{figure}

In Fig.~\ref{fig:wilkin} we compare $R(0)/R(90)$ with the analytical solution
for a bow shock with a thin
shell~\citep[$R(0)/R(90)\approx1/\sqrt{3}\approx 0.58$;][]{wilkin_459_apj_1996}.
$R(0)/R(90)$ at the contact discontinuity decreases as a function of $v_{\star}$,
e.g. models MS2020 and MS2070 have $R(0)/R(90) \approx0.59$ and $\approx0.56$,
respectively. $R(0)/R(90)$ at the forward shock increases with $v_{\star}$ and
$\dot{M}$ (see Figs.~\ref{fig:m10ms} to~\ref{fig:m40ms}). The contact
discontinuity is the appropriate measure to match the analytical 
solution~\citep[see][]{mohamed_aa_541_2012}. $R(0)/R(90)$ is within $<10$ per cent of
Wilkin's solution but does not satisfy it at both discontinuities, except for
MS4070 with $R(0)/R(90) \approx0.59$ at the contact discontinuity and 
$R(0)/R(90) \approx0.60$ at the forward shock. Model MS4070 is the most
compressive bow shock and it has a thin unstable shell bounded by the contact
discontinuity and forward shock. Fig.~\ref{fig:anal} shows good agreement
between model MS4070 and Wilkin's solution for angles $\theta > 90\degree$. Our
model MS4020 is the most deviating simulation at the forward shock, because the
brevizy of its main sequence phase prevents the bow shock from reaching a
steady state.

\begin{table}
	\centering
	\caption{Bow shock morphological properties at the contact discontinuity. 
	 The parameter $R(0)$ (in pc) is the stand-off distance of the 
	 bow shock at the contact discontinuity
	 and $R(0)/R(90)$ is the ratio plotted in Fig.~\ref{fig:wilkin}, 
	 with $R(90)$ the radius perpendicular to the direction of motion.}
	\begin{tabular}{cccc}
	\hline
	\hline
	${\rm {Model}}$ &   $R(0) \, (\rm pc)$                              
 			 &   $R(0)/R(90)$
			\\ \hline   
	MS1020   &  $0.13$      &  $0.595$ \\             
	MS1040   &  $0.06$      &  $0.587$  \\        
	MS1070   &  $0.03$      &  $0.586$  \\ 
	\hline 
	RSG1020  &  $0.30$       &  $0.625$  \\             
	RSG1040  &  $0.22$       &  $0.594$  \\        
	RSG1070  &  $0.15$       &  $0.576$  \\  
	\hline   
	MS2020   &  $1.40$       &  $0.590$  \\         
	MS2040   &  $0.69$       &  $0.582$  \\     
	MS2070   &  $0.38$       &  $0.563$  \\
	\hline 
	RSG2020  &  $1.35$       &  $0.600$ \\         
	RSG2040  &  $0.65$       &  $0.590$  \\     
	RSG2070  &  $0.31$       &  $0.578$  \\
	\hline 
	MS4020   &  $5.60$       &  $0.598$  \\              
	MS4040   &  $2.85$       &  $0.593$  \\   
	MS4070   &  $1.72$       &  $0.587$  \\    
	\hline 
	\end{tabular}
\label{tab:params}
\end{table}


\subsection{Thermal conduction}
\label{subsect:mixms}

\begin{figure*}
	\begin{minipage}[b]{0.46\textwidth}
          \centering 
	  \includegraphics[width=1.0\textwidth,angle=0]{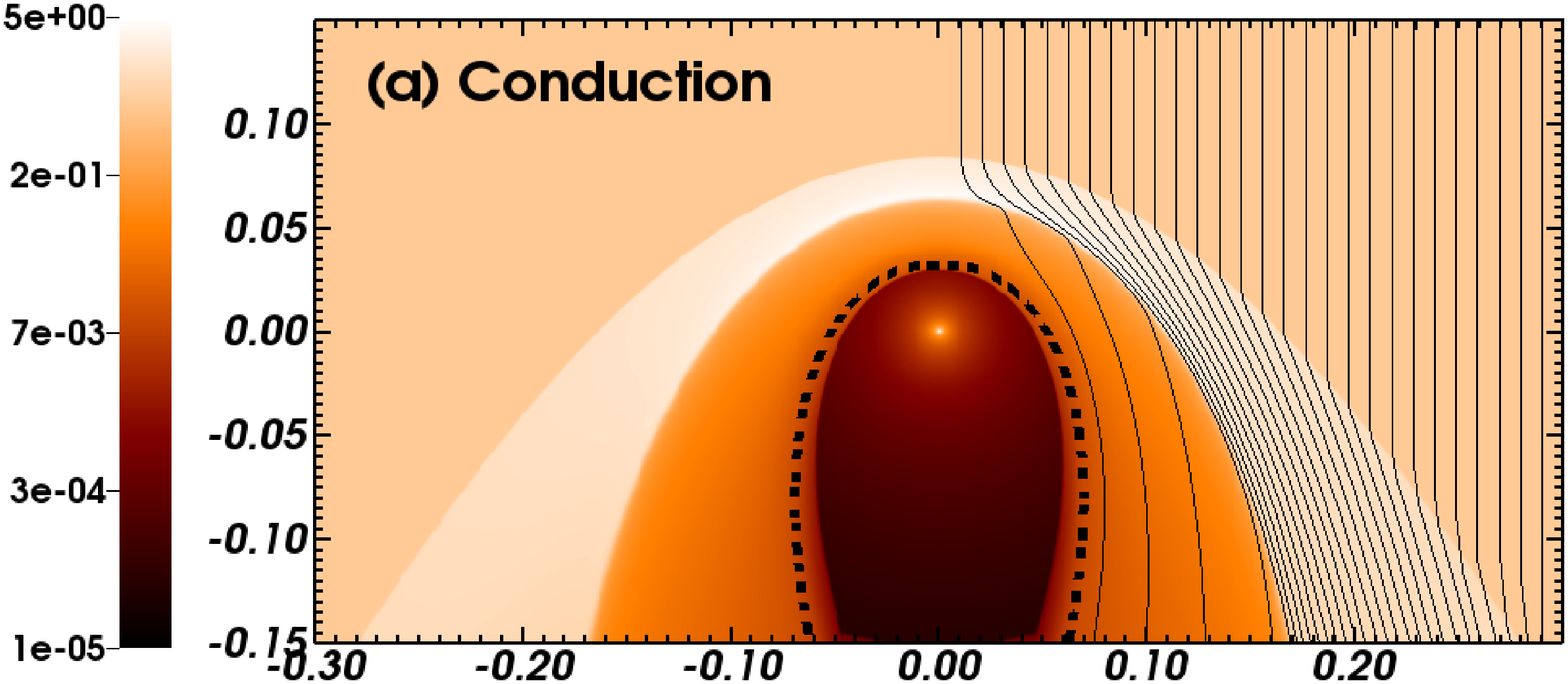} 
   	\end{minipage}
	\begin{minipage}[b]{0.46\textwidth}
          \centering 
	  \includegraphics[width=1.0\textwidth,angle=0]{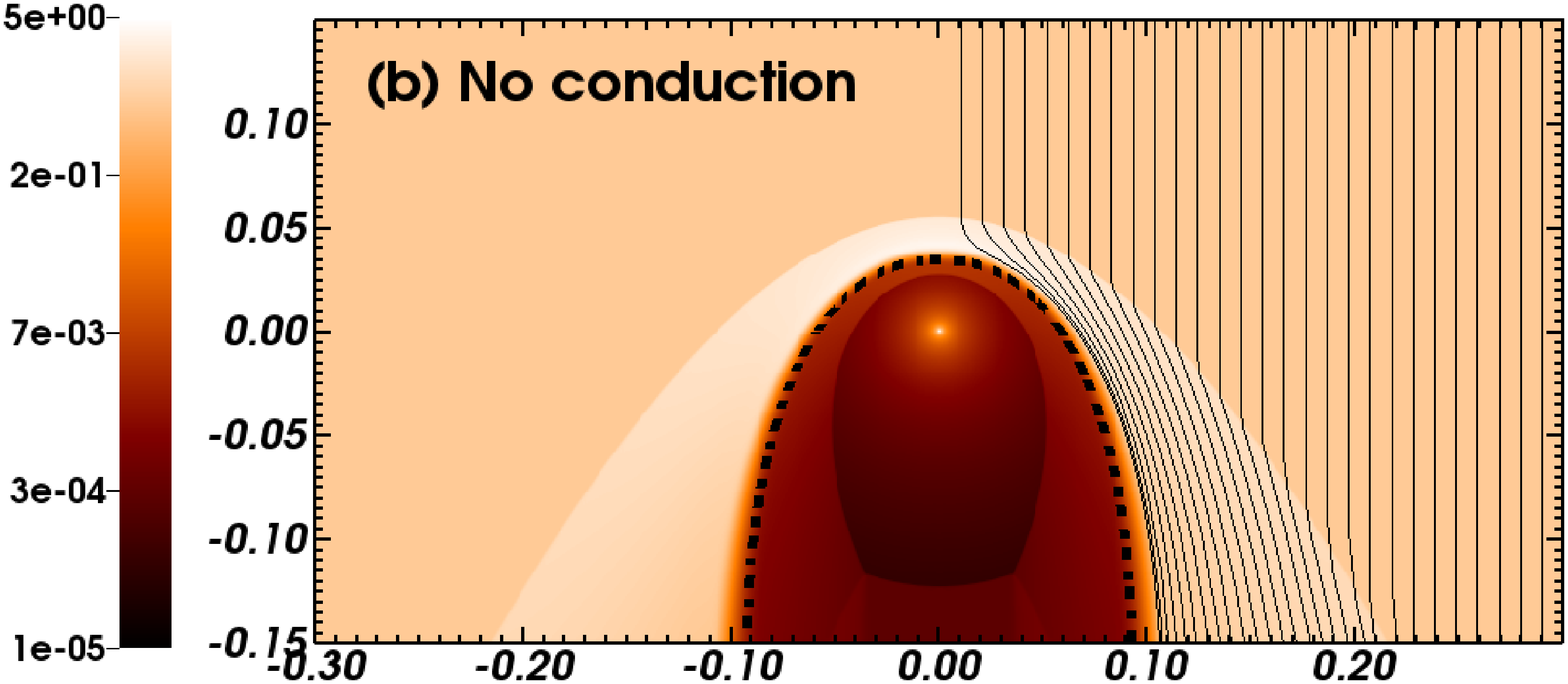}
   	\end{minipage}
	  \caption{ Changes in the location of ISM and wind material induced by thermal conduction in the hot bubble of a bow shock.
		    Figures show gas number density (in $\rm cm^{-3}$) for model $\mathrm{MS1040}$ (a) and for the same setup 
		    run without heat conduction (b).
		    For each figure the dotted thick line traces the material discontinuity, $Q(\bmath{r})=1/2$.
 		    The right part of each figure overplots ISM flow streamlines.
 		    It highlights the penetration of ISM material into the hot layer of the bow shock because of heat conduction.
		    Comparing the two figures illustrates its effects, increasing the density inside 
		    the region of shocked wind and enlarging the global size of the bow shock.
		    The $x$-axis represents the radial direction and the $y$-axis the direction of stellar motion (in $\mathrm{pc}$). 
		    Only part of the computational domain is shown. }
          \label{fig:mixing}
\end{figure*}

Fig.~\ref{fig:mixing} illustrates the effects of heat conduction on the shape of
a bow shock. Panel (a) shows the density field of model MS2040, and panel (b)
shows the same model but without thermal conduction. The dashed contour traces
the border between wind and ISM gas. The streamlines show the
penetration of ISM material into the hot bubble. The bow shock including thermal
conduction is larger by a factor $\approx 1.4$ in both the directions normal and 
parallel to the direction of motion of the star. Its shell is
denser and splits into two layers of hot and cold shocked ISM, whereas the model
without thermal conduction has a single and less compressed region of ISM
material.

\begin{figure}
          \centering 
	  \includegraphics[width=0.33\textwidth,angle=270]{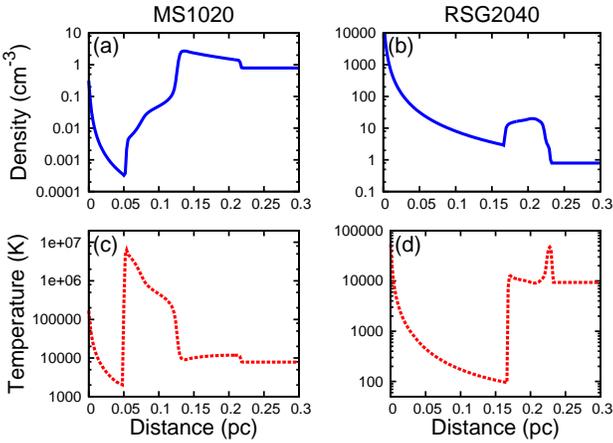}    
	  \caption{Total number density (solid blue lines, in $\mathrm{cm}^{-3}$) and temperature (dotted red lines, 
		   in $\mathrm{K}$) profiles for two typical bow shocks of a main sequence and a red supergiant \textcolor{black}{star}. 
		   The profiles are plotted for the model MS1020 in panels (a) and (c) and for the model RSG2040 in 
		   panel\textcolor{black}{s} (b) and (d) as a function of the distance to the star along the direction of motion.}
          \label{fig:profile}
\end{figure}

The position of the reverse shock is insensitive to thermal conduction because
heat lost at the material discontinuity is counterbalanced by the large wind
ram pressure (see panels (a) and (b) of Fig.~\ref{fig:mixing}).
Fig.~\ref{fig:profile} illustrates that the shocked regions of a bow shock with
heat conduction have smooth density profiles around the contact discontinuity
(see panels (a) and (c) of Fig.~\ref{fig:profile}). This is consistent with
previous models of a steady star~\citep[see fig. 3 of][]{weaver_apj_218_1977}
and of moving stars~\citep[see Fig. 7 of][]{comeron_aa_338_1998}. Electrons
carry internal energy from the hot shocked wind to the shocked ISM, e.g. the
$10\, M_{\odot}$ models have a temperature jump amplitude of $\itl{ \Delta} T
\approx 10^{7} \, \rm K$ across the contact discontinuity.

\textcolor{black}{
Our simulation of model MS1040 (see Fig.~\ref{fig:m10ms}) provides us with the 
parameters of the hot bubble ($T \approx 10^{7}\, \rm K$, $n\approx 0.02\, \rm cm^{-3}$) 
and the shell ($T \approx 10^{4}\, \rm K$, $n\approx 3.3\, \rm cm^{-3}$). 
The shocked ISM gas has a velocity $v\approx 25\, \rm km\, \rm s^{-1}$ and $\mu=0.61$.
Using Eq.~(\ref{eq:time_dyn})$-$(\ref{eq:time_cool}), we find that the hot gas in the
inner ($t_{\rm cool} \approx 1.11 \times 10^{2} \gg t_{\rm dyn} \approx 1.4 \times 10^{-3}\, \rm Myr$ ) 
and outer ($t_{\rm cool} \approx 2.94 \times 10^{-3} \gtrsim t_{\rm dyn} \approx 1.0\times 10^{-3}\, \rm Myr$ ) 
layers of the bow shock are adiabatic and slightly radiative, respectively. The radiative character
of the shell is more pronounced for models with $v_{\star}>40\, \rm km\, \rm s^{-1}$.
Note that the hot bubble never cools, i.e. $t_{\rm cool}$ refers here to the timescale of the losses of 
internal energy by optically-thin radiative processes, which are compensated by 
the conversion of kinetic energy to heat at the reverse shock. The thermal conduction timescale is,
\begin{equation}
  t_{\rm cond} = \frac{7pl^{2}}{2(\gamma-1)\kappa(T)T}, 
\end{equation}
where $\kappa(T)$ is the heat conduction coefficient and
$l$ a characteristic length along which heat transfer happens~\citep{orlando_apj_678_2008}. Because
$\kappa(T) \propto T^{5/2}$~\citep{cowie_apj_211_1977}, $t_{\rm cond} \propto
T^{-7/2}$, i.e. heat conduction is a fast process in a hot medium. Consequently,
we have $t_{\rm dyn}/t_{\rm cond}\approx1.46\times 10^{5}\gg 1$ 
and $t_{\rm cool}/t_{\rm cond}\approx1.16\times 10^{10}\gg 1$ in 
the hot bubble ($l=0.035\, \rm pc$) whereas we find $t_{\rm dyn}/t_{\rm cond}\approx1.71\times 10^{-5}\ll 1$ and 
$t_{\rm cool}/t_{\rm cond}\approx5.03\times 10^{-5}\ll 1$ in the shell ($l=0.025\, \rm pc$) of the model
MS1040, which explains the differences between the models shown in
Fig.~\ref{fig:mixing}. All of our simulations of the main sequence phase 
behave similarly because their hot shocked wind layers have similar temperatures. Heat transfer across 
the bubble is always faster than the dynamical timescale of the gas.}


As a consequence, the pressure increases in the shocked ISM, pushing both the contact discontinuity inwards
and the forward shock outwards. The region of shocked wind conserves its mass
but loses much of its pressure. To balance the external pressure, its volume
decreases and the gas becomes denser. Two concentric substructures of shocked
ISM form: an inner one with high temperature and low density adjoining the material
discontinuity, and an outer one with low temperature and high density.
\textcolor{black}{Previous investigations about the effects of heat conduction inside circumstellar
nebulae around runaway hot stars are available in section 4.6 of~\citet{comeron_aa_338_1998}.}

\subsection{Bow shock emissivity}
\label{subsect:mslum}

\subsubsection{Luminosities}
\label{subsubsect:mslumscalar}

\begin{figure}
         \centering 
	 \includegraphics[width=0.33\textwidth,angle=270]{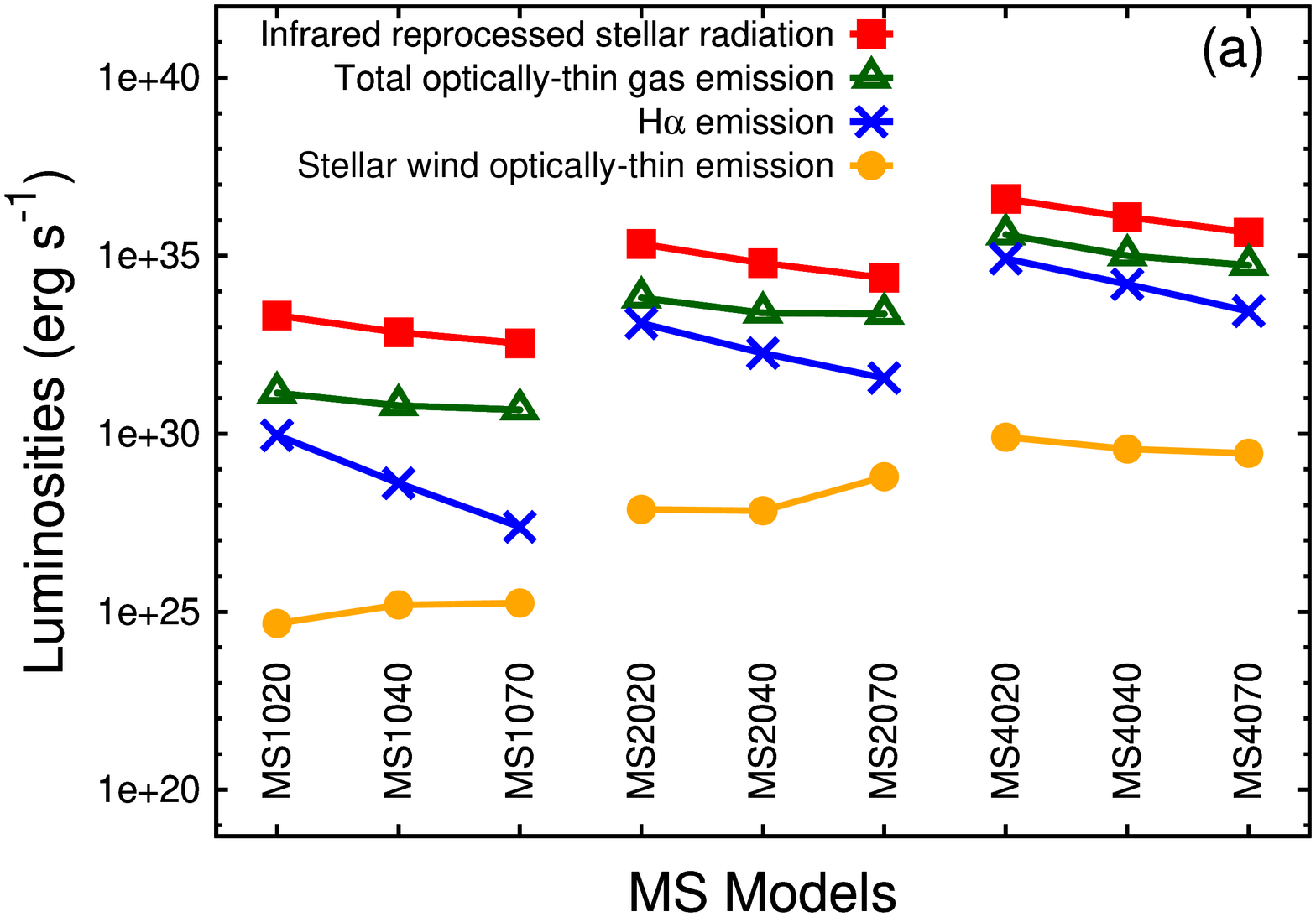}   
	 \includegraphics[width=0.33\textwidth,angle=270]{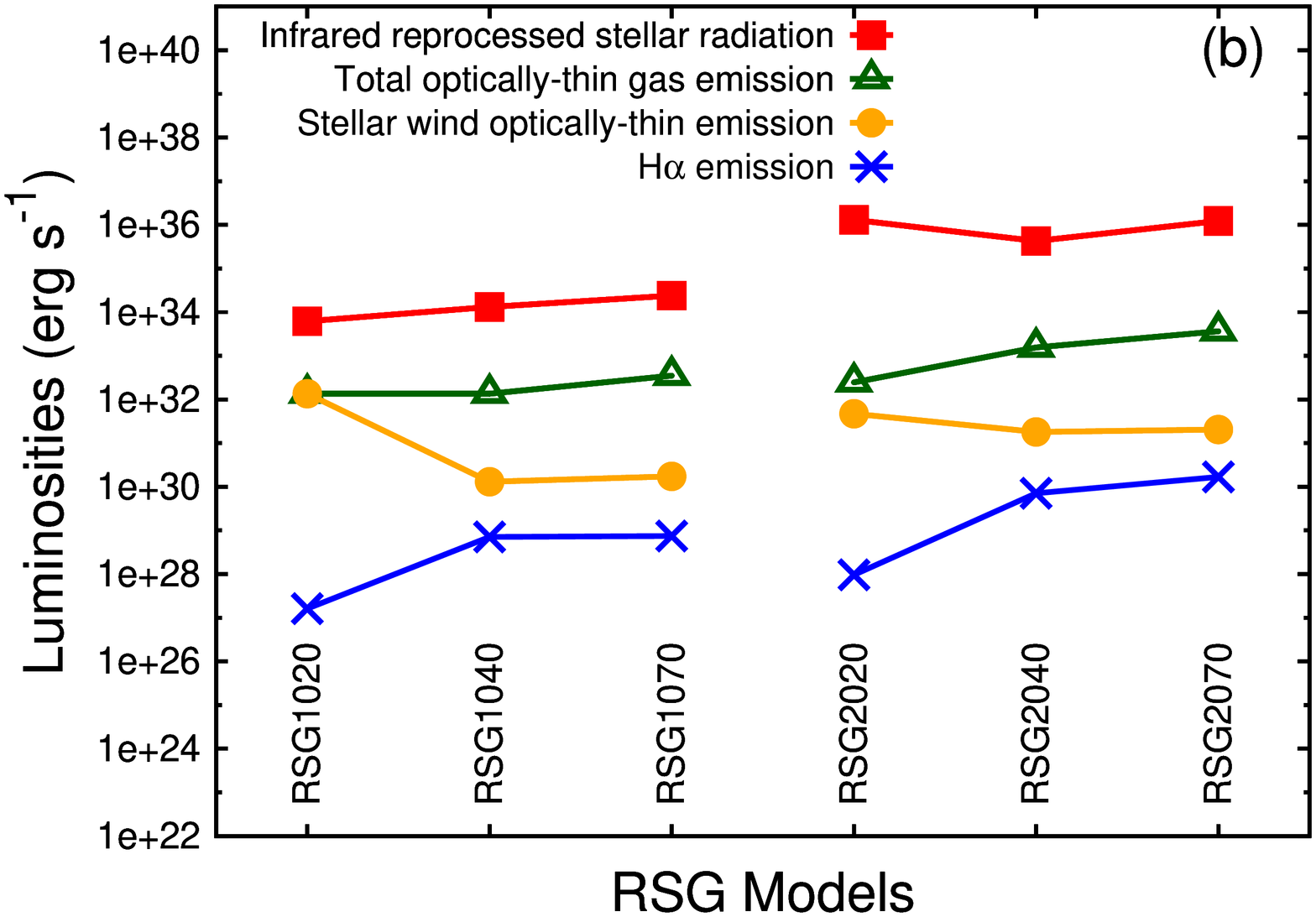}   	 
	 \caption{ Bow shock luminosities and reprocessed stellar infrared radiation for main sequence (a) and red supergiant models (b). 
		   The total bow shock luminosity of optically-thin gas radiation (\textcolor{black}{green} triangles) is distinguished from 
		   the contribution due to the wind material only (orange dots). 
		   The luminosity of H$\alpha$ emission (\textcolor{black}{blue} crosses) and the reprocessed 
		   infrared stellar radiation (\textcolor{black}{red} squares) are also plotted.
		   The infrared radiation is not considered in the simulations 
		   and is therefore not included in the total optically-thin gas radiation. 
		   The simulation labels are written vertically under each triplet
		   related to a given stellar model (see Table~\ref{tab:ms}).}
        \label{fig:bslum}
\end{figure}

The bow shock luminosities of all our models are plotted in panel (a) of Fig.~\ref{fig:bslum}. 
It shows the emitted light as a function of mass-loss $\dot{M}$ and space velocity $v_{\star}$ (i.e. by
model). $L_{\rm gas}$ is the bow shock luminosity from optically-thin cooling of
the gas and the part of this which originates from the wind material is
designated as $L_{\rm wind}$. The bow shock luminosities are calculated taking
into account the cylindrical symmetry of the models by integrating the radiated
energy in the $ z \ge 0$ region~\citep{mohamed_aa_541_2012}. The optically-thin
gas radiation is therefore computed as,
\begin{equation}
	\itl{ L}_{\rm gas} = 2\pi \iint_{\mathcal D} \itl{\Lambda} n_{\rm H}^{2} R dRdz,
\label{eq:bslum}
\end{equation}
where $\mathcal D$ represents the considered volume. The heating terms are estimated with a similar method, as,
\begin{equation}
	\widetilde{\itl{\Gamma}}_{\alpha} = 2\pi \iint_{\mathcal D} \itl{ \Gamma}_{\alpha} n_{\rm H}^{\alpha} R dRdz,
\label{eq:bsheat}
\end{equation}
where $\widetilde{\itl{\Gamma}}_{\alpha=1}$ is the heating rate per unit volume for UV heating of
grains, and $\widetilde{\itl{\Gamma}}_{\alpha=2}$ is the heating rate per unit volume square for
photoionization heating. Inserting the quantities $Q(\bmath{r})$ or $1$$-$$Q(\bmath{r})$ 
in the integrant of Eq.~(\ref{eq:bslum}) or (\ref{eq:bsheat}) allows us to separate the
contributions from wind and ISM material. The panels of Fig.~\ref{fig:bslum}
also specify the luminosity from H${\alpha}$ emission $L_{\rm H\alpha}$
(calculated using the prescriptions by~\citet{osterbrock_1989}, our
Appendix~\ref{maps_Ha}) and the infrared luminosity of reprocessed starlight 
by dust grains $L_{\rm IR}$ (calculated treating the dust as in~\citet{mackey_apjlett_751_2012}, 
our Appendix~\ref{maps_IR}). Nonetheless, $L_{\rm IR}$ does not contribute
to the thermal physics of the plasma and is not included in the calculations of
either $L_{\rm gas}$ or $L_{\rm wind}$. The luminosities $L_{\rm gas}$, $L_{\rm
wind}$, $L_{\rm H\alpha}$, $L_{\rm IR}$, the heating rates $\widetilde{\itl{\Gamma}}_{\alpha}$,
and the stellar luminosity $L_{\star}$, provided by the stellar
evolution models~\citep{brott_aa_530_2011a}, are detailed in Table~\ref{tab:lum_val}.

The bow shock luminosity of optically thin gas $L_{\rm gas}$
decreases by an order of magnitude
between the models with $v_{\star}=20$ to $70\, \rm km\, s^{-1}$, but increases
by several orders of magnitude with $\dot{M}$, e.g. $L_{\rm gas} \approx 1.4
\times 10^{31}$ and $\approx 3.9 \times 10^{35}\, \rm erg\, \rm s^{-1}$ for the
models MS1020 and MS4020, respectively. $L_{\rm gas}$ is influenced by i)
$v_{\star}$ which governs the compression factor of the shell, and ii) by the
size of the bow shock which increases with $\dot{M}$ and decreases with
$v_{\star}$. Moreover, we find that emission by optically-thin cooling is
principally caused by optical forbidden lines such as [O\,{\sc ii}] and
[O\,{\sc iii}] which is included in the cooling curve in the range $\approx 8000
\le T \le 6.0\times 10^{4}\, \rm K$ (see estimate of the luminosity 
$L_{\rm FL}$ produced by optical forbidden lines in Table~\ref{tab:lum_val}).

The contribution of optically-thin emission from stellar wind material, $L_{\rm
wind}$, to the total luminosity of optically-thin gas radiation is negligible 
e.g. $L_{\rm wind}/L_{\rm gas}\approx 10^{-6}$ for model MS2020. The variations 
of $L_{\rm wind}$ roughly follows the variations of $L_{\rm gas}$. 
The volume occupied by the shocked wind material is reduced by heat transfer (see black
contours in Figs.~\ref{fig:m10ms} to~\ref{fig:m40ms}) and this prevents $L_{\rm
wind}$ from becoming important relative to $L_{\rm gas}$. It implies that most
of the emission by radiative cooling comes from shocked ISM gas which cools as
the gas is advected from the forward shock to the contact discontinuity.

$L_{\rm H\alpha}$ is smaller than $L_{\rm gas}$ by about $1$$-$$3$ orders of
magnitude and larger than $L_{\rm wind}$ by $2$$-$$5$ orders of magnitude, e.g.
model MS2040 has $L_{\rm H\alpha}/L_{\rm gas} \approx 10^{-1}$ and $L_{\rm
H\alpha}/L_{\rm wind} \approx 10^{4}$. The H$\alpha$ emission therefore mainly
comes from ISM material. More precisely, we suggest that $L_{\rm H\alpha}$ originates 
from the cold innermost shocked ISM since the H$\alpha$ emissivity $\propto T^{-0.9}$ 
(our Appendix~\ref{maps_Ha}). The variations of $L_{\rm H\alpha}$ follow the global
variations of $L_{\rm gas}$, i.e. the H$\alpha$ emission is fainter at high
$v_{\star}$, e.g. $L_{\rm H\alpha} \approx 1.3\times 10^{33}$ and $\approx 3.6
\times 10^{31}\, \mathrm{erg}\, \mathrm{s}^{-1}$ for models MS2020 and MS2070,
respectively. The gap between $L_{\rm gas}$ and $L_{\rm H\alpha}$ increases
with $v_{\star}$ because the luminosities are calculated for $z>0$ whereas the
H$\alpha$ maximum is displaced to $z<0$ as $v_{\star}$ increases (see further
discussion in Section~\ref{subsubsect:mslummaps}).

$L_{\rm IR}$ is larger than $L_{\rm gas}$ by about $1$$-$$3$ orders of magnitude
in all our simulations. We find that $L_{\rm IR} \gg L_{\rm H\alpha}$, with a gap increasing with
$v_{\star}$ at a considered $\dot{M}$, e.g. $L_{\rm IR }/L_{\rm H\alpha} \approx
10^{2}$ and $\approx 10^{3}$ for models MS2020 and MS2070, respectively. These
large $L_{\rm IR}$ suggest that bow shocks around main sequence stars should be
much more easily observed in the infrared than at optical wavelength. We draw
further conclusions on the detectability of bow shocks generated by a runaway
main sequence star\textcolor{black}{s} moving through the Galactic plane in Section~\ref{subsect:observability}.

\subsubsection{Synthetic emission maps}
\label{subsubsect:mslummaps}

\begin{figure*}
	\begin{minipage}[b]{ 0.48\textwidth}
		\includegraphics[width=1.0\textwidth]{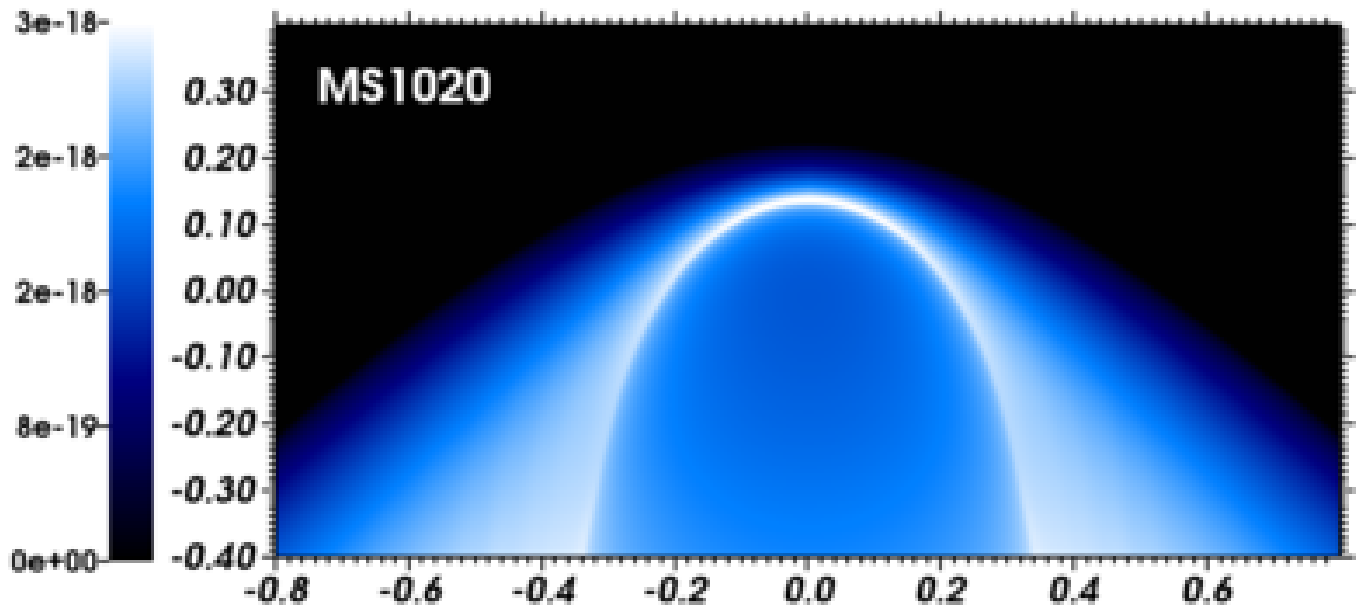}
	\end{minipage}
	\begin{minipage}[b]{ 0.48\textwidth}
		\includegraphics[width=1.0\textwidth]{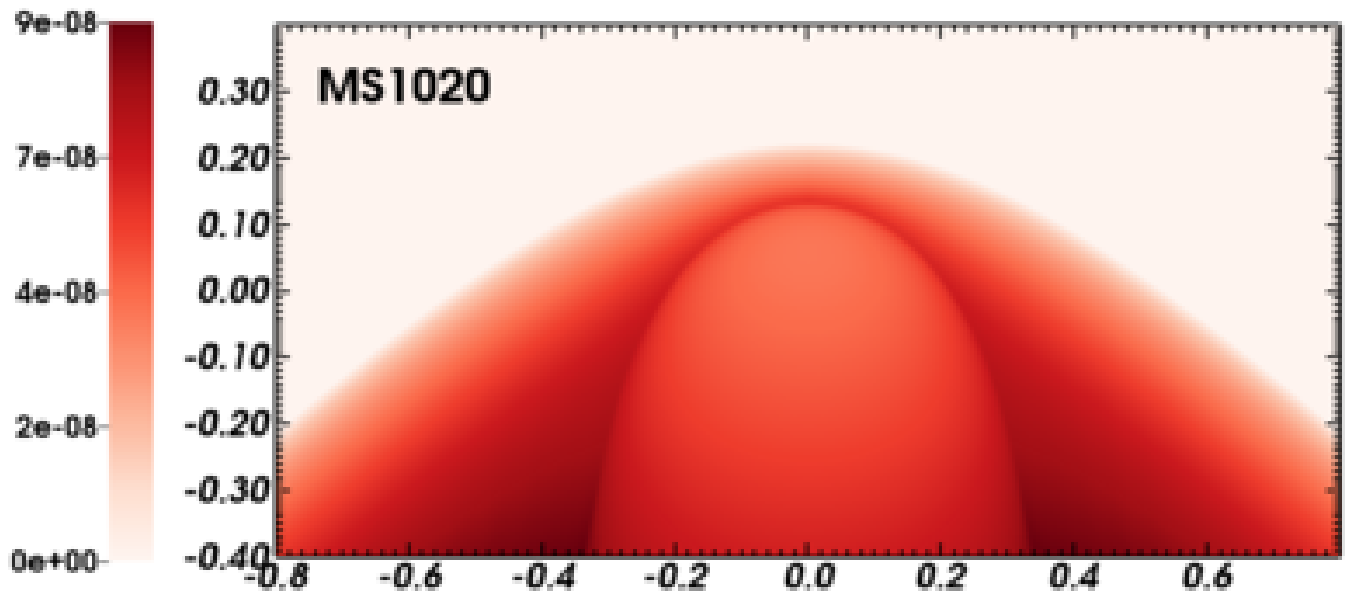}
	\end{minipage} \\

	\begin{minipage}[b]{ 0.48\textwidth}
		\includegraphics[width=1.0\textwidth]{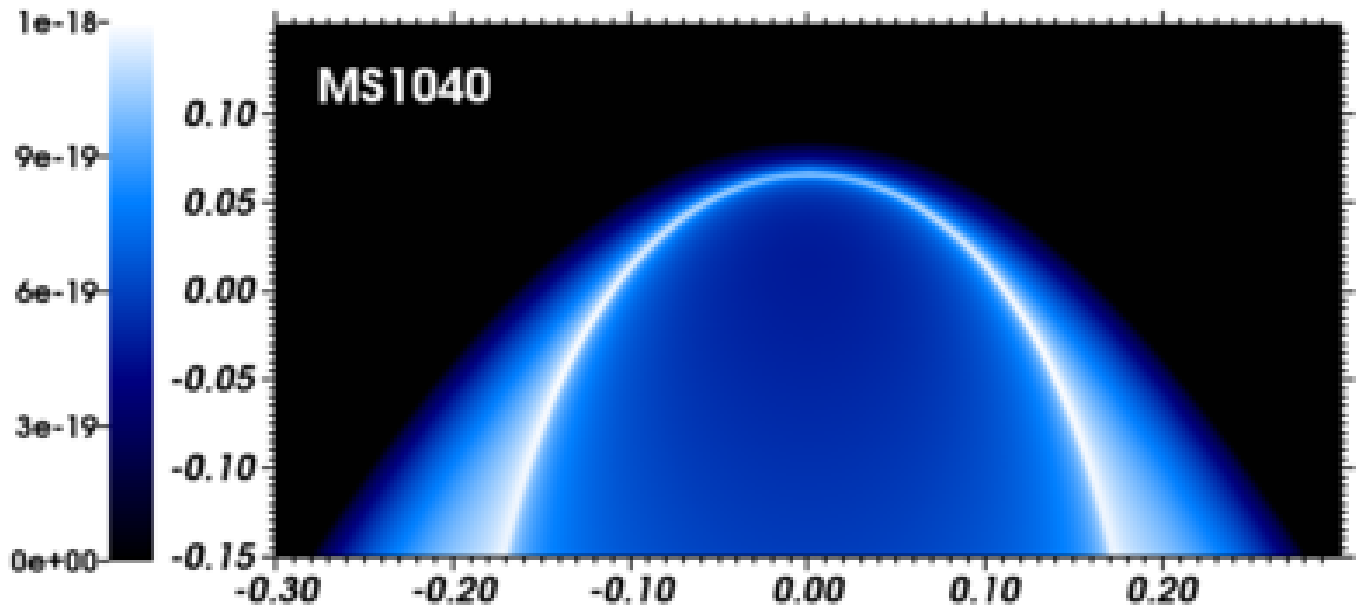}
	\end{minipage}
	\begin{minipage}[b]{ 0.48\textwidth}
		\includegraphics[width=1.0\textwidth]{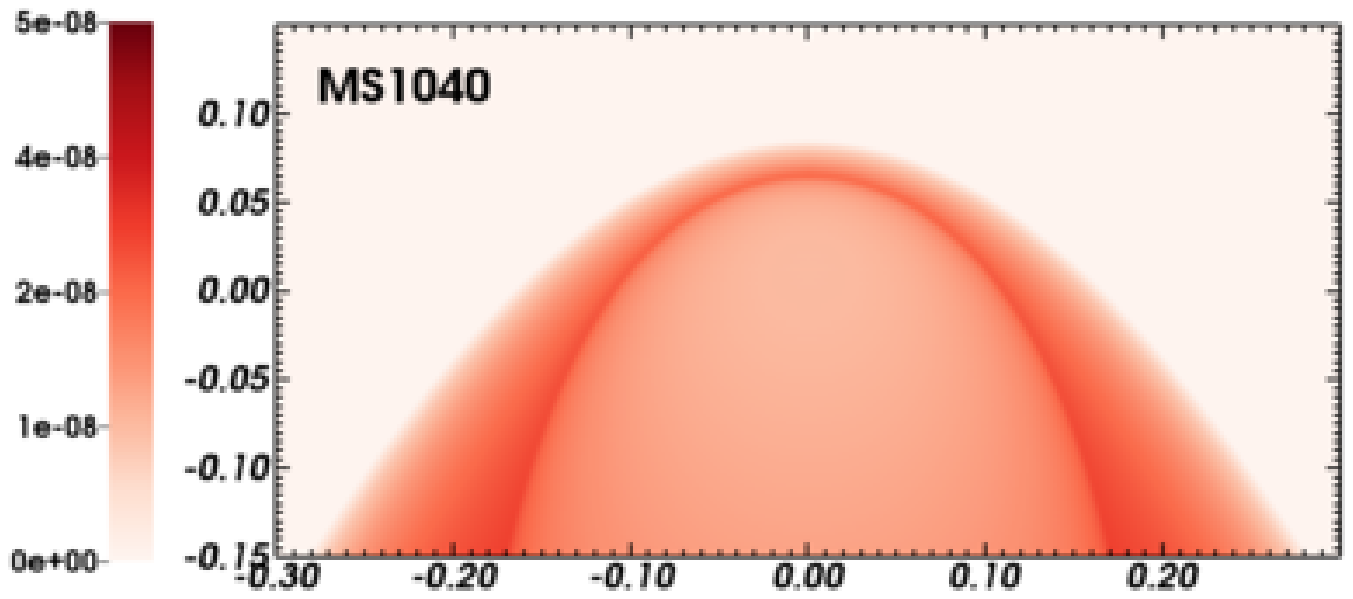}
	\end{minipage} \\
	
	\begin{minipage}[b]{ 0.48\textwidth}
		\includegraphics[width=1.0\textwidth]{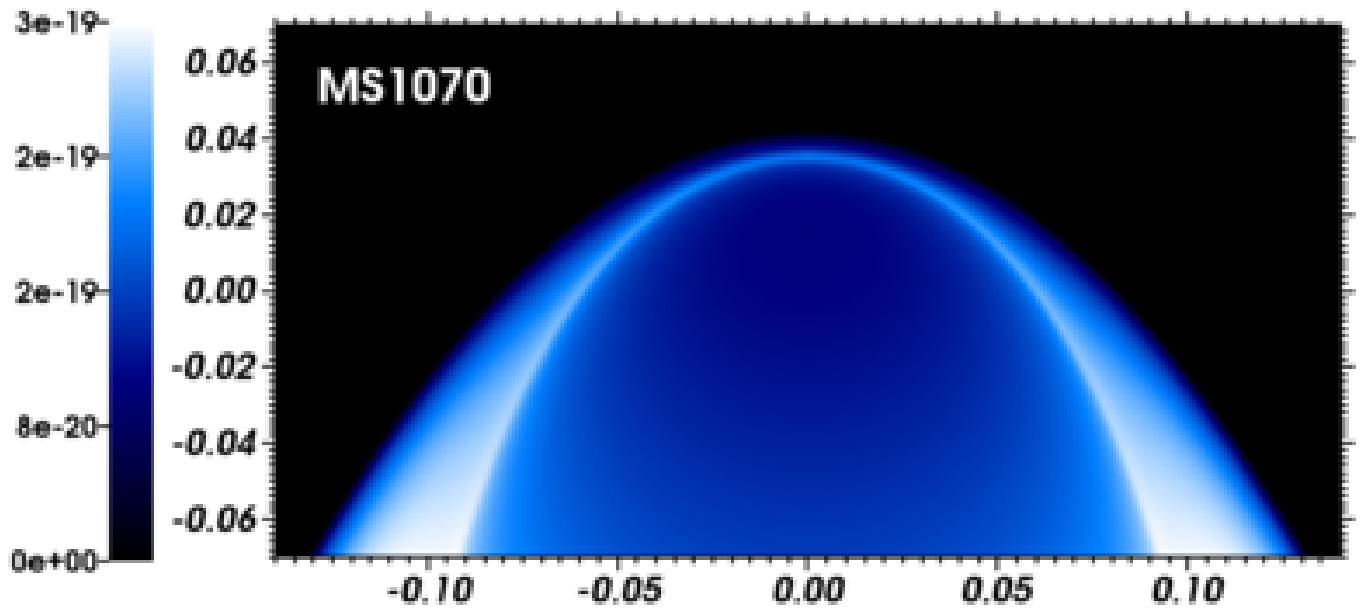}
	\end{minipage}
	\begin{minipage}[b]{ 0.48\textwidth}
		\includegraphics[width=1.0\textwidth]{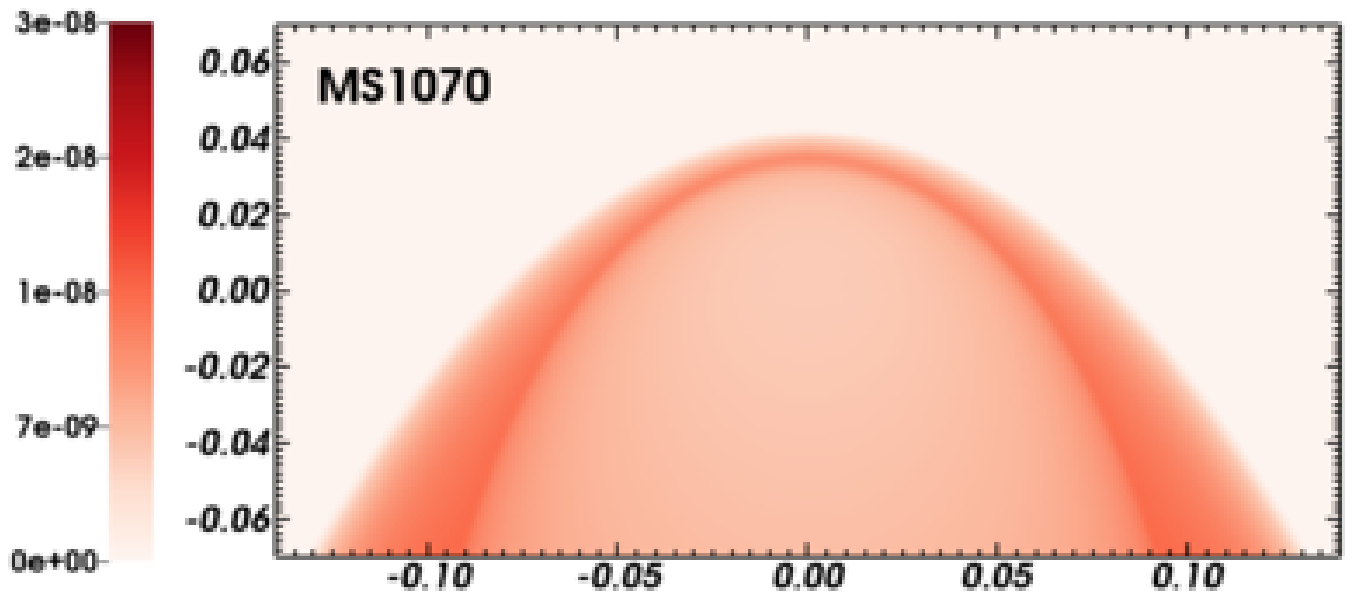}
	\end{minipage} \\  
	\caption{The figures show the H$\alpha$ surface brightness (left, in 
		 $\mathrm{erg}\, \mathrm{s}^{-1}\, \mathrm{cm}^{-2}\, \mathrm{arcsec}^{-2}$) and the dust surface mass density 
		 (right, in $\rm g\, \mathrm{cm}^{-2}$) for the bow shocks from the main sequence phase 
		 of our $10\, \mathrm{M}_{\odot}$ initial mass star.
		 Quantities are calculated excluding the undisturbed ISM and plotted in the linear scale, 
		 as a function of the considered space velocities.
		 The $x$-axis represents the radial direction and the $y$-axis the direction of stellar motion (in $\mathrm{pc}$). 
		 Only part of the computational domain is shown.
		 }	
	\label{fig:projemms10}  
\end{figure*}

\begin{figure*}
	\begin{minipage}[b]{ 0.48\textwidth}
		\includegraphics[width=1.0\textwidth]{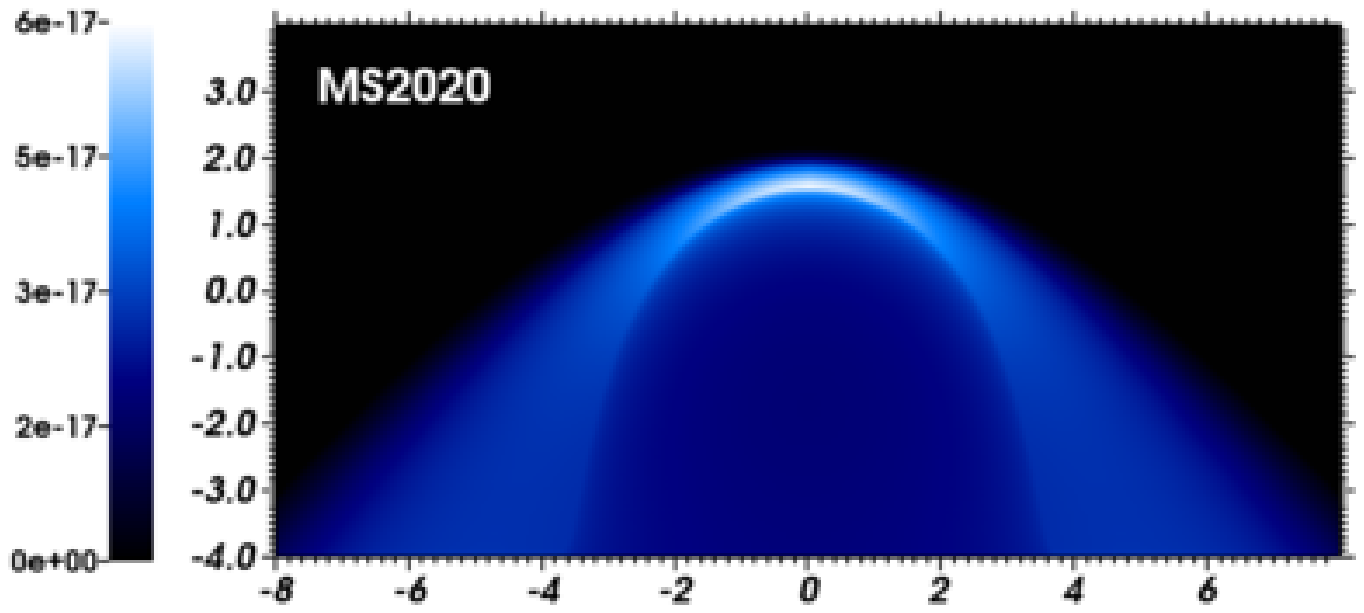}
	\end{minipage}
	\begin{minipage}[b]{ 0.48\textwidth}
		\includegraphics[width=1.0\textwidth]{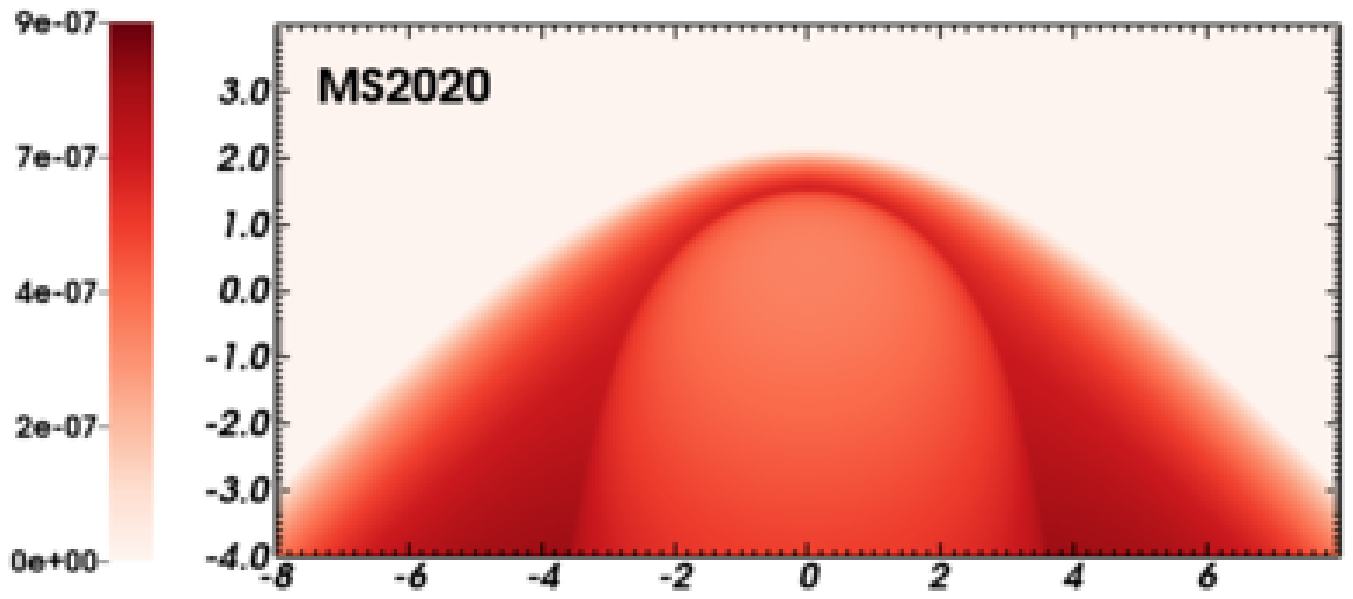}
	\end{minipage} \\

	\begin{minipage}[b]{ 0.48\textwidth}
		\includegraphics[width=1.0\textwidth]{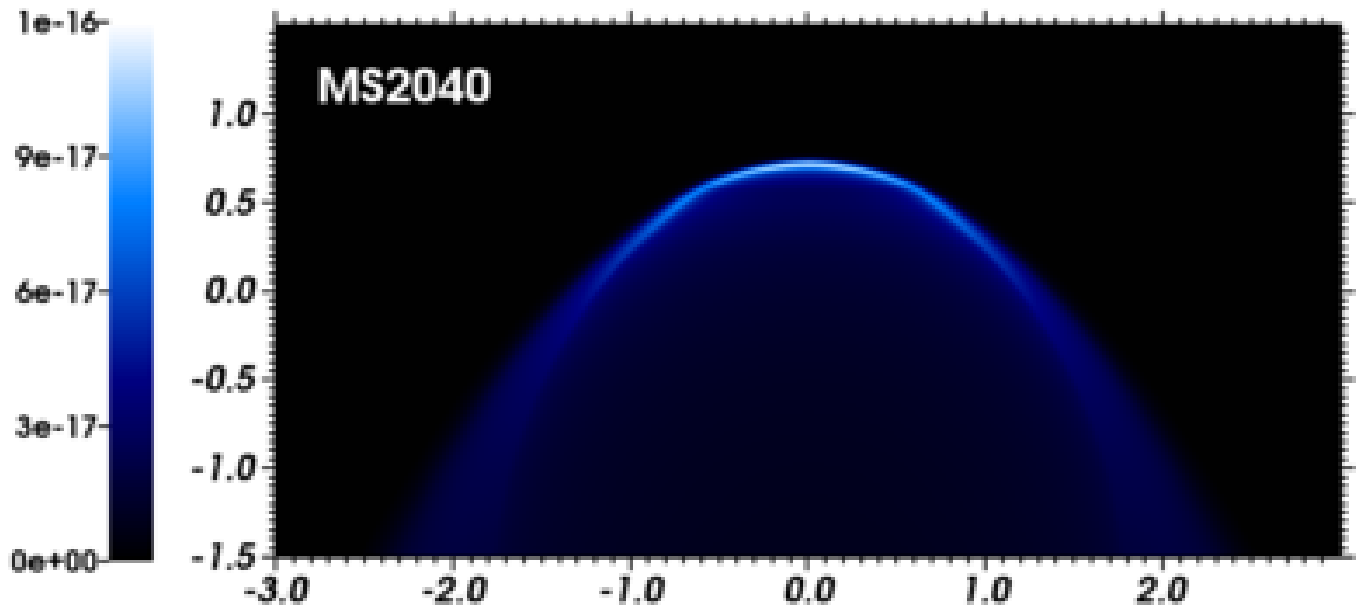}
	\end{minipage}
	\begin{minipage}[b]{ 0.48\textwidth}
		\includegraphics[width=1.0\textwidth]{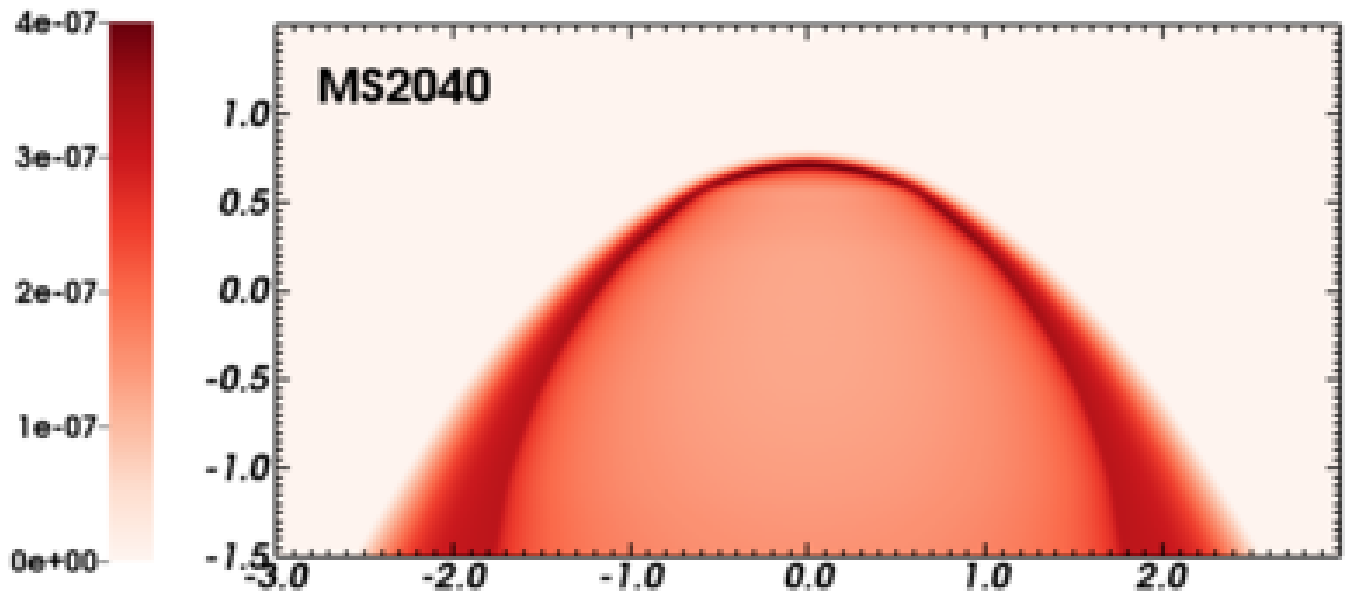}
	\end{minipage} \\

	\begin{minipage}[b]{ 0.48\textwidth}
		\includegraphics[width=1.0\textwidth]{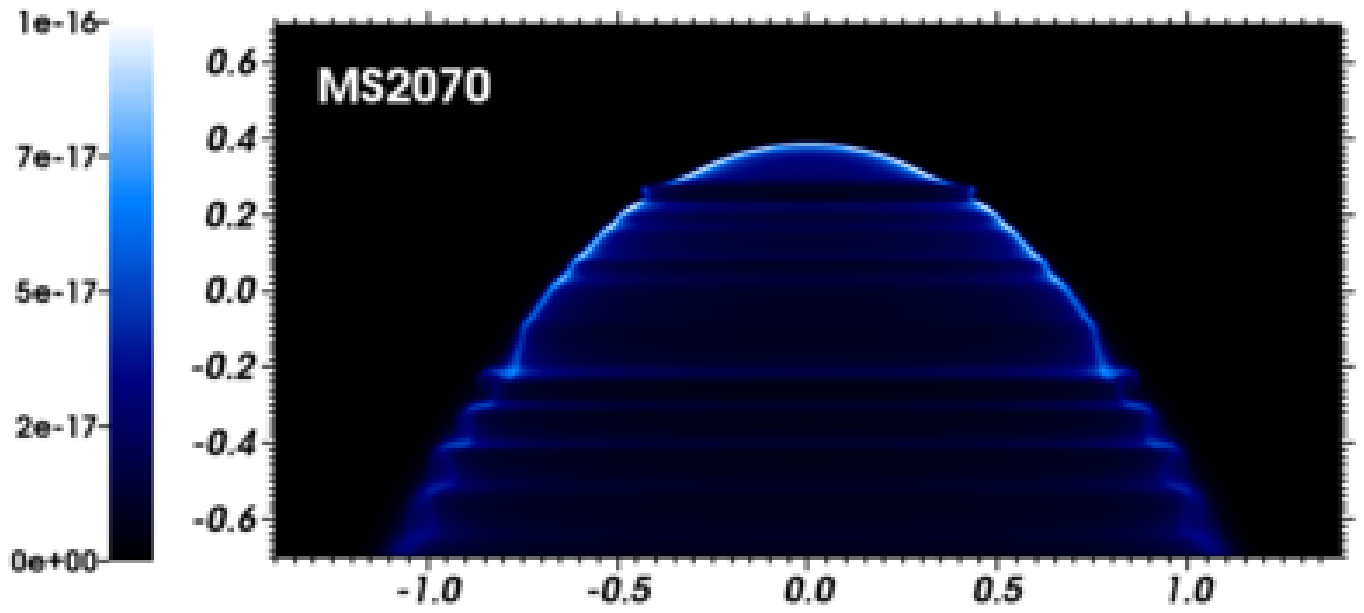}
	\end{minipage}
	\begin{minipage}[b]{ 0.48\textwidth}
		\includegraphics[width=1.0\textwidth]{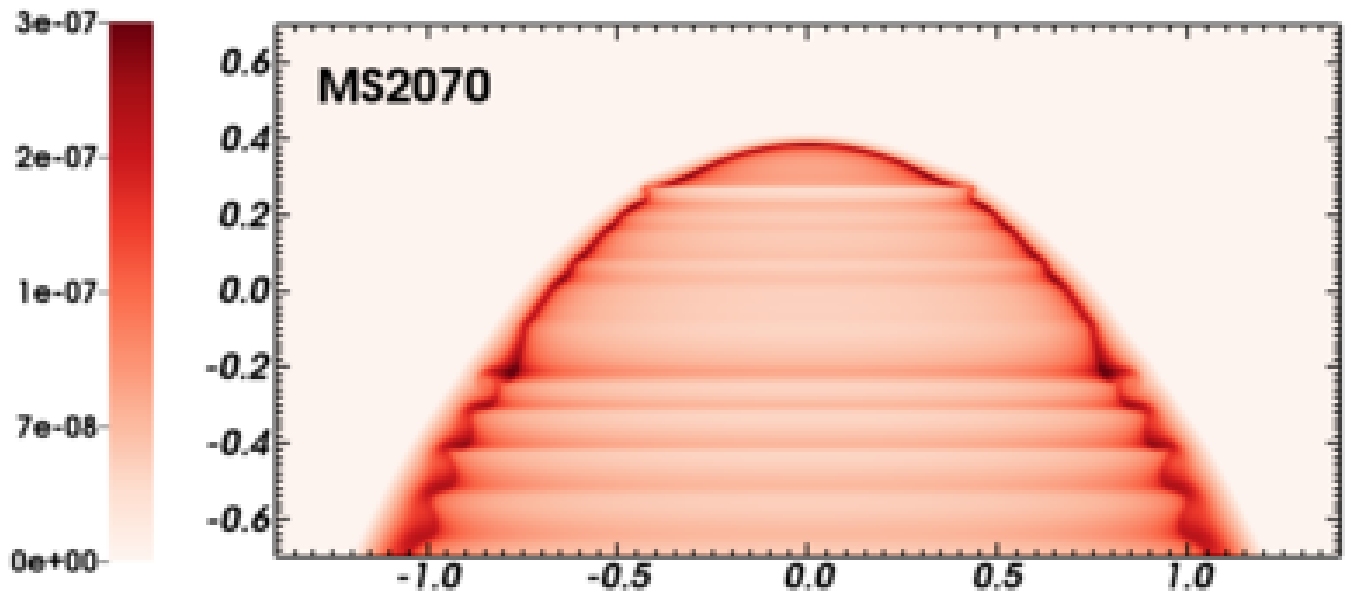}
	\end{minipage} \\  
	\caption{As Fig.~\ref{fig:projemms10}, with an initial stellar mass of $20\, M_{\odot}$.  }
	\label{fig:projemms20}  
\end{figure*}

\begin{figure*}
	\begin{minipage}[b]{ 0.48\textwidth}
		\includegraphics[width=1.0\textwidth]{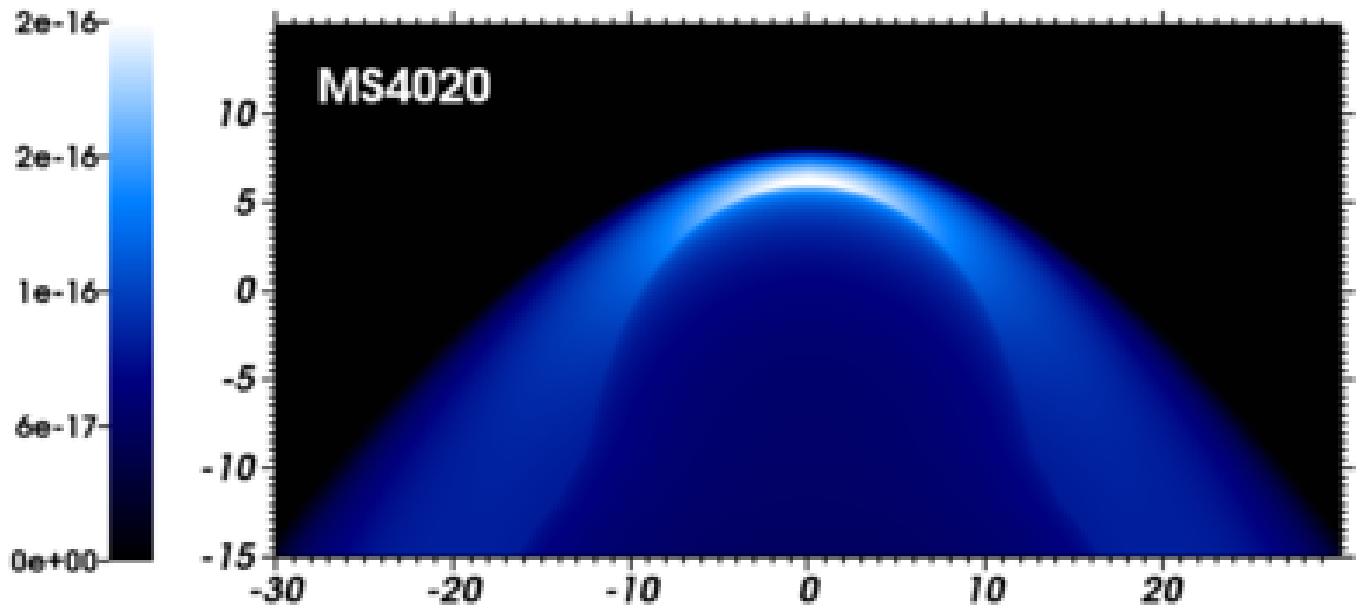}
	\end{minipage}
	\begin{minipage}[b]{ 0.48\textwidth}
		\includegraphics[width=1.0\textwidth]{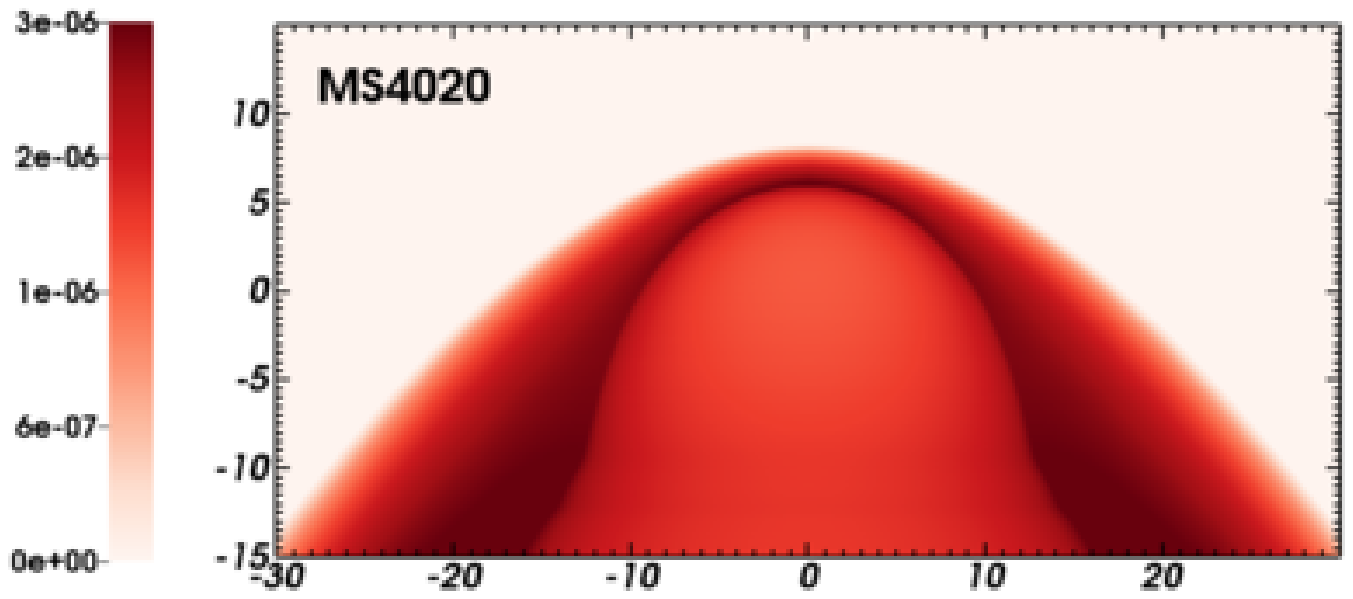}
	\end{minipage} \\

	\begin{minipage}[b]{ 0.48\textwidth}
		\includegraphics[width=1.0\textwidth]{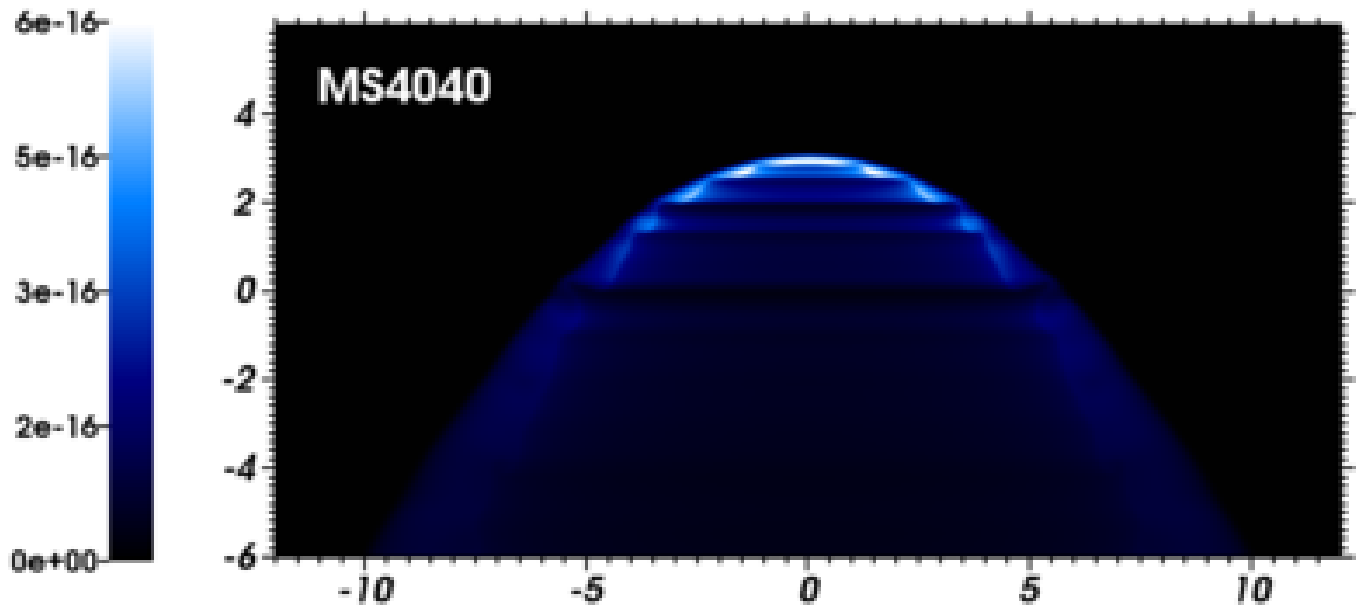}
	\end{minipage}
	\begin{minipage}[b]{ 0.48\textwidth}
		\includegraphics[width=1.0\textwidth]{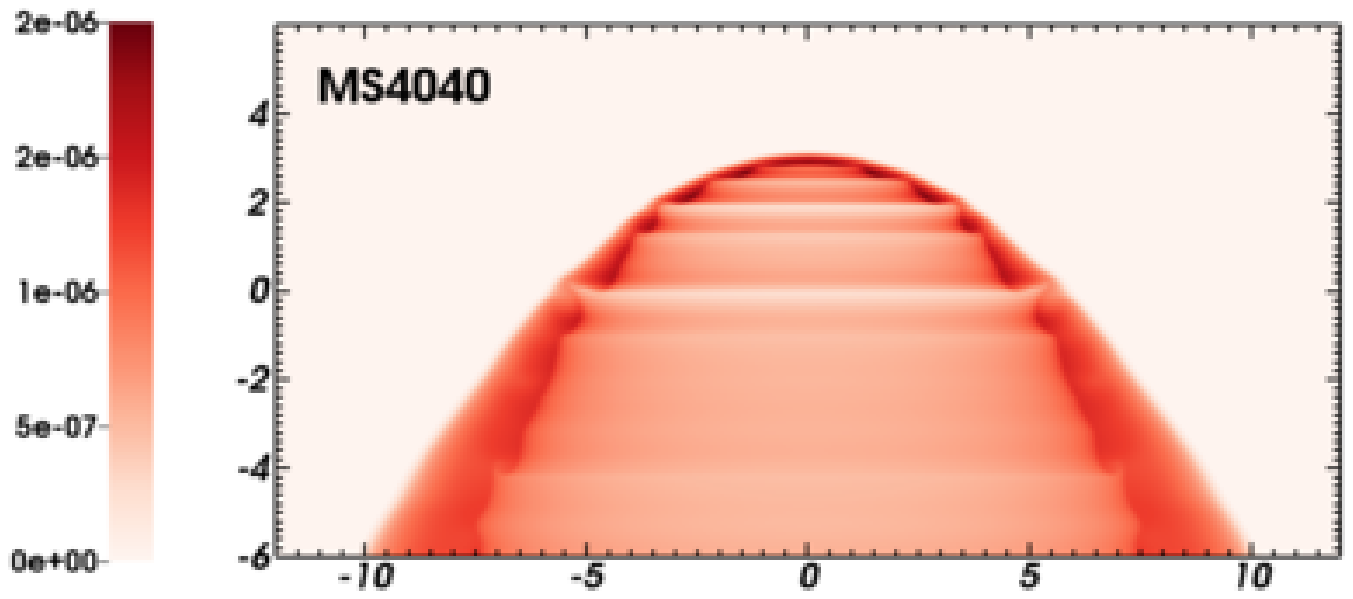}
	\end{minipage} \\

	\begin{minipage}[b]{ 0.48\textwidth}
		\includegraphics[width=1.0\textwidth]{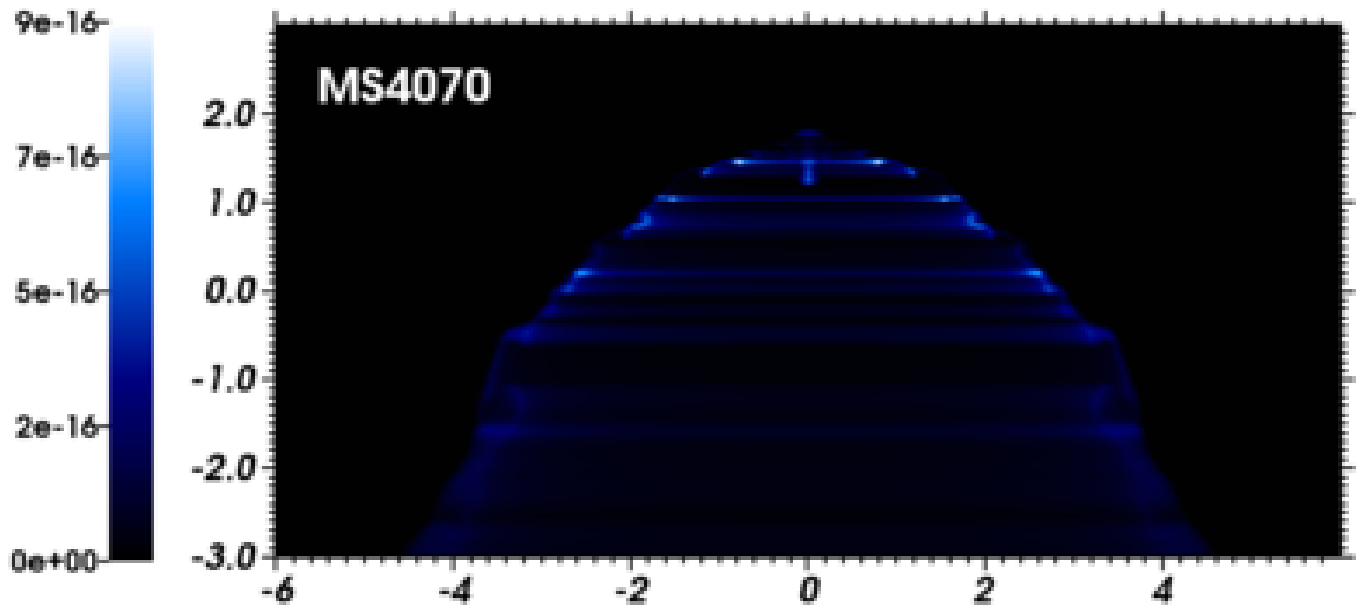}
	\end{minipage}
	\begin{minipage}[b]{ 0.48\textwidth}
		\includegraphics[width=1.0\textwidth]{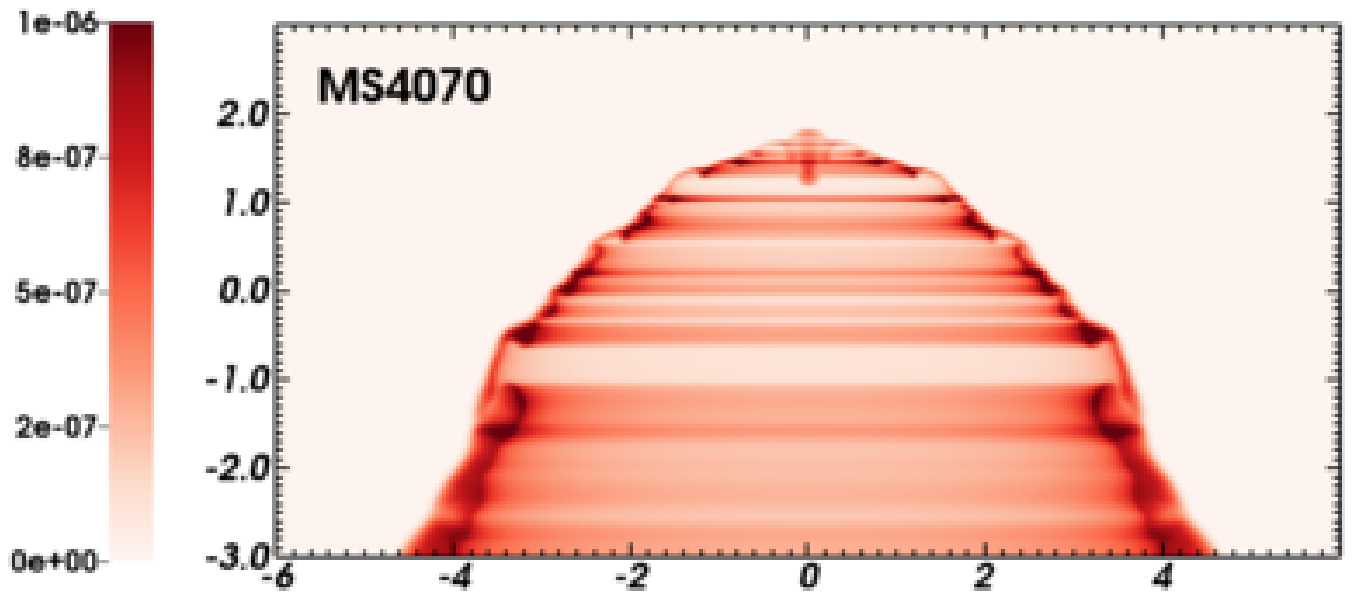}
	\end{minipage} \\  
	\caption{As Fig.~\ref{fig:projemms10}, with an initial stellar mass of $40\, M_{\odot}$.  }
	\label{fig:projemms40}  
\end{figure*}

Figs.~\ref{fig:projemms10},~\ref{fig:projemms20} and~\ref{fig:projemms40} show
synthetic H$\alpha$ emission maps of the bow shocks (left) together with dust
surface mass density maps (right), from the slowest ($v_{\star}=20\, \rm km\,
\rm s^{-1}$, top panels) to the highest ($v_{\star}=70\, \rm km\, \rm s^{-1}$,
bottom panels) models, respectively. These maps take into account the rotational
symmetry of the coordinate system (our Appendix~\ref{maps_Ha}). The
ISM background is ignored, i.e. we set its density to zero in the computation
of the projected emissivity and dust density, so that the surface brightness and
the surface mass density only refer to the bow shocks. The dust
surface density is calculated by projecting the shocked ISM gas weighted by a 
gas-to-dust ratio (our Appendix~\ref{maps_IR}), i.e. we considered that the 
wind material of a star is dust free during the main sequence.

The region of maximum H$\alpha$ surface brightness is located at the apex of the
bow shocks in the simulations with $v_{\star}=20\, \rm km\, \rm s^{-1}$ and
extends or displaces to its tail (i.e. $z<0$) as $v_{\star}$ increases. As the
ISM gas enters a bow shock generated by a main sequence star, its density increases and the material is heated
by thermal conduction towards the contact discontinuity, so its H$\alpha$
emissivity decreases (see panels (a) and (c) of Fig.~\ref{fig:profile}). The
competition between temperature increase and gas compression produces the
maximum emission at the contact discontinuity which separates hot and cold
shocked ISM gas. The reverse shock and the hot bubble are not seen because
of both their low density and their high post-shock temperature. Simulations with $v_{\star}
\ge 40\, \rm km\, \rm s^{-1}$ have their peak emissivity in the tail of the bow
shock because the gas does not have time to cool at the apex before it is advected
downstream. Simulations with high $v_{\star}$ and strong $\dot{M}$ (e.g. model
MS4070) have bow shocks shining in H$\alpha$ all along their contact
discontinuity, i.e. the behaviour of the H$\alpha$ emissivity with respect to
the large compression factor in the shell ($\propto n^{2}$) overwhelms that
of the post-shock temperature ($\propto T^{-0.9}$).

\begin{figure}
	\begin{minipage}[b]{1.0\textwidth}
		\includegraphics[width=0.335\textwidth,angle=270]{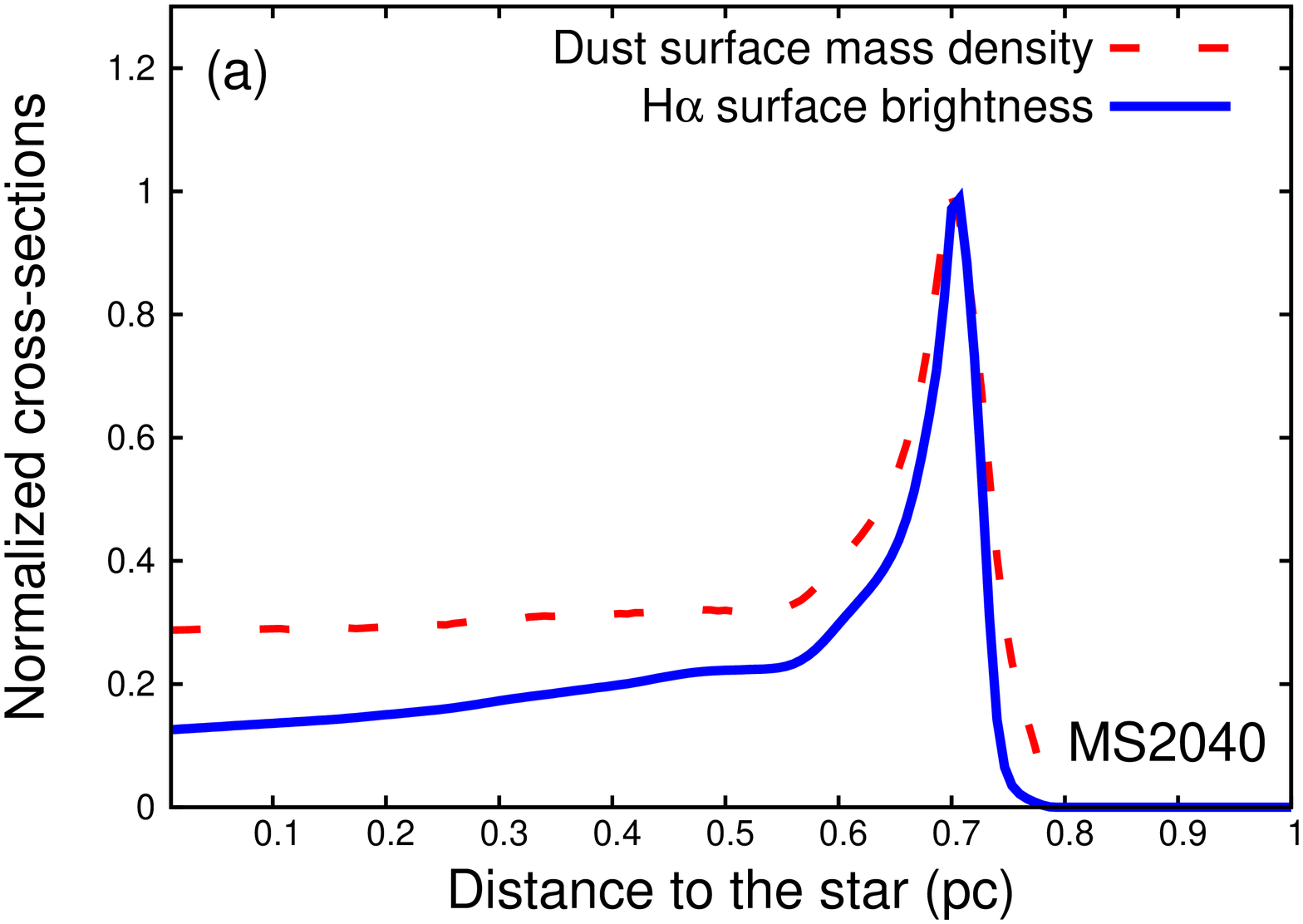}
	\end{minipage}  \\
	\begin{minipage}[b]{1.0\textwidth}
		\includegraphics[width=0.335\textwidth,angle=270]{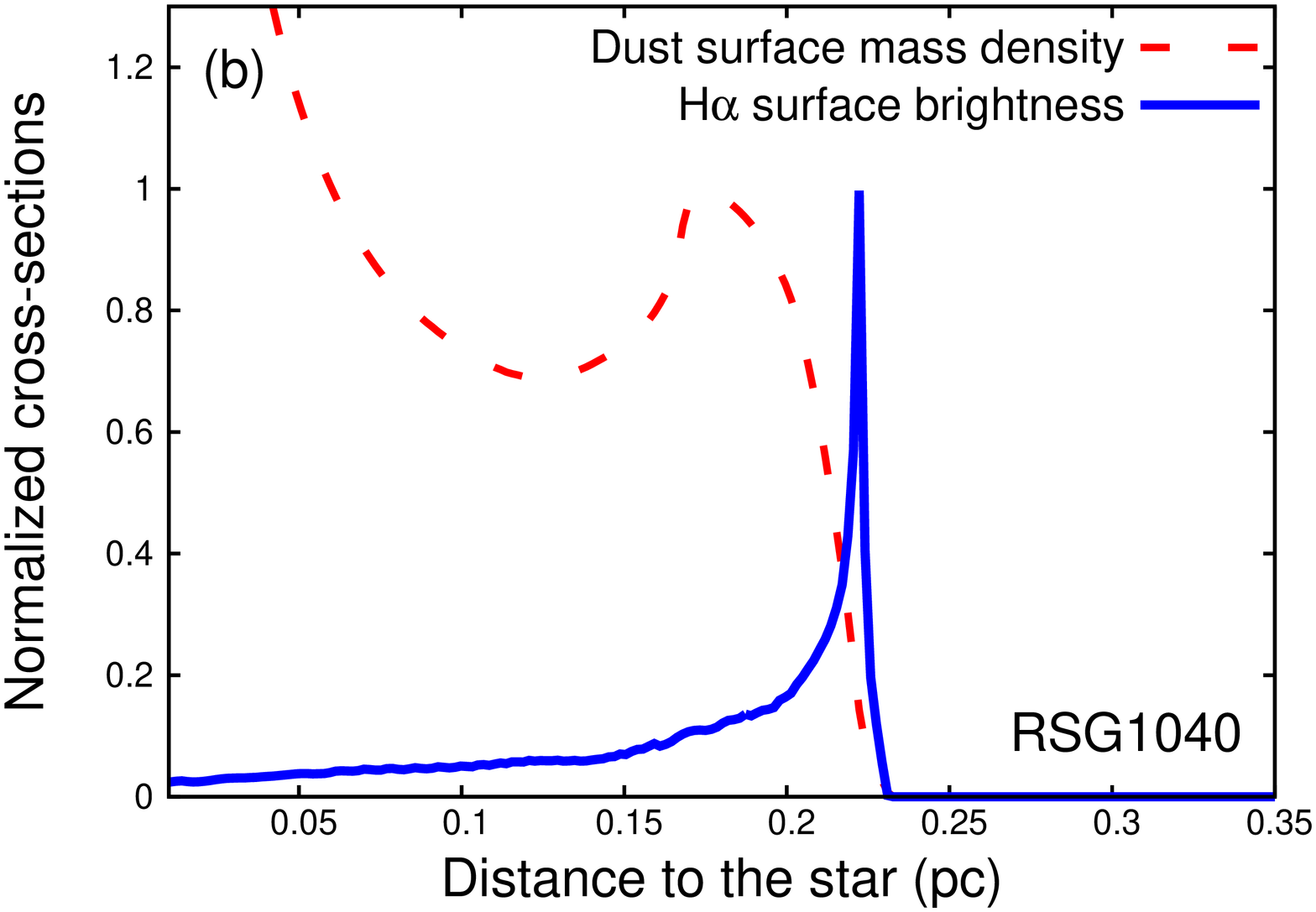}
	\end{minipage} 
	\caption{Normalized cross-sections taken along the direction of motion of the star, through the H$\alpha$ surface brightness
	and the dust surface mass density of the bow shock models MS2040 (a) and RSG1040 (b).
		 }
	\label{fig:cs}  
\end{figure}

The dust surface mass density increases towards the contact discontinuity (see
left panels of Figs.~\ref{fig:projemms10} to~\ref{fig:projemms40}). Panel (a) of
Fig.~\ref{fig:cs} shows that normalized cross-sections of both the H$\alpha$
surface brightness and the dust surface mass density of model MS2040, taken
along the direction of motion of the star in the $z \ge 0$ region of the bow
shock,  peak at the same distance from the star. We find a similar behaviour for
all our bow shock models of hot stars. This suggests that both maximum H$\alpha$
and infrared emission originate from the same region, i.e. near the contact
discontinuity in the cold region of shocked ISM material constituting the
outermost part of a bow shock generated \textcolor{black}{by} a main sequence star.

The maximum H$\alpha$ surface brightness of the brightest models (e.g. model
MS2020) is $\ge 6\times 10^{-17}\, \mathrm{erg}\, \mathrm{s}^{-1}\,
\mathrm{cm}^{-2}\, \mathrm{arcsec}^{-2}$, which is above the diffuse emission 
sensitivity limit of the SuperCOSMOS H-alpha Survey
(SHS;~\citealt{parker_mnras_362_2005}) of $1.1$$-$$2.8 \times 10^{-17}\,
\mathrm{erg}\, \mathrm{s}^{-1}\, \mathrm{cm}^{-2}\, \mathrm{arcsec}^{-2}$ and
could therefore be observed. The bow shocks around a central star less massive than 
$20\, M_{\odot}$ are fainter and could be screened by the $\HII$
region generated by their driving star. This could explain why we do not see many
stellar wind bow shocks around massive stars in H$\alpha$.

\section{ The stellar phase transition }
\label{sect:result_transition}

\begin{figure}
	\begin{minipage}[b]{0.46\textwidth}
		\includegraphics[width=1.0\textwidth]{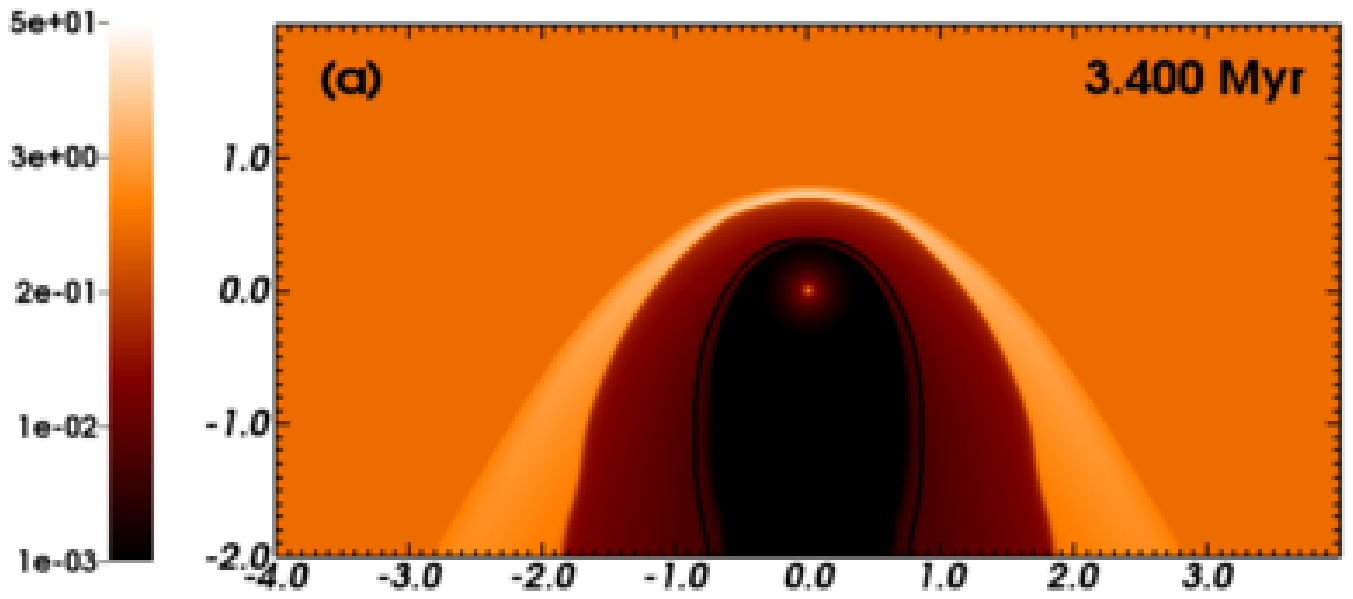}
	\end{minipage} \\
	\begin{minipage}[b]{0.46\textwidth}
		\includegraphics[width=1.0\textwidth]{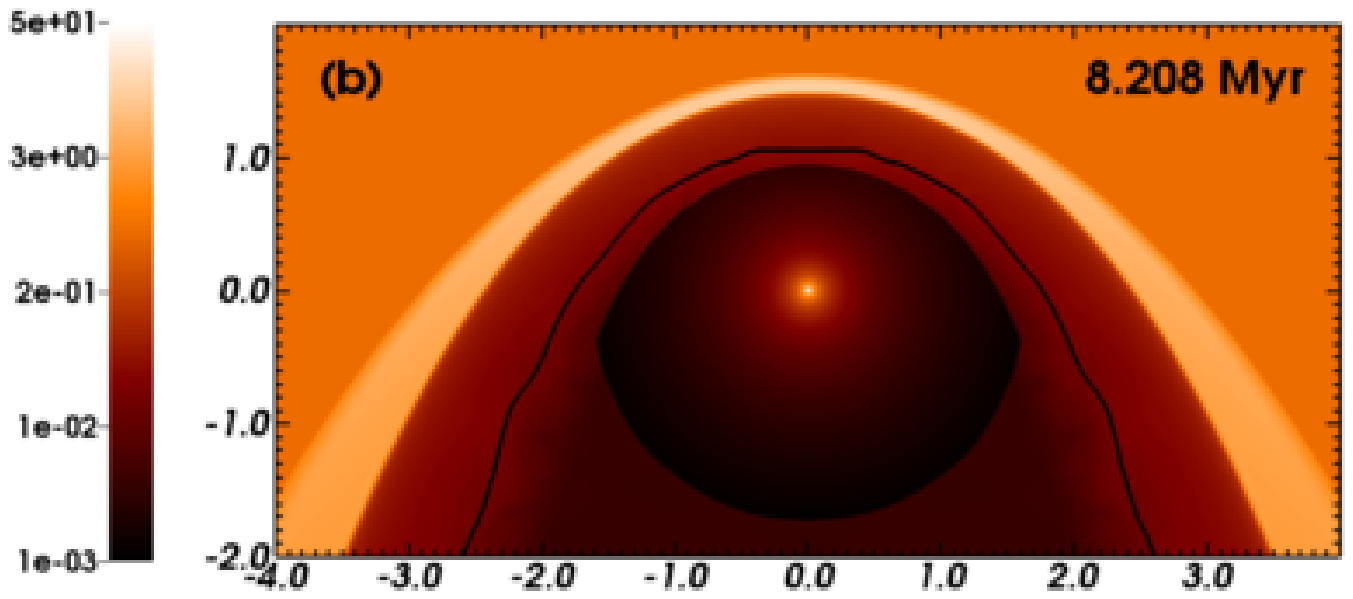}
	\end{minipage} \\
	\begin{minipage}[b]{0.46\textwidth}
		\includegraphics[width=1.0\textwidth]{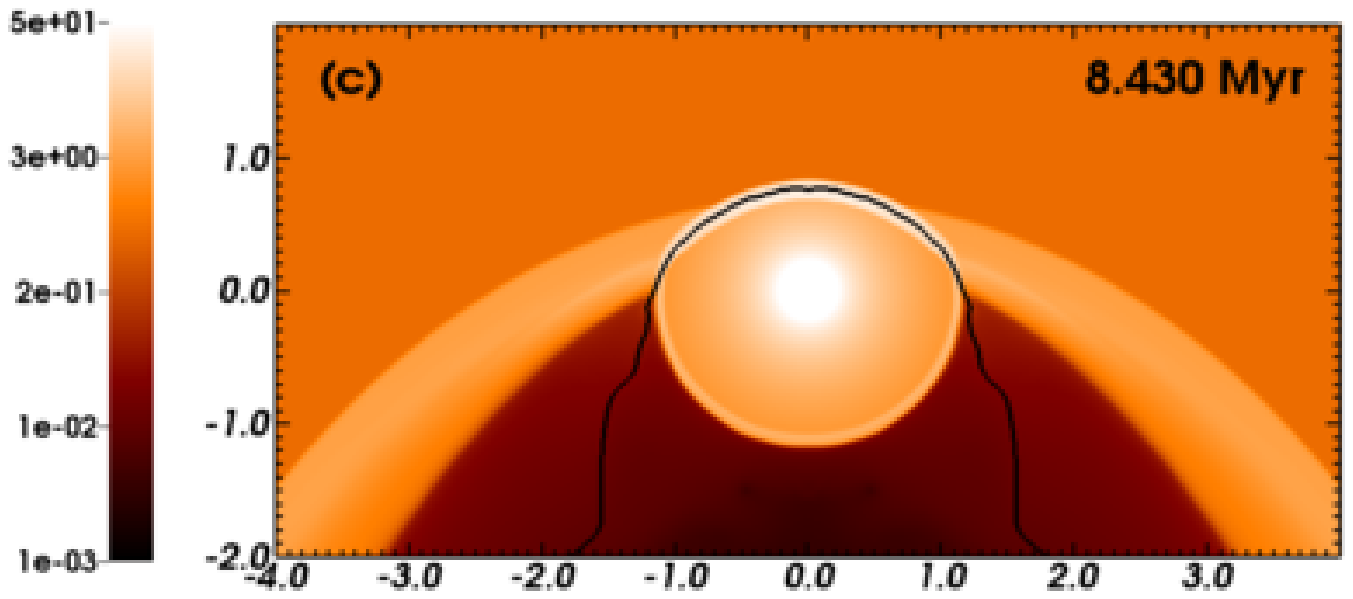}
	\end{minipage}\\
	\begin{minipage}[b]{0.46\textwidth}
		\includegraphics[width=1.0\textwidth]{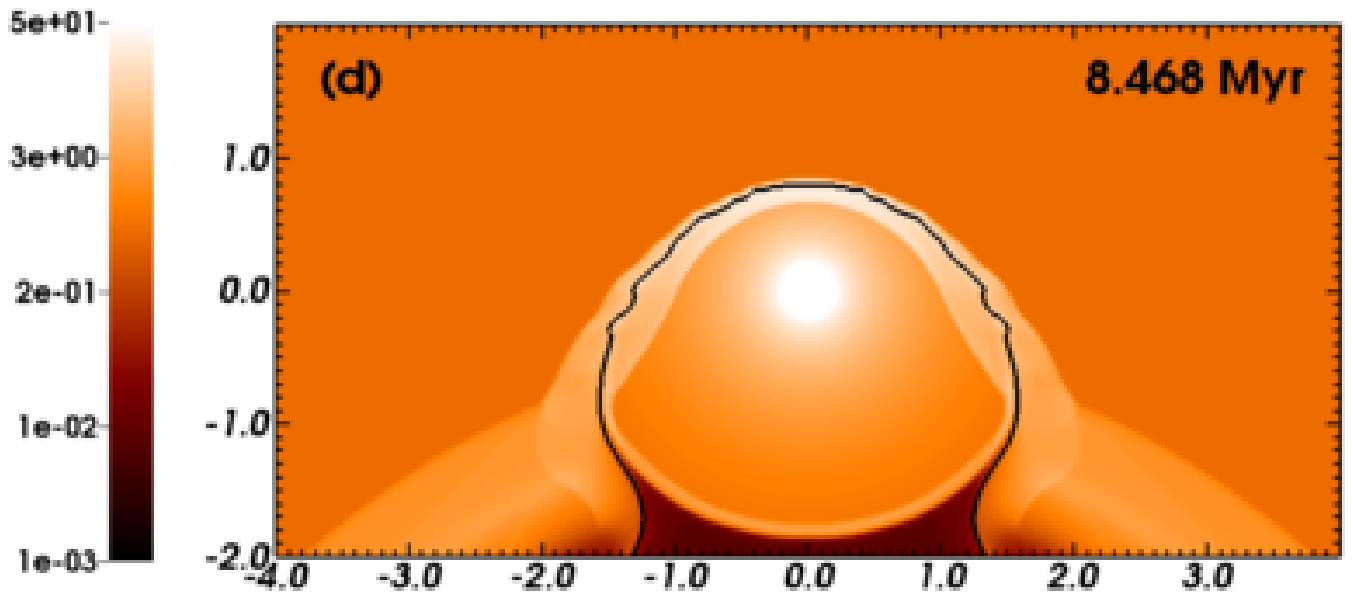}
	\end{minipage}\\
	\begin{minipage}[b]{0.46\textwidth}
		\includegraphics[width=1.0\textwidth]{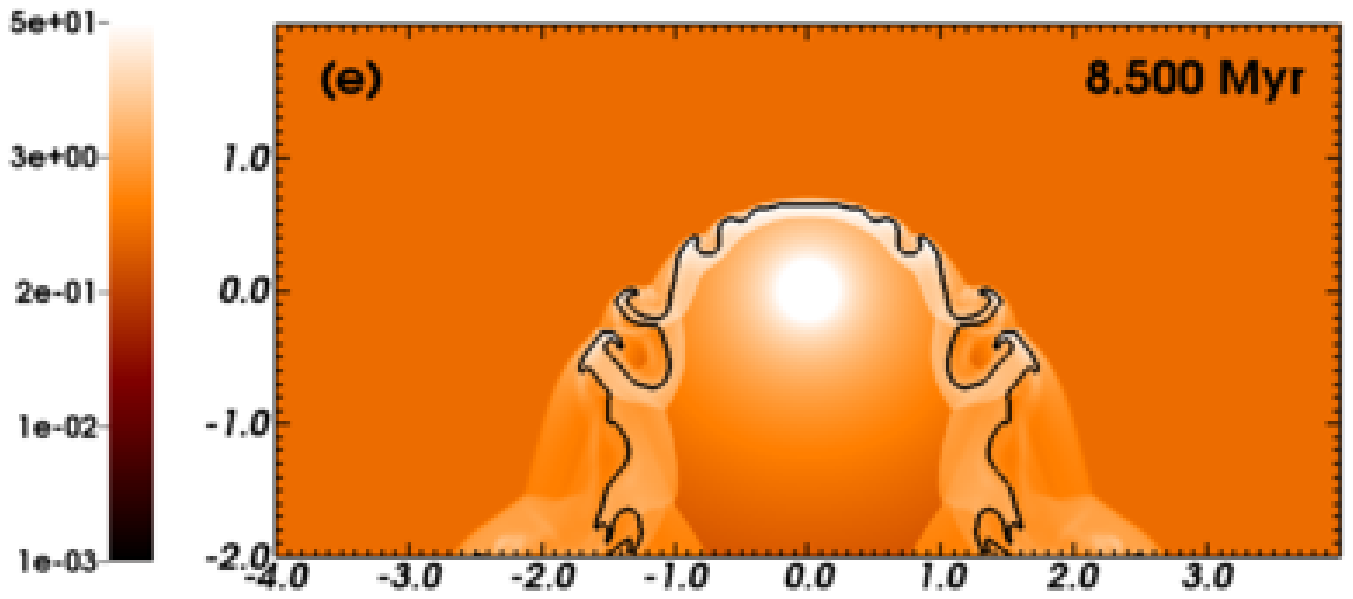}
	\end{minipage}
	\caption{
		\textcolor{black}{ Time sequence of the stellar phase transition of the initially $20\,
M_{\odot}$ star moving with $40\, \mathrm{km}\, \mathrm{s}^{-1}$. The
figures show the transition from the main sequence phase (top panel) to the red
supergiant phase (bottom panel) of the star. The gas number density is shown
with a density range from $10^{-3}$ to $5.0\, \mathrm{cm}^{-3}$ in the
logarithmic scale. The solid black contour
traces the boundary between wind and ISM material $Q(\bmath{r})=1/2$. The
$x$-axis represents the radial direction and the $y$-axis the direction of
stellar motion (in $\mathrm{pc}$). }
}
	\label{fig:m20trans}  
\end{figure}

\textcolor{black}{ 
In Fig.~\ref{fig:m20trans}, we show the gas density field in our bow shock 
model of our initially $20\, M_{\odot}$ star moving with velocity  
$v_{\star}=40\, \mathrm{km}\, \mathrm{s}^{-1}$ during the stellar phase 
transition from the main sequence phase (top panel) to the red supergiant phase 
(bottom panel). The figures correspond to times $3.400$, $8.208$, $8.430$, $8.468$ 
and $8.500\, \rm Myr$, respectively.  
}

\textcolor{black}{ 
The panel (a) of Fig.~\ref{fig:m20trans} shows the density field of the circumstellar medium
during the main-sequence phase of our star (as in the middle panel of Fig.~\ref{fig:m20ms}).
When the main sequence phase ends, both the stellar mass-loss rate $\dot{M}$ and wind
density $n_{\rm w}$ increase by more than an order of magnitude 
(see panel (e) of Fig.~\ref{fig:models_ms_rsg}) so that the
bow shock inflates and its stand-off distance doubles to reach about $1.7\, \rm pc$
(see panel (b) of Fig.~\ref{fig:m20trans}). At about $8.350\, \rm Myr$, the wind velocity decreases rapidly
and a shell of dense and slow red supergiant wind develops inside the
bow shock from the main sequence phase (see panel (c) of Fig.~\ref{fig:m20trans}). A
double-arced structure forms at its apsis, as shown in the study detailing a model of \textcolor{black}{Betelgeuse} 
returning to the red supergiant phase after undergoing a blue loop  
~\citep{mackey_apjlett_751_2012}. Under the influence of 
the stellar motion, the colliding shells expand beyond the forward shock 
of the main sequence bow shock and penetrate into the undisturbed ISM. 
The former bow shock recedes downwards from the direction of 
stellar motion because it is not supported by 
the ram pressure of the hot gas, whereas the new-born red supergiant bow
shock adjusts itself to the changes in the wind parameters and a new contact
discontinuity is established (see panel (d) of Fig.~\ref{fig:m20trans}). 
After the phase transition, only the bow shock from the red
supergiant phase remains in the domain (see panel (e) of Fig.~\ref{fig:m20trans}). 
}

\textcolor{black}{
As the star leaves the main sequence phase, the modifications of its wind
properties affect the strengths of its termination and forward shocks. 
The decelerating wind slows the gas velocity by about 2 orders of 
magnitude in the post-shock region at the reverse shock. The hot bubble cools rapidly 
($t_{\rm cool} \ll t_{\rm dyn} \ll t_{\rm cond}$) while the
region of shocked wind becomes thicker and denser (see panels (c)-(d) of 
Fig.~\ref{fig:models_ms_rsg}). The transfer of thermal energy by
heat conduction ceases because there is no longer a sharp temperature 
change $\Delta T \ge 10^{7}\, \rm K$ across the contact discontinuity.
Consequently, the position of the material discontinuity migrates from near the
reverse shock to be coincident with the contact discontinuity (see the 
solid black line in panels (a) to (c) of Fig.~\ref{fig:m20trans}). 
It sets up a dense and cold bow shock whose layer of shocked wind is thicker than 
the outer region of ISM gas (see panel (d) of Fig.~\ref{fig:m20trans}). }

\textcolor{black}{ 
The above described young bow shock of our initially $20\, M_{\odot}$ 
star is typical of the circumstellar medium of a runaway 
star undergoing a transition from a hot to a cold evolutionary phase.
The phase transition timescale is longer for small
$v_{\star}$ and shorter for high $v_{\star}$.
The bow shocks generated by lower mass stars, e.g. our initially $10\, M_{\odot}$ 
star may be more difficult to observe because of their smaller and fainter shells.   
The wind parameters of our initially $10\, M_{\odot}$ star change more abruptly 
($\sim 10^{4}\, \rm yr$, see panels (a) and (d) of
Fig.~\ref{fig:models_ms_rsg}), i.e. the preliminary increase
of $\dot{M}$ and $n_{\rm w}$ is quicker and the subsequent inflation 
of their bow shock is much less pronounced. The slightly inflated 
bow shock from the main sequence phase has no time to reach a steady 
state before the transition happens (as in panel (b) in
Fig.~\ref{fig:m20trans}). Our slowly moving star with velocity $20\, \rm km\, \rm s^{-1}$ 
(i.e. the model RSG2020) has a supergiant phase that is shorter than the
advection time of the hot bow shock, i.e. the former bow shock has not 
progressed downstream when the star ends its life (Section~\ref{sect:result_rsg}). 
}

\textcolor{black}{
Our stellar phase transitions last $10^{4}-10^{5}\, \mathrm{yr}$, i.e. they are much
shorter than both the main sequence and the red supergiant phases 
(see Fig.~\ref{fig:models_ms_rsg}). This makes the direct observation
of interacting bow shocks of stars in the field a rare event. Changes in 
the ambient medium can also affect the properties of bow shocks and wind 
bubbles, e.g. the so-called Napoleon's hat which surrounds the remnant 
of the supernova SN1987A~\citep{wampler_apj_362_1990,wang_aa_262_1992} 
and highlights the recent blue loop of its progenitor~\citep{wang_MNRAS_261_1993}. 
}

\section{The red supergiant phase}
\label{sect:result_rsg}

\subsection{Physical characteristics of the bow shocks}
\label{subsect:introduction_models_rsg}

\begin{figure}
	\begin{minipage}[b]{0.48\textwidth}
		\includegraphics[width=1.0\textwidth]{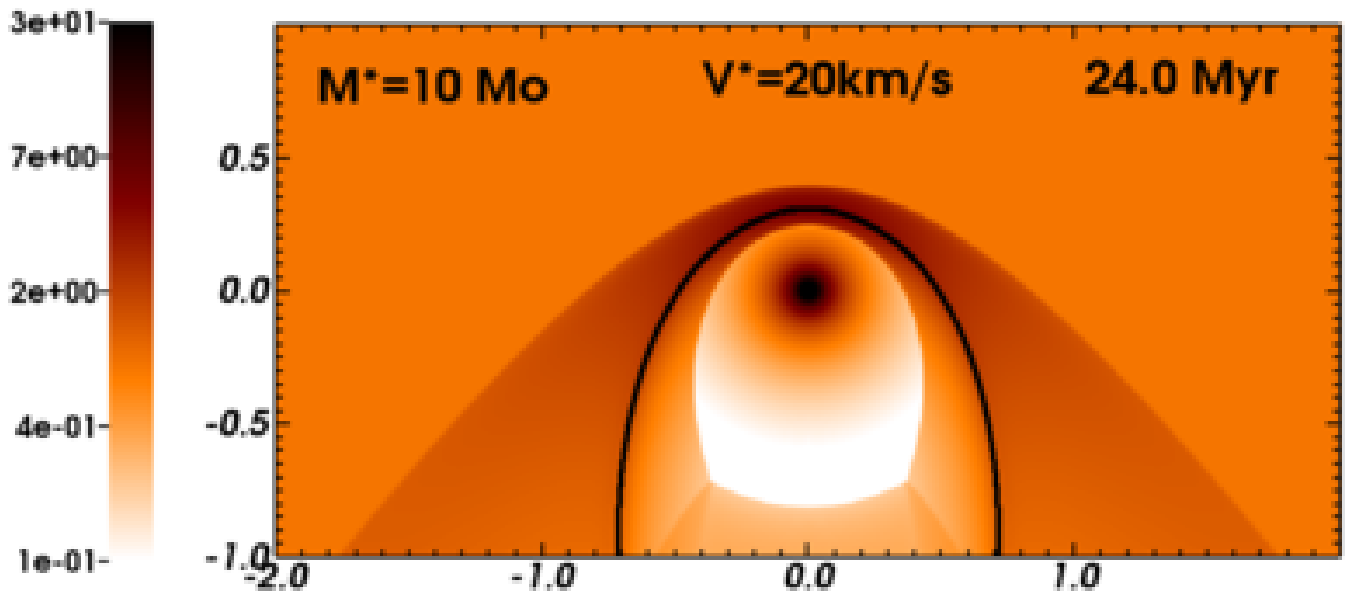}
	\end{minipage}  \\
	\begin{minipage}[b]{0.48\textwidth}
		\includegraphics[width=1.0\textwidth]{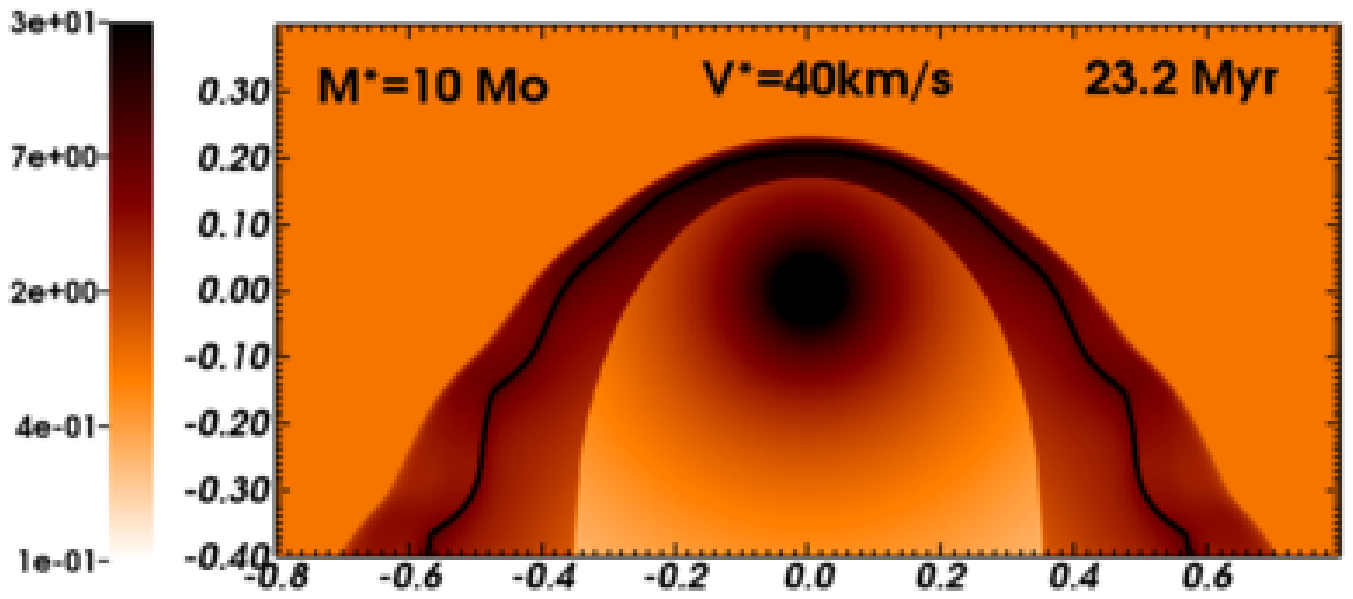}
	\end{minipage} \\
	\begin{minipage}[b]{0.48\textwidth}
		\includegraphics[width=1.0\textwidth]{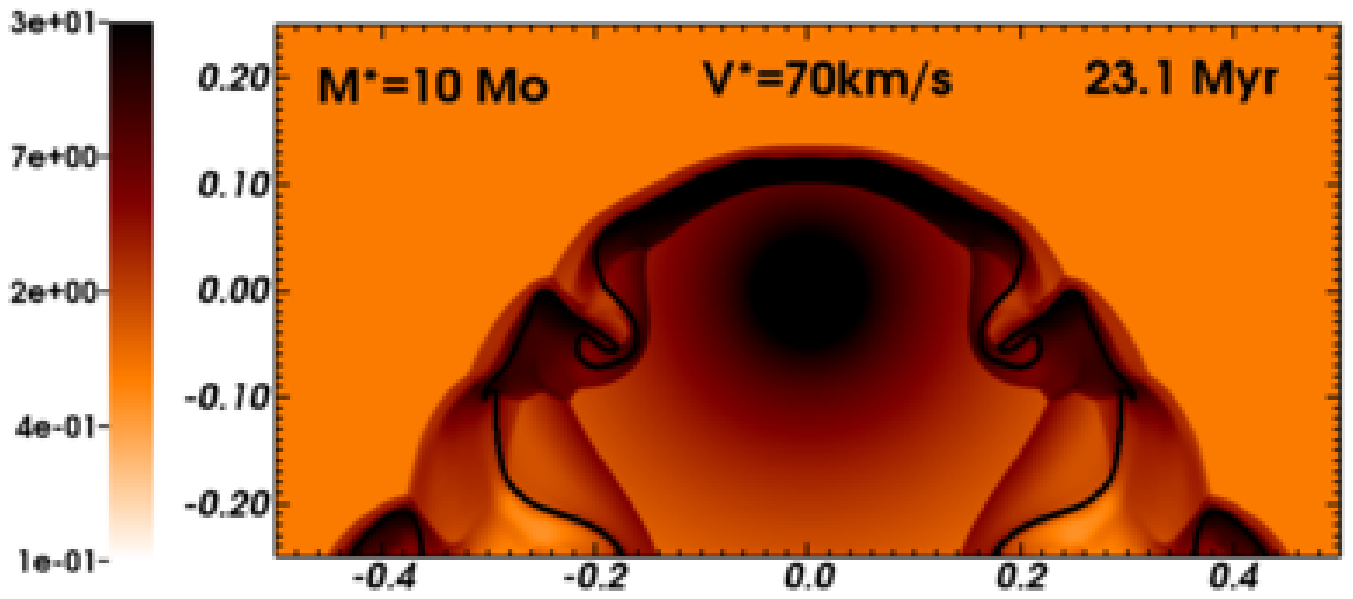}
	\end{minipage}
	\caption{Grid of stellar wind bow shocks from the red supergiant phase of the $10\, M_{\odot}$ initial mass star according 
		 to the space velocity with respect to the ISM, with $20\, \mathrm{km}\, \mathrm{s}^{-1}$ (top panel), 
		 $40\, \mathrm{km}\, \mathrm{s}^{-1}$ (middle panel) and $70\, \mathrm{km}\, \mathrm{s}^{-1}$ (bottom panel).
		 Models nomenclature follows Table~\ref{tab:ms}.  
		 Gas number density is shown with a density range from $0.1$ to $30.0\, {\rm cm}^{-3}$ in the logarithmic scale.
		 Note that the color scale is upset compared to Figs.~\ref{fig:m10ms},~\ref{fig:m20ms} and~\ref{fig:m40ms}. 
		 The solid black contours trace the boundary between wind and ISM, $Q(\bmath{r})=1/2$.
		 The $x$-axis represents the radial direction and the $y$-axis the direction of stellar motion (in $\mathrm{pc}$).
		 Only part of the computational domain is shown.  }
	\label{fig:m10rsg}  
\end{figure}

\begin{figure}
	\begin{minipage}[b]{0.48\textwidth}
		\includegraphics[width=1.0\textwidth]{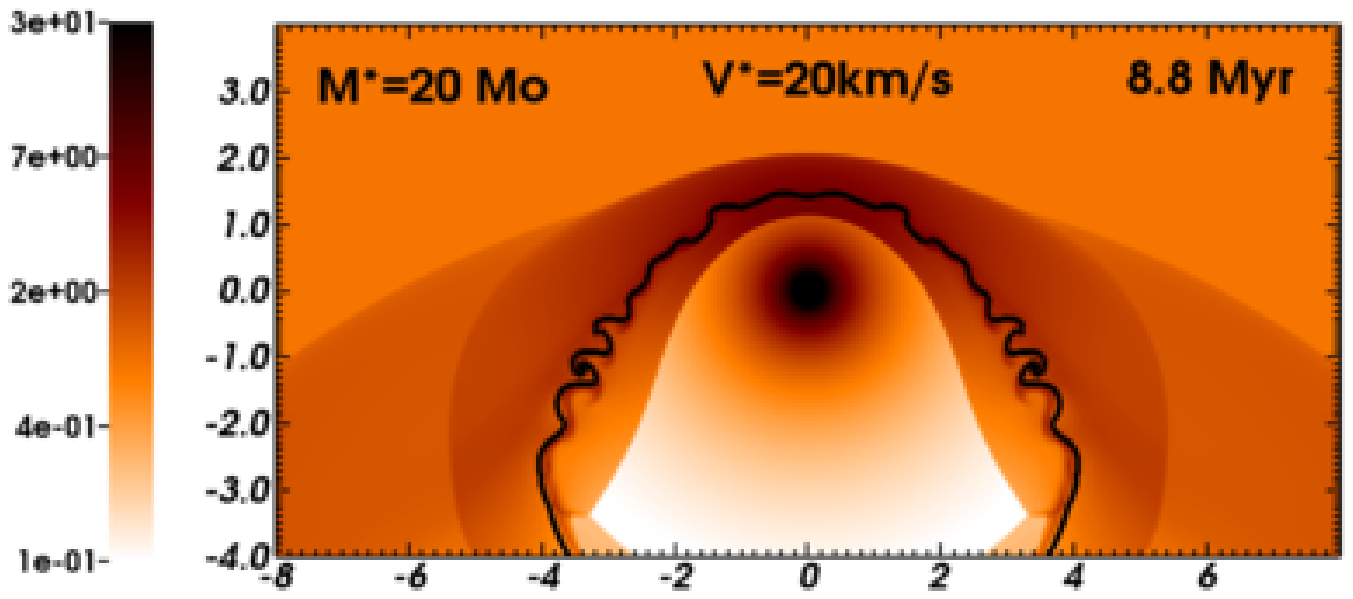}
	\end{minipage}  \\
	\begin{minipage}[b]{0.48\textwidth}
		\includegraphics[width=1.0\textwidth]{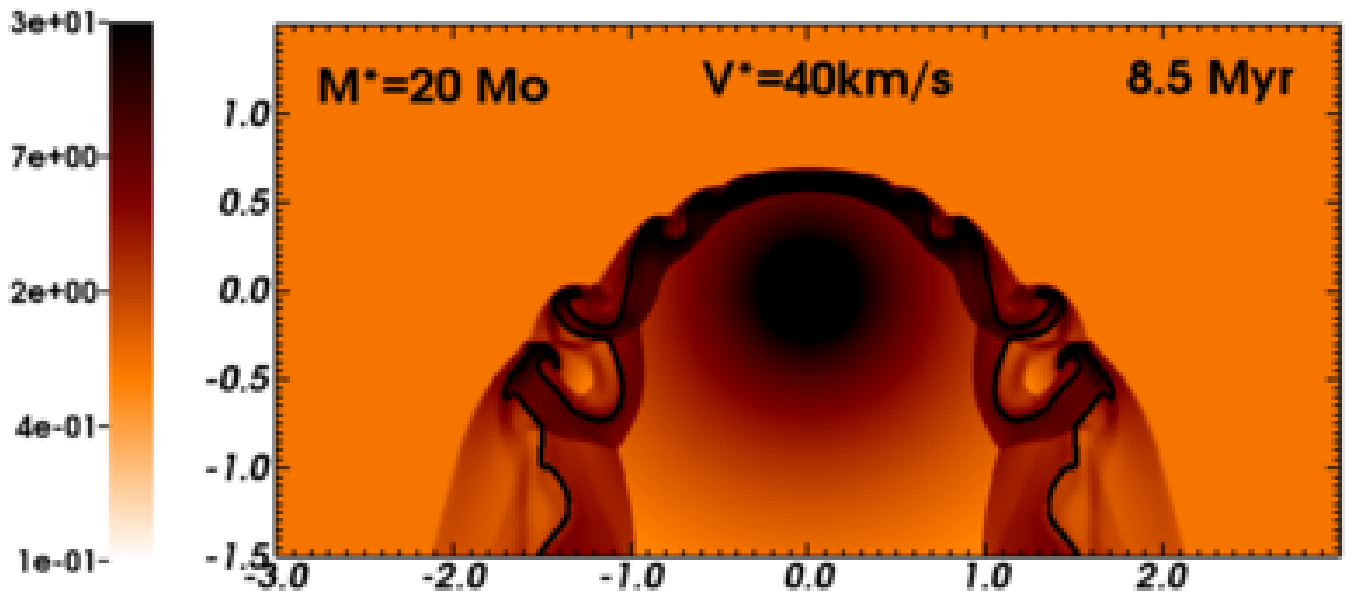}
	\end{minipage} \\
	\begin{minipage}[b]{0.48\textwidth}
		\includegraphics[width=1.0\textwidth]{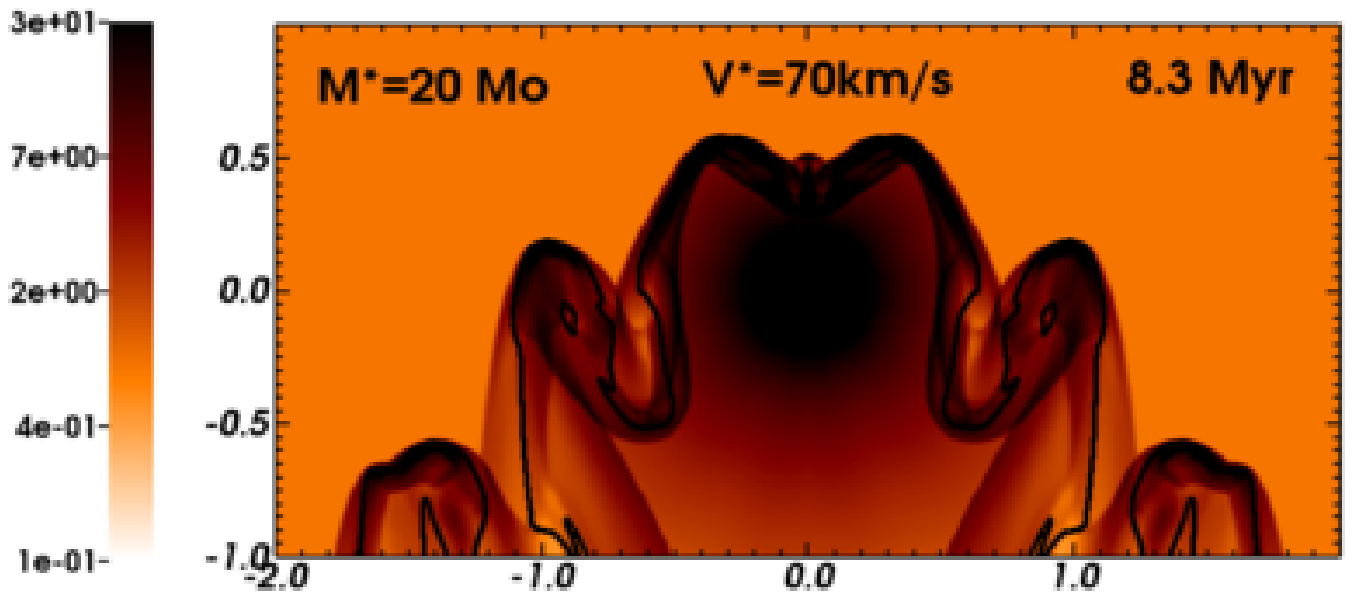}
	\end{minipage}
	\caption{As Fig.~\ref{fig:m10rsg}, with an initial stellar mass of $20\, M_{\odot}$.  }
	\label{fig:m20rsg}  
\end{figure}

We show the gas density field in our bow shock models of the red supergiant phase RSG1020 
($10\, M_{\odot}$ initial stellar mass, 
$v_{\star}=20\, \mathrm{km}\, \mathrm{s}^{-1}$, upper panel), 
RSG1040 ($10\, M_{\odot}$, $40\, \mathrm{km}\, \mathrm{s}^{-1}$, middle panel)
and RSG1070 ($10\, M_{\odot}$, $70\, \mathrm{km}\, \mathrm{s}^{-1}$, lower panel)
in Fig.~\ref{fig:m10rsg}. Fig.~\ref{fig:m20rsg} is similar for the $20\, M_{\odot}$ initial mass star.
Figs.~\ref{fig:m10rsg} and~\ref{fig:m20rsg} show the contour
$Q(\bmath{r})=1/2$ which traces the discontinuity between the wind and the ISM gas.
$R(0)$ and $R(0)/R(90)$ are summarised for each panel in Table~\ref{tab:params}.
The simulations were run until at least 40$\, t_{\rm cross}$ after the stellar
phase transition, i.e. after the abrupt increase of $\dot{M}$ accompanied by a
steep decrease of $v_{\rm w}$ (see panels (d)$-$(f) of Fig.~\ref{fig:models_ms_rsg}).

The size of the bow shocks is predicted to scale as $\dot{M}^{1/2}$, $v_{\rm
w}^{1/2}$ and $v_{\star}^{-1}$ according to Eq.~(\ref{eq:Ro}) 
and~\citet{baranov_sphd_15_1971}. The scaling between simulations with
$v_{\star} = 40\, \rm km\, s^{-1}$ and $v_{\star} = 70\, \rm km\, s^{-1}$
follows the prediction well, but deviations occur in the $v_{\star}=20\, \rm
km\, s^{-1}$ simulations (see Table~\ref{tab:params}). The most deviating
simulations either have a very weak shock preventing the forward shock from
cooling and forming a thin shell (e.g. model RSG1020), or have not reached a
steady state after the phase transition and consist of two interacting bow
shocks (e.g. model RSG2020).

The thickness of the shocked layers depends on the cooling physics of the gas.
Our simulations with $v_{\star} = 20\, \rm km\, \rm s^{-1}$ have a roughly
constant density across the material discontinuity. The reverse and forward shocks
are weak without much heating and both layers can cool to about the same temperature.
In models with $v_{\star} = 40\, \rm km\, \rm s^{-1}$ the post-shock temperature at the
forward shock is larger than for $v_{\star}=20\, \rm km\, s^{-1}$ and rapid
cooling to $T\approx 10^{4}\, \rm K$ leads to a stronger compression of the
material (see panel (b) and (d) of Fig.~\ref{fig:profile}). At $v_{\star}=70\,
\rm km\, s^{-1}$ the shocked ISM is a thin layer that has much lower density than
the shocked wind (e.g. models RSG1070 and RSG2070). The forward shock is strong,
therefore the hot shocked ISM has insufficient time to cool before it is advected
downstream.

Our model RSG1020 with the weakest shocks is stable. Model RSG2020 has an
expanding red supergiant wind that is replacing the previous main sequence
shell. This simulation still has the remainder of the main sequence wind bow
shock interacting with the bow shock from the red supergiant wind at the end of
the star life. The contact discontinuity of the supergiant shell shows
Rayleigh-Taylor fingers because of the density gradient between the old and new bow
shocks. Our models with $v_{\star} \ge 40\, \rm km\, \rm s^{-1}$ have $v_{\rm w}
\ll v_{\star}$ and so their bow shocks develop instabilities which
distort dense and thin shells~\citep{dgani_apr_461_1996}. 
The density field of the model RSG2070 resembles an isothermal bow shock with a distortion
of the forward shock typical of the non-linear thin shell and transverse
acceleration instabilities~\citep{blondin_na_57_1998}. This instability arises 
because $R(0)$ is much larger than the cooling length in the shocked ISM and shocked
wind.

$R(0)/R(90)$ decreases at the contact discontinuity as a function of $v_{\star}$,
e.g. $R(0)/R(90) \approx0.6$ and $\approx 0.58$ for models RSG2020 and
RSG2070, respectively. $R(0)/R(90)$ at the forward shock increases with
$v_{\star}$ and $\dot{M}$, e.g. model RSG2020 and RSG2070 have $R(0)/R(90)
\approx0.46$ and $\approx0.59$, respectively. These measures do not perfectly
satisfy Wilkin's solution, except for the models with $v_{\star}=70\, \rm km\,
\rm s^{-1}$, although the ratios for the contact discontinuity are all within
$10$ per cent of the analytic solution. Only the $v_{\star}=70\,
\mathrm{km}\, \mathrm{s}^{-1}$ simulations, with their thin bow shocks that come
closest to the isothermal limit, have forward shocks that satisfy
$R(0)/R(90)\approx 1/\sqrt{3}$ (see Fig.~\ref{fig:m20rsg}).

Because the temperature jumps are small across the interfaces and shocks in the
bow shocks around red supergiants, e.g. $\itl{ \Delta} T \approx
10^{3}\, \rm K$ at the reverse shock and $\itl{ \Delta} T \approx 4 \times
10^{4}\, \rm K$ at the forward shock of model RSG1040, thermal conduction
is not important. The bow shocks around red supergiants
therefore have coincident contact and material discontinuities (see 
black contours in Figs.~\ref{fig:projemrsg1} and~\ref{fig:projemrsg2}).

\begin{table*}
	\centering
	\caption{Stellar and bow shock luminosities. 
	 $L_{\star}$ represents the stellar luminosity at the end of each simulation,
	 $L_{\rm gas}$ is the bow shock luminosity from optically-thin cooling of the gas, 
	 and $L_{\rm wind}$ the part of $L_{\rm gas}$ originating from the wind material. 
	 $L_{\rm H\alpha}$ is the luminosity of H$\alpha$ emission and $L_{\rm FL }$ is the luminosity 
	 generated for photoionized bow shocks by cooling from [O\,{\sc ii}] and [O\,{\sc iii}] forbidden lines emission
	 is the range of $\approx8000 \le T \le 6.0\times 10^{4}\, \rm K$.
	 $L_{\rm IR}$ is the infrared luminosity, calculated on the basis of reemission of starlight by the dust grains 
	 (our Appendix~\ref{maps_IR}). $\widetilde{\itl{\Gamma}}_{\alpha}$ represents the radiative heating of the gas (see Eq.~\ref{eq:bsheat}).
	 }
	\begin{tabular}{cccccccc}
	\hline
	\hline
	${\rm {Model}}$  &   $L_{\star}\, (\rm erg\, s^{-1})$
			 &   $L_{\rm gas}\, (\rm erg\, s^{-1})$
			 &   $L_{\rm wind}\, (\rm erg\, s^{-1})$
			 &   $L_{\rm H\alpha}\, (\rm erg\, s^{-1})$ 
			 &   $L_{\rm FL}\, (\rm erg\, s^{-1})$ &   $L_{\rm IR}\, (\rm erg\, s^{-1})$
 			 &   $\widetilde{\itl{\Gamma}}_{\alpha}\, (\rm erg\, s^{-1})$                                        
			\\ \hline   
	MS1020   &  $2.42 \times 10^{37}$     & $1.39 \times 10^{31}$ & $4.66 \times 10^{24}$ &  $9.00 \times 10^{29}$ & $1.26\times 10^{31}$ & $2.10  \times 10^{33}$ & $5.32\times 10^{30}$  \\                    
	MS1040   &  $2.42 \times 10^{37}$     & $6.17 \times 10^{30}$ & $1.55 \times 10^{25}$ &  $4.10 \times 10^{28}$ & $5.75\times 10^{30}$ & $7.00  \times 10^{32}$ & $2.53\times 10^{29}$  \\     
	MS1070   &  $2.42 \times 10^{37}$     & $4.70 \times 10^{30}$ & $1.76 \times 10^{25}$ &  $2.40 \times 10^{27}$ & $3.53\times 10^{30}$ & $3.50  \times 10^{32}$ & $1.21\times 10^{28}$  \\      
	\hline 
	RSG1020  &  $7.66 \times 10^{37}$     & $1.36 \times 10^{32}$ & $1.35 \times 10^{32}$ &  $1.60 \times 10^{27}$ & $-$ & $6.20  \times 10^{33}$ & $2.27\times 10^{30}$  \\
	RSG1040  &  $7.32 \times 10^{37}$     & $1.35 \times 10^{32}$ & $1.30 \times 10^{30}$ &  $7.10 \times 10^{28}$ & $-$ & $1.30  \times 10^{34}$ & $3.62\times 10^{28}$  \\       
	RSG1070  &  $7.32 \times 10^{37}$     & $3.50 \times 10^{32}$ & $1.73 \times 10^{30}$ &  $7.50 \times 10^{28}$ & $-$ & $2.40  \times 10^{34}$ & $2.30\times 10^{28}$  \\ 
	\hline   
	MS2020   &  $2.59 \times 10^{38}$     & $6.60 \times 10^{33}$ & $7.50 \times 10^{27}$ &  $1.30 \times 10^{33}$ & $5.57\times 10^{33}$ & $2.10  \times 10^{35}$ & $7.90 \times 10^{33}$  \\       
	MS2040   &  $2.16 \times 10^{38}$     & $2.48 \times 10^{33}$ & $6.83 \times 10^{27}$ &  $1.90 \times 10^{32}$ & $2.17\times 10^{33}$ & $6.20  \times 10^{34}$ & $1.00 \times 10^{33}$  \\ 
	MS2070   &  $1.64 \times 10^{38}$     & $2.32 \times 10^{33}$ & $6.18 \times 10^{28}$ &  $3.60 \times 10^{31}$ & $1.69\times 10^{33}$ & $2.40  \times 10^{34}$ & $2.20\times 10^{32}$  \\
	\hline 
	RSG2020  &  $5.94 \times 10^{38}$     & $2.46 \times 10^{32}$ & $4.70 \times 10^{31}$ &  $9.70 \times 10^{27}$ & $-$ & $1.30  \times 10^{36}$ & $1.34\times 10^{31}$  \\      
	RSG2040  &  $5.18 \times 10^{38}$     & $1.56 \times 10^{33}$ & $1.80 \times 10^{31}$ &  $7.10 \times 10^{29}$ & $-$ & $4.30  \times 10^{35}$ & $1.42\times 10^{31}$  \\  
	RSG2070  &  $5.95 \times 10^{38}$     & $3.65 \times 10^{33}$ & $2.04 \times 10^{31}$ &  $1.70 \times 10^{30}$ & $-$ & $1.20  \times 10^{36}$ & $9,98\times 10^{30}$  \\ 
	\hline 
	MS4020   &  $1.30 \times 10^{39}$     & $3.90 \times 10^{35}$ & $8.00 \times 10^{29}$ &  $8.30 \times 10^{34}$ & $3.30\times 10^{35}$ & $4.00  \times 10^{36}$ & $4.76\times 10^{35}$  \\                 
	MS4040   &  $1.03 \times 10^{39}$     & $1.00 \times 10^{35}$ & $3.70 \times 10^{29}$ &  $1.60 \times 10^{34}$ & $8.74\times 10^{34}$ & $1.20  \times 10^{36}$ & $9.10\times 10^{34}$  \\       
	MS4070   &  $9.00 \times 10^{38}$     & $5.40 \times 10^{34}$ & $2.80 \times 10^{29}$ &  $2.80 \times 10^{33}$ & $4.46\times 10^{34}$ & $4.50  \times 10^{35}$ & $1.60\times 10^{34}$  \\          
	\hline 
	\end{tabular}
\label{tab:lum_val}
\end{table*}


\subsection{Bow shock emissivity}
\label{subsect:rsglum}

\subsubsection{Luminosities}
\label{subsubsect:rsglumscalar}

The luminosities $L_{\rm gas}$, $L_{\rm wind}$, $L_{\rm H\alpha}$ and $L_{\rm
IR}$ of the bow shocks generated by our red supergiant models are plotted as a function of
$\dot{M}$ and $v_{\star}$ in panel (b) of Fig.~\ref{fig:bslum}. As is the case for bow
shocks produced by main sequence stars, $L_{\rm gas}$ is influenced by
$v_{\star}$ and by the size of the bow shock. $L_{\rm gas}\, \propto n^{2}$
and slightly increases with $v_{\star}$ because the compression factor of the
shell is larger for high $v_{\star}$. The variations in size drive the increase
of $L_{\rm gas}$ as a function of $\dot{M}$ if $v_{\star}$ is fixed. 
In contrast to the bow shocks around main sequence stars, the increase of
$L_{\rm gas}$ seen in panel (b) of Fig.~\ref{fig:bslum} for a given model triplet
shows that the luminosity is more influenced by the density than by the volume
of the bow shocks.

$L_{\rm wind}$ is several orders of magnitude dimmer than $L_{\rm gas}$,
e.g. $L_{\rm wind}/L_{\rm gas}\approx 10^{-2}$ for model RSG1040, i.e. the wind
contribution is negligible compared to the luminosity of the shocked ISM gas.
The difference between $L_{\rm wind}$ and $L_{\rm gas}$ is less than
in our main sequence models because the gas cooling behind the slow red
supergiant reverse shock is efficient. Model RSG1020 behaves differently
because even though it scales in volume with model RSG1040, its small
$v_{\star}$ \textcolor{black}{results in} a weak forward shock which is cool 
so there is little cooling in the shocked ISM ($L_{\rm wind}
\sim L_{\rm gas}$). The total bow shock luminosity of optically-thin
radiation of model RSG2020 is \textcolor{black}{increased by a contribution from}
the former main sequence bow shock around the forming red supergiant shell 
(see upper panel of Fig.~\ref{fig:m20rsg}).

The bow shock luminosity of H$\alpha$ emission is negligible compared to the
total bow shock luminosity, e.g. $L_{\rm H\alpha}/L_{\rm gas} \approx
10^{-3}$$-$$10^{-5}$, see lower panel of Fig.~\ref{fig:bslum}. $L_{\rm H\alpha}$
increases with $v_{\star}$, e.g. $L_{\rm H\alpha}\approx 7.1 \times
10^{29}$ and $\approx 1.7 \times 10^{30}\, \mathrm{erg}\, \mathrm{s}^{-1}$ for
model RSG2040 and RSG2070, respectively. The H$\alpha$ emission of the bow
shocks for the $10$ and $20\, \rm M_{\odot}$ stars differs by $\approx 1$ order
of magnitude. Models RSG1020 and RSG2020 have little H$\alpha$ emission
because their weak forward shocks do not ionize the gas significantly
and prevent the formation of a dense shell.

The infrared luminosity is such that $L_{\rm IR} \gg L_{\rm gas}$. This is because
of the fact that $L_{\rm IR}$ provides an upper limit for the infrared light
(our Appendix~\ref{maps_IR}) and because the circumstellar medium around red
supergiants is denser than that during the main sequence phase, i.e. there is a
lot of dust from the stellar wind in these bow shocks that can reprocess the
stellar radiation. $L_{\rm IR}$ increases by about two orders of magnitude
between the $10$ and $20\, \rm M_{\odot}$ models if $v_{\star}$ is considered
fixed, which is explained by their different wind and bow shock densities (see
Figs.~\ref{fig:m10rsg} and~\ref{fig:m20rsg}). Model RSG2020 does not fit this
trend because the huge mass of the bow shock of the previous evolutionary phase
affects its luminosity $L_{\rm IR}\approx 1.3 \times 10^{36}\,
\mathrm{erg}\, \mathrm{s}^{-1}$. The enormous infrared luminosity of bow
shocks around red supergiants compared to their optically-thin gas radiation
suggests that they should be more easily observed in the infrared than in the optical
bands and partly explains why the bow shock around Betelgeuse was discovered in
the infrared.


\subsubsection{Synthetic emission maps}
\label{subsubsect:rsglummaps}

\begin{figure*}
	\begin{minipage}[b]{ 0.48\textwidth}
		\includegraphics[width=1.0\textwidth]{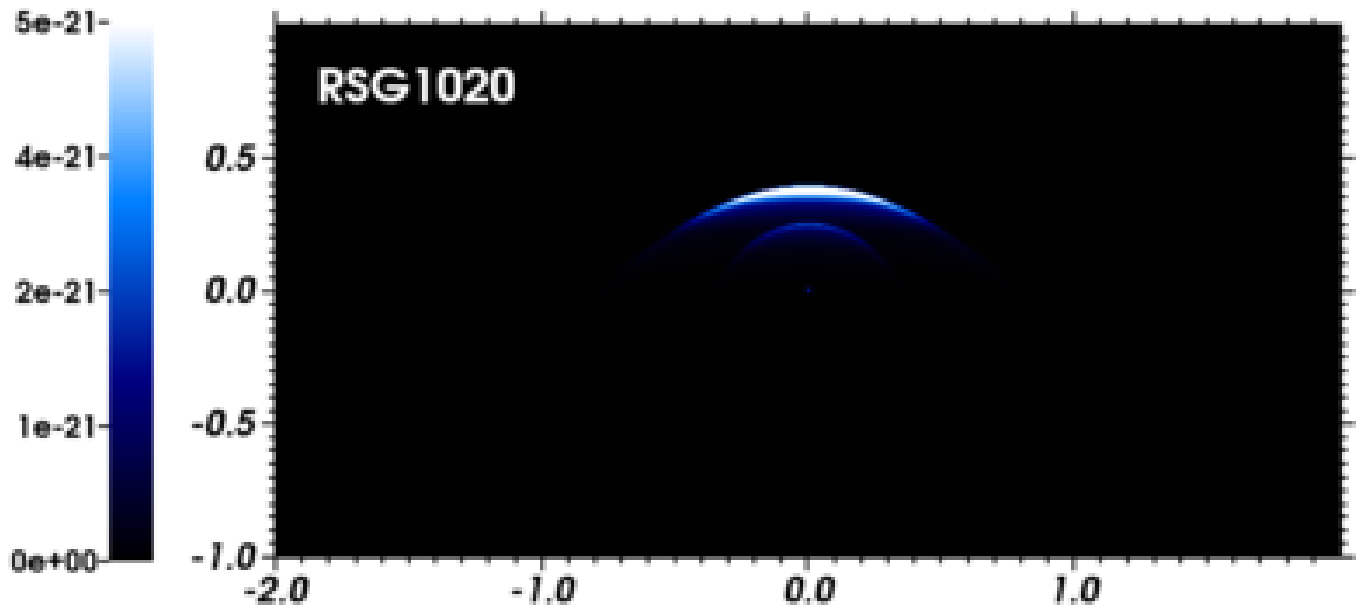}
	\end{minipage}
	\begin{minipage}[b]{ 0.48\textwidth}
		\includegraphics[width=1.0\textwidth]{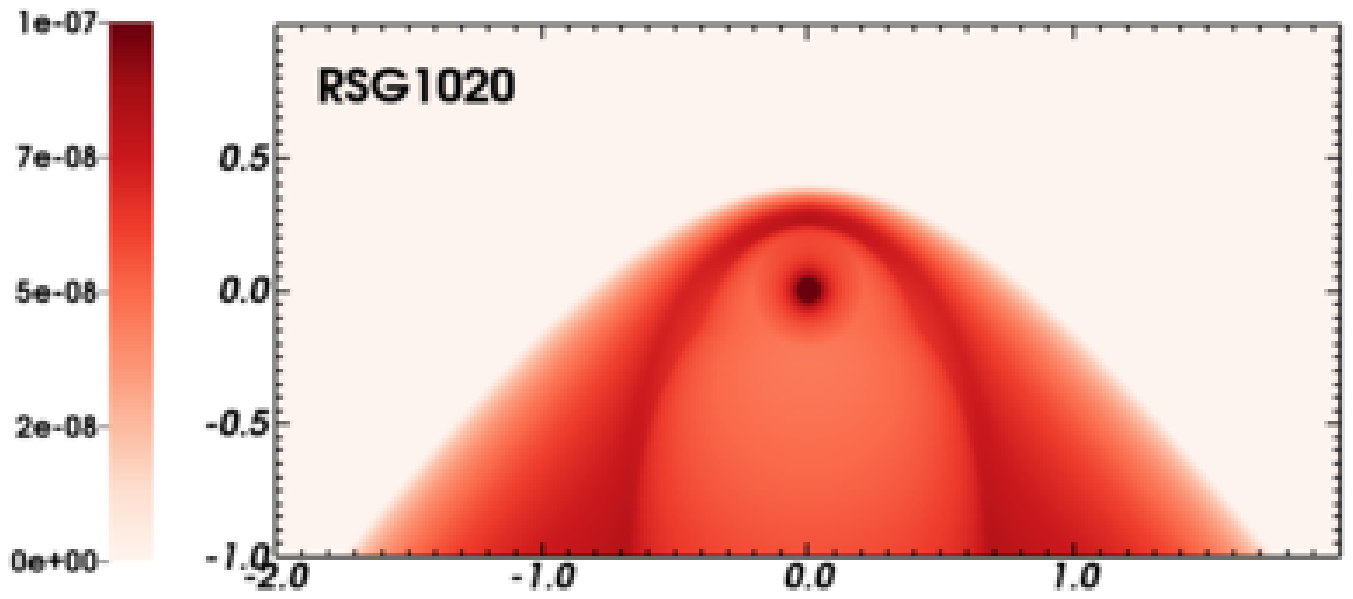}
	\end{minipage} \\
	\begin{minipage}[b]{ 0.48\textwidth}
		\includegraphics[width=1.0\textwidth]{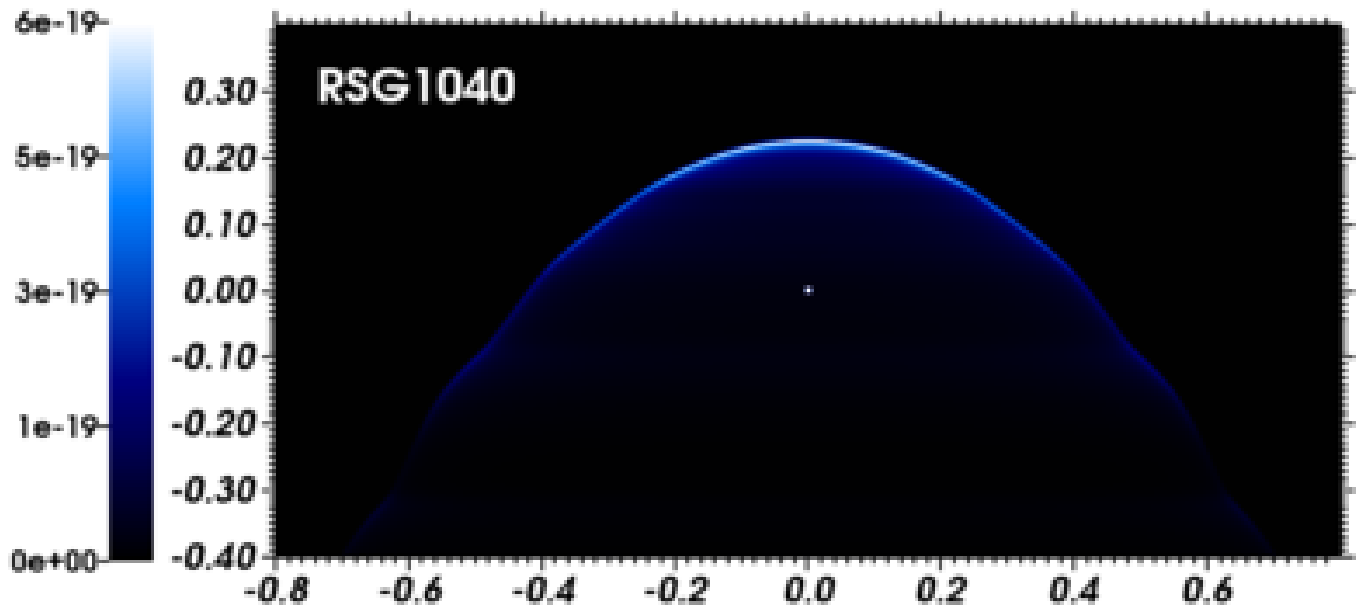}
	\end{minipage}
	\begin{minipage}[b]{ 0.48\textwidth}
		\includegraphics[width=1.0\textwidth]{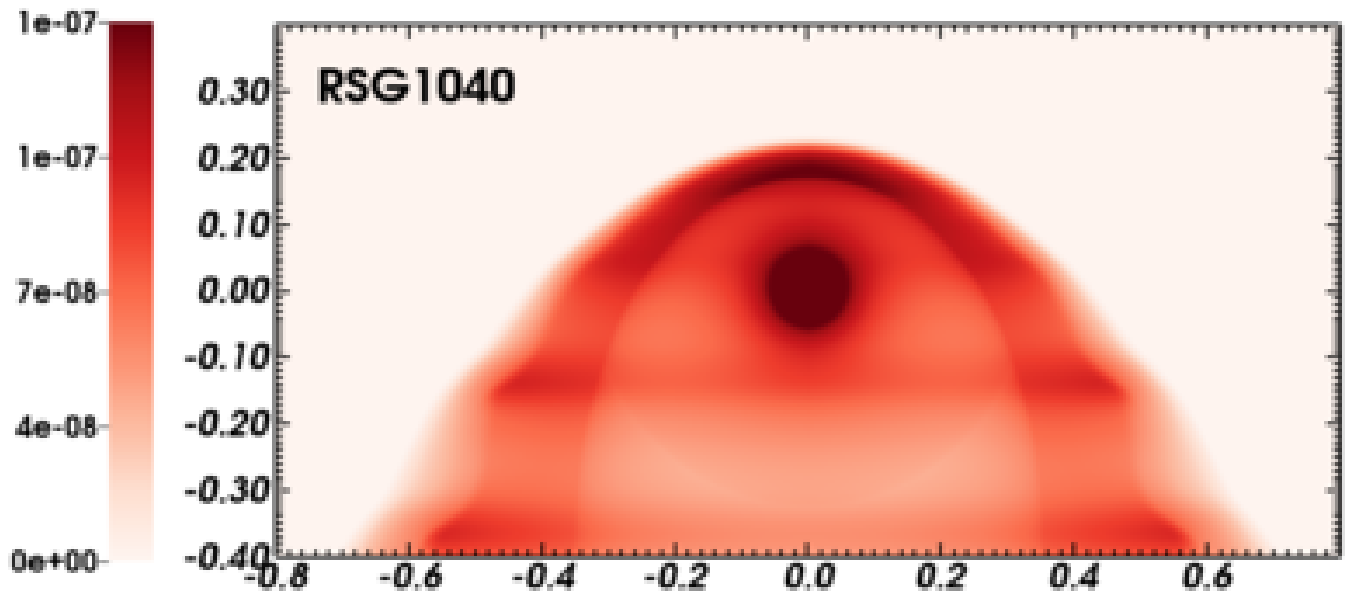}
	\end{minipage} \\
	\begin{minipage}[b]{ 0.48\textwidth}
		\includegraphics[width=1.0\textwidth]{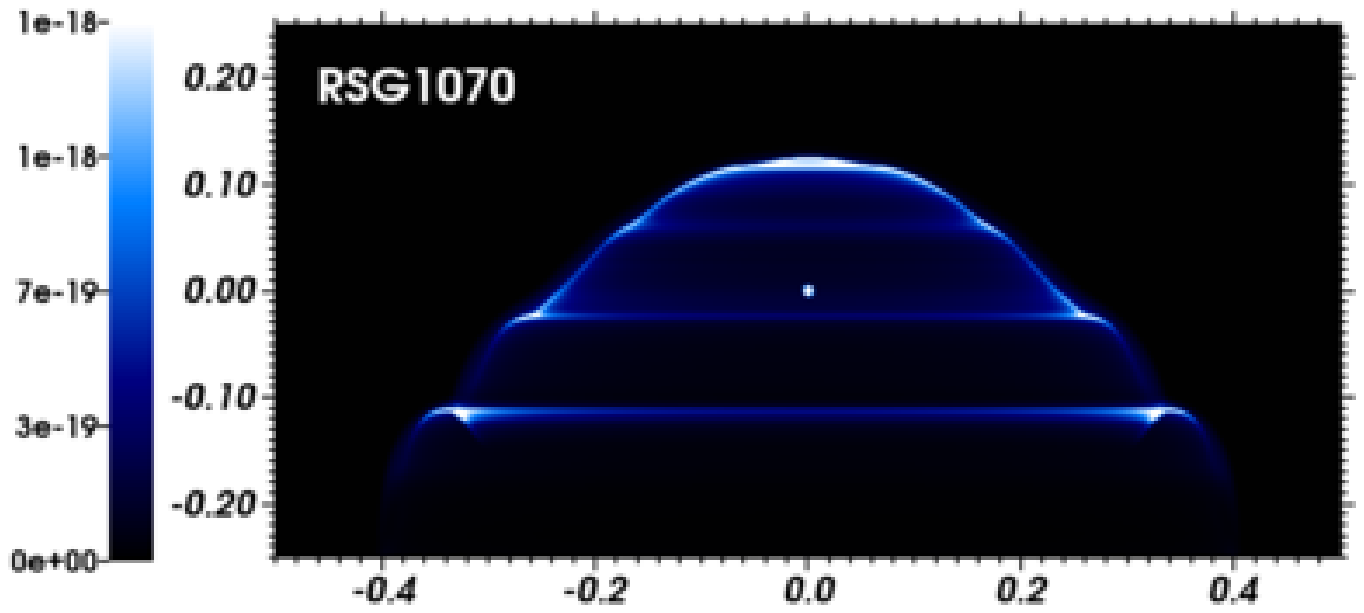}
	\end{minipage}
	\begin{minipage}[b]{ 0.48\textwidth}
		\includegraphics[width=1.0\textwidth]{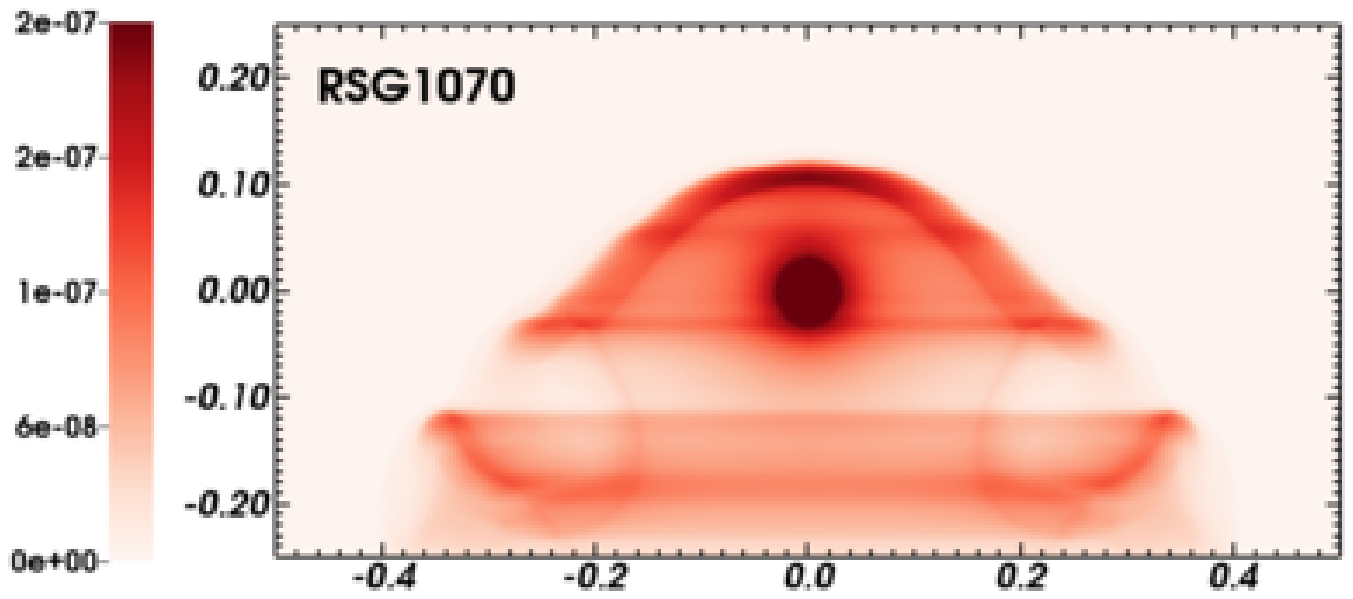}
	\end{minipage} \\ 
	\caption{The figures show the H$\alpha$ surface brightness (left, in 
		 $\mathrm{erg}\, \mathrm{s}^{-1}\, \mathrm{cm}^{-2}\, \mathrm{arcsec}^{-2}$) and the dust surface mass density 
		 (right, in $\rm g\, \mathrm{cm}^{-2}$) for the bow shocks from the red supergiant 
		 phase of our $10\, \mathrm{M}_{\odot}$ initial mass star.
		 Quantities are calculated excluding the undisturbed ISM and plotted in the linear scale, 
		 as a function of the considered space velocities.
		 The $x$-axis represents the radial direction and the $y$-axis the direction of stellar motion (in $\mathrm{pc}$). 
		 Only part of the computational domain is shown.
		 }
	\label{fig:projemrsg1}  
\end{figure*}

\begin{figure*}
	\begin{minipage}[b]{ 0.48\textwidth}
		\includegraphics[width=1.0\textwidth]{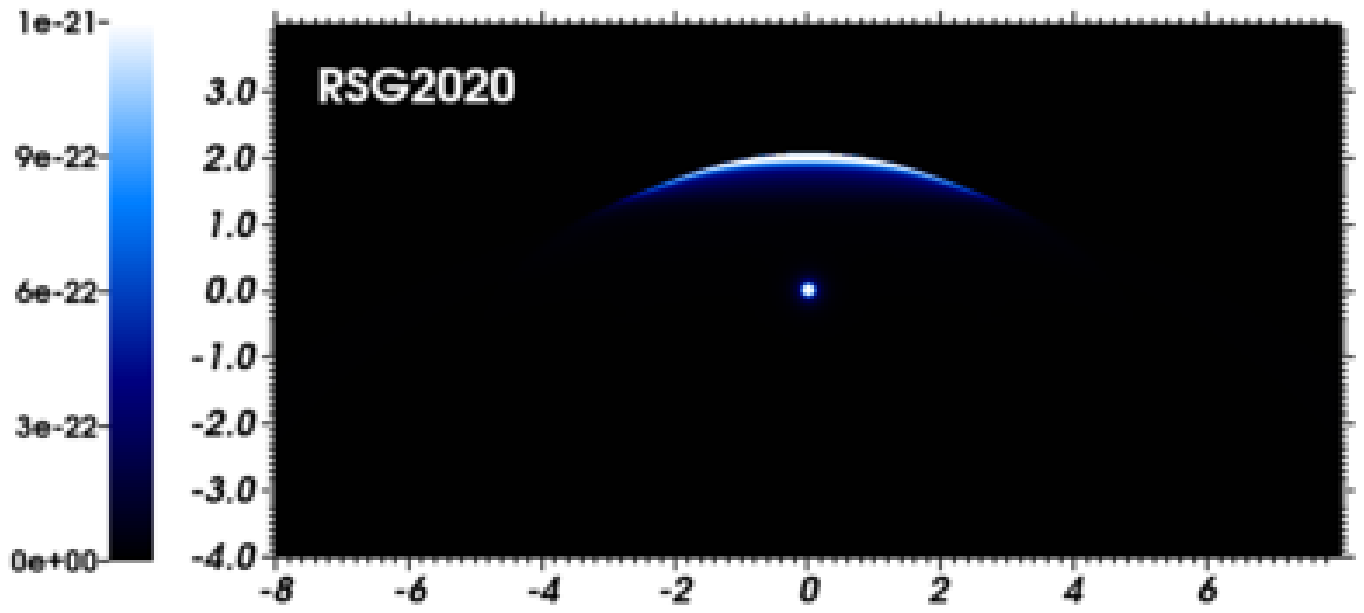}
	\end{minipage}
	\begin{minipage}[b]{ 0.48\textwidth}
		\includegraphics[width=1.0\textwidth]{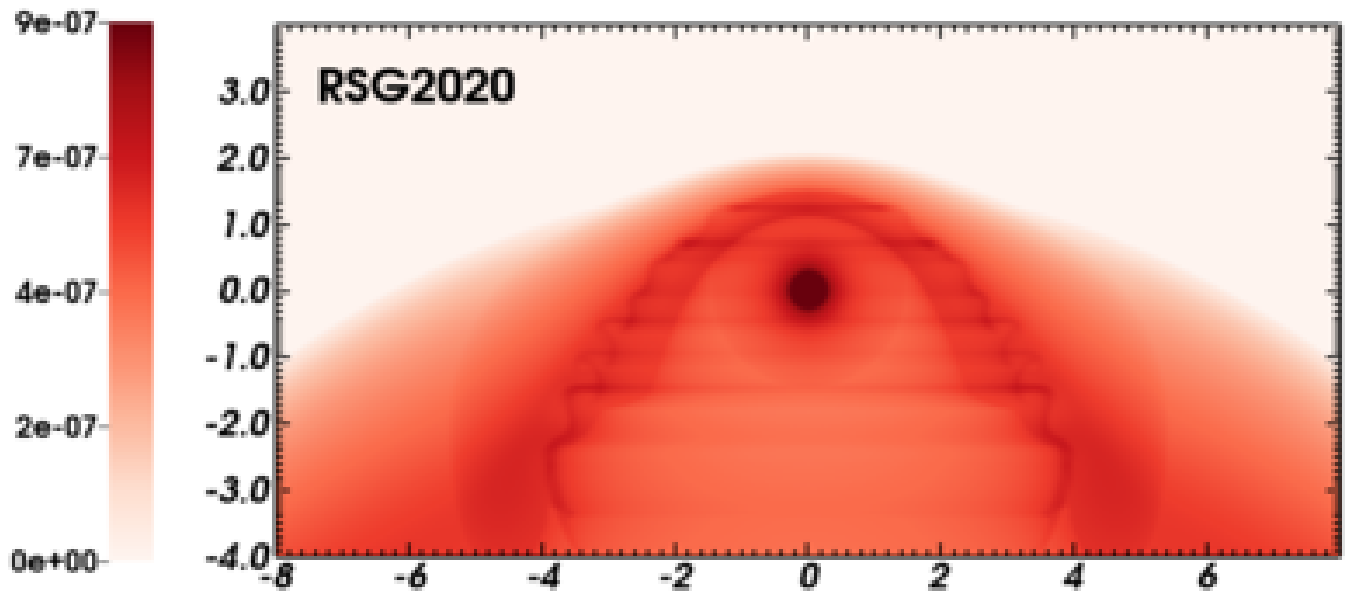}
	\end{minipage} \\
	\begin{minipage}[b]{ 0.48\textwidth}
		\includegraphics[width=1.0\textwidth]{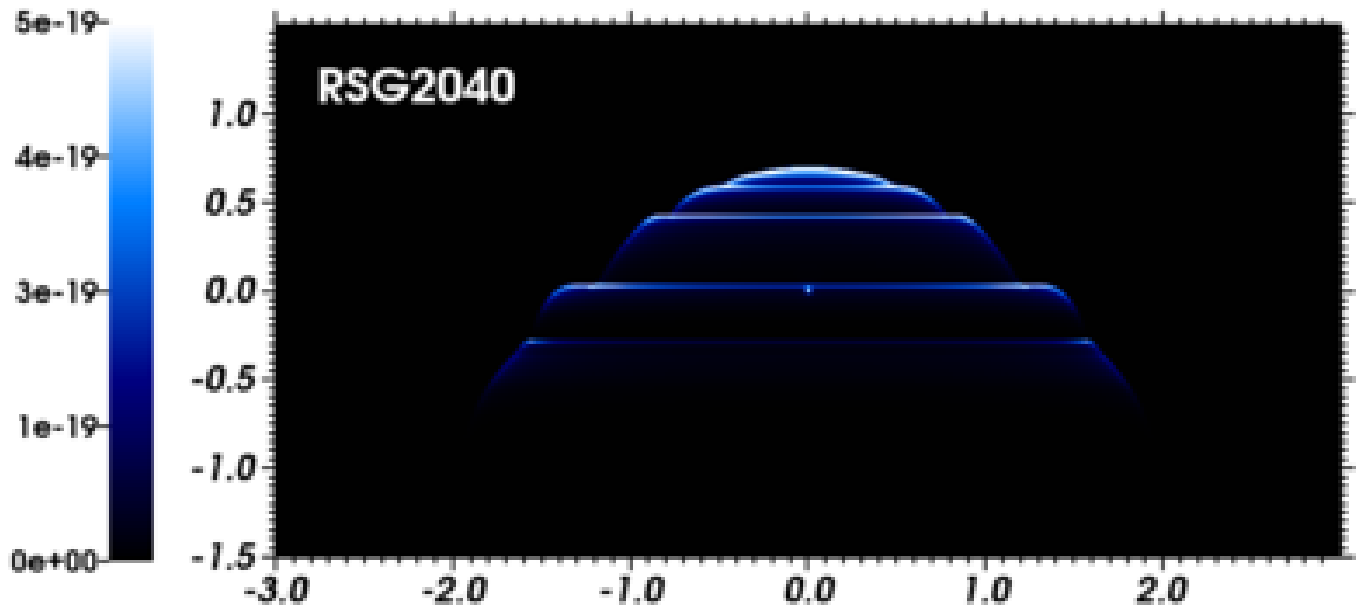}
	\end{minipage}
	\begin{minipage}[b]{ 0.48\textwidth}
		\includegraphics[width=1.0\textwidth]{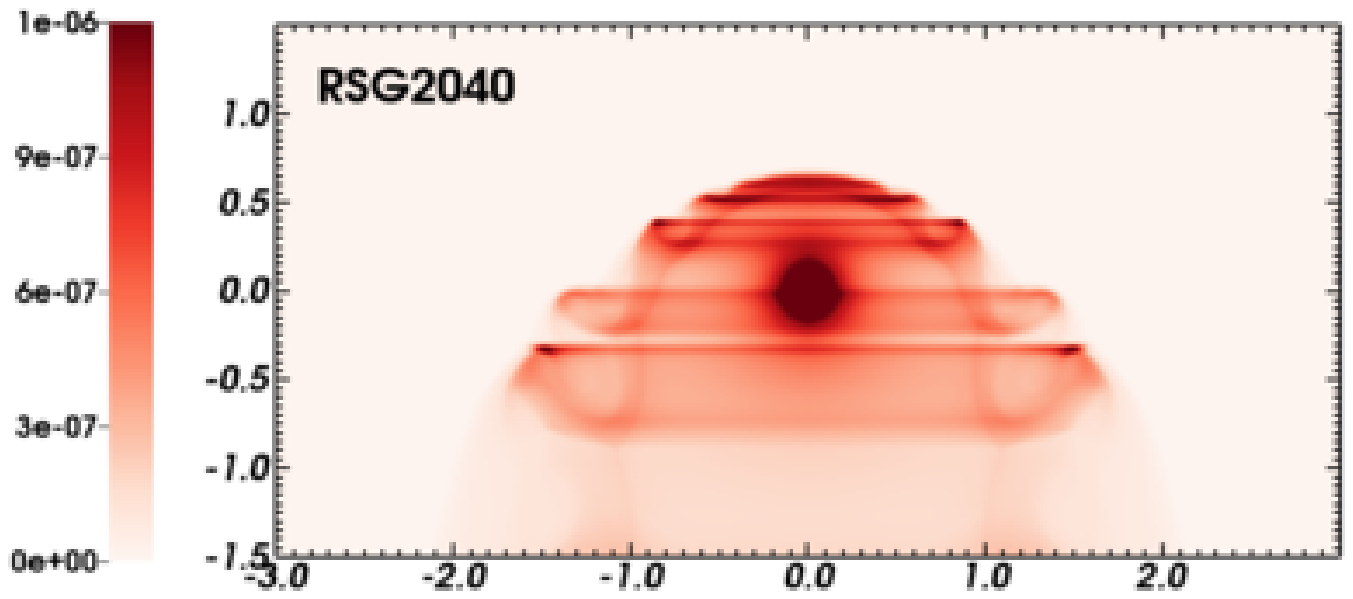}
	\end{minipage} \\
	\begin{minipage}[b]{ 0.48\textwidth}
		\includegraphics[width=1.0\textwidth]{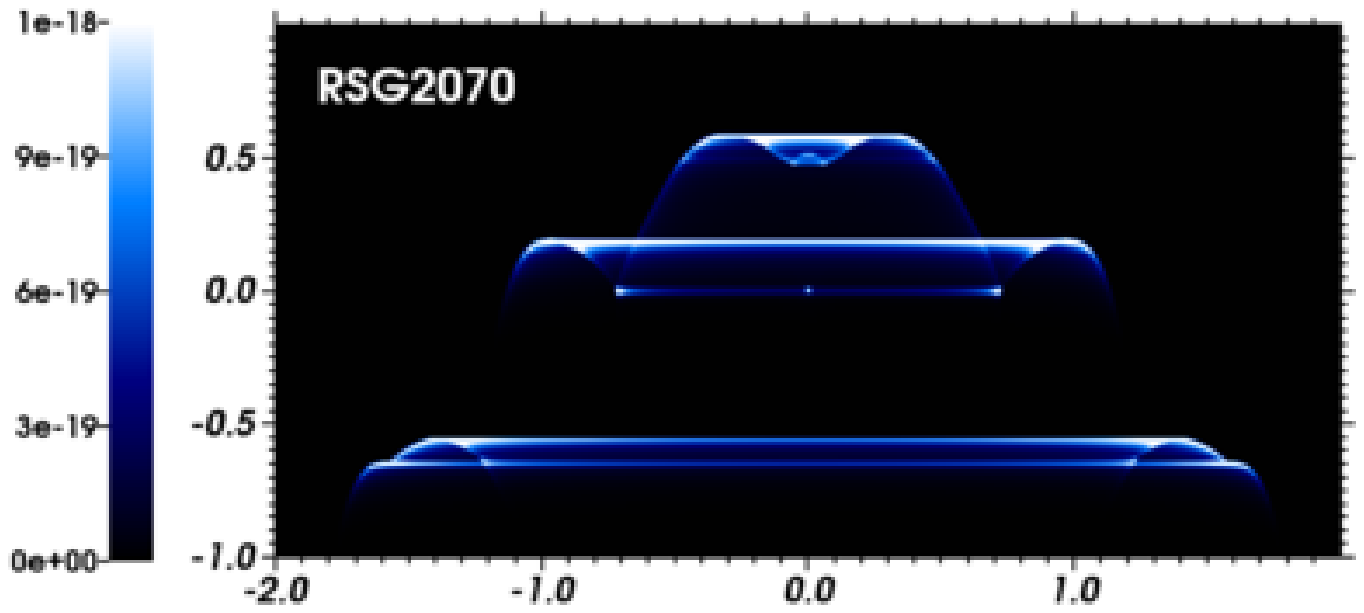}
	\end{minipage}
	\begin{minipage}[b]{ 0.48\textwidth}
		\includegraphics[width=1.0\textwidth]{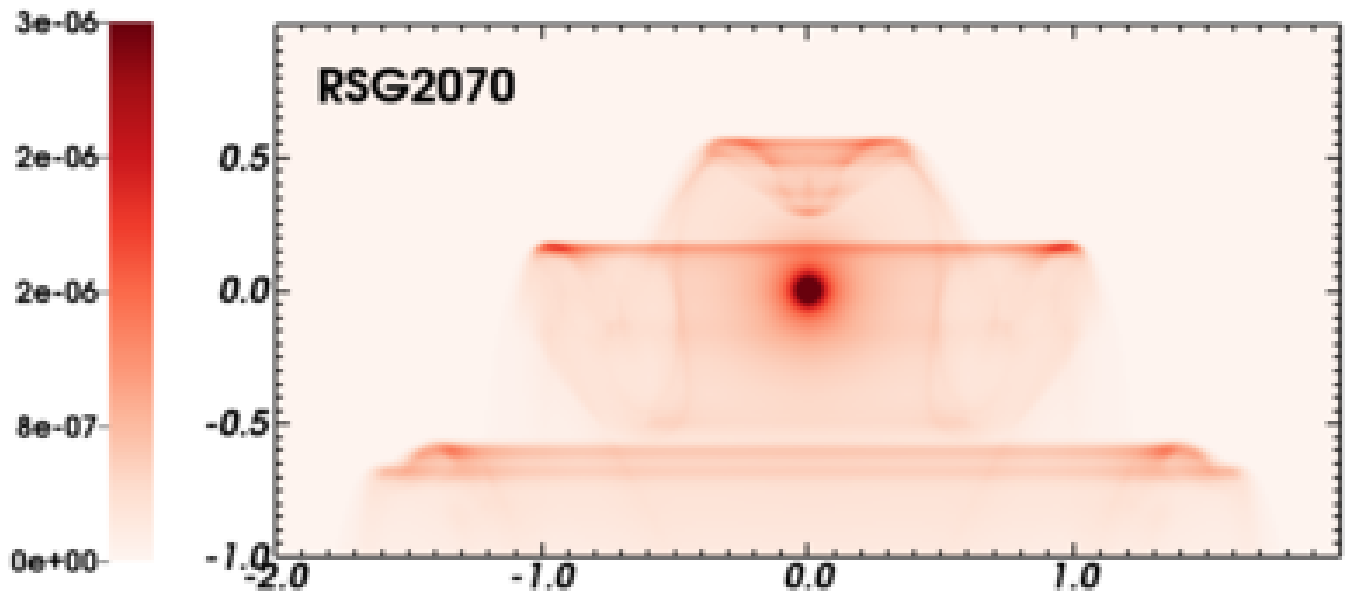}
	\end{minipage} \\ 
	\caption{As Fig.~\ref{fig:projemrsg1}, with an initial stellar mass of $20\, M_{\odot}$.  }
	\label{fig:projemrsg2}  
\end{figure*}

Figs.~\ref{fig:projemrsg1} and~\ref{fig:projemrsg2} show the bow shock H$\alpha$
surface brightness (left panels) and dust surface mass density (right
panels) for our $10$ and $20\, M_{\odot}$ models, respectively. Each figure
shows $v_{\star}=20\, \rm km\, \rm s^{-1}$ (top), $v_{\star}=40\,
\rm km\, \rm s^{-1}$ (middle) and $v_{\star}=70\, \rm km\, \rm s^{-1}$ (bottom).
For red supergiants we assume that both the stellar wind and the ISM 
gas include dust (our Appendix~\ref{maps_IR}).

In our models the H$\alpha$ emission of bow shocks produced by red
supergiants originates from the shocked ISM in the post-shock region at
the forward shock. The region of maximum emission is at the apex of the
structure for simulations with $v_{\star}=20\, \rm km\, \rm s^{-1}$ and is
extended to the tail as $v_{\star}$ increases, e.g. for the model RSG1040. 
The surface brightness increases with $v_{\star}$ and $\dot{M}$ because the
post-shock temperature at the forward shock increases when the shocks are
stronger. However, the H$\alpha$ emission is fainter by several orders of
magnitude than our bow shock models for hot stars (see
Figs.~\ref{fig:projemms10} and~\ref{fig:projemrsg1}). As a consequence, these bow
shocks are not likely to be observed in H$\alpha$ because their
H$\alpha$ surface brightnesses is below the detection sensitivity of the
SHS~\citep{parker_mnras_362_2005}.

Panel (b) of Fig.~\ref{fig:cs} plots the normalized cross-section\textcolor{black}{s} taken from the
H$\alpha$ surface brightness and the dust surface mass density of the bow shock
model RSG1020. The H$\alpha$ emission is maximum in the post-shock region at the
forward shock, whereas the dust surface density peaks in the post-shock region
at the reverse shock of the bow shock. All our models for bow shocks
for red supergiants exhibit such comportment which suggests that
H$\alpha$ and infrared emission do not originate from the same region of the
bow shock. Because the red supergiant wind is denser than the ISM, most of the
infrared emission probably originates from the \textcolor{black}{shocked wind}.

\section{Discussion}
\label{sect:discussion}

\subsection{Comparison with previous works}
\label{sect:discussion}

\subsubsection{Bow shocks around main sequence stars}
\label{subsect:msphase}

We carried out tests with two numerical methods to integrate the
parabolic term associated with heat conduction: the explicit method used
in~\citet{comeron_aa_338_1998} and the Super-Time-Step
method~\citep{alexiades_cnme_12_1996}. The results are consistent between the
two methods, except that the explicit scheme is less diffusive but also
extremely time consuming. We adopt the Super-Time-Step algorithm given that
the spatial resolution of our models is better than in~\citet{comeron_aa_338_1998}.

We tested this method using the code {\sc pluto} with respect to the models
in~\citet{comeron_aa_338_1998}. Our simulations support their study in that all
the bow shocks are reproduced reasonably well. Our simulations that aim to
reproduce the highly unstable simulation cases C (bow shock with strong
wind) and E (bow shock in high density ambient medium)
in~\citet{comeron_aa_338_1998} are slightly more affected by the development
of overdensities at the apex of the structure which later govern the shape of
the instabilities which distort the whole bow shocks. Our results vary 
depending on the chosen coordinate system and the interpolation 
scheme used at the symmetry axis. We conclude that 
instabilities growing at the apsis are artificially confined near $R=0$ by
the rotational symmetry imposed by the coordinate system.

Our models with $v_{\star}=20\, \rm km\, \rm s^{-1}$ produce weak bow shocks.
Such bow shocks correspond to the Case A model in~\citet{comeron_aa_338_1998},
which uses a similar wind velocity ($\sim 1000\, \rm km\, \rm s^{-1}$), and a
mass-loss rate of $10^{-7}\, \rm M_{\odot}\, \rm yr^{-1}$ (i.e. 1.5 orders
of magnitude larger, similar and one order of magnitude smaller than our $10$,
$20$ and $40\, \rm M_{\odot}$ stars, respectively), a less dense ISM ($0.1\, \rm
cm^{-3}$) and a much higher $v_{\star}$ ($\approx 100\, \rm km\, \rm s^{-1}$).
Our models include cooling by forbidden collisionally excited lines and assume
the same $T_{\rm ISM}\approx 8000\, \rm K$ as their Case A. These models are
similar because their weak forward shocks do not allow the gas to cool rapidly
and they all have a region of shocked ISM thicker than the hot bubble along the
direction of motion of the star, as signified by the absence of a sharp density peak
in the region of shocked ISM in panel (a) of Fig.~\ref{fig:profile}, compared
to lower panel of fig. 7 in~\citet{comeron_aa_338_1998}.

Our models MS4040 and MS4070 have strong shocks and are similar to the Case C model
in~\citet{comeron_aa_338_1998}. Their case C uses a higher $v_{\rm w}\approx 3000\, \rm
km\, \rm s^{-1}$, a slightly larger $\dot{M}\sim 10^{-6}\, \rm M_{\odot}\, \rm
yr^{-1}$, a less dense ISM ($0.1\, \rm cm^{-3}$) and a higher $v_{\star}$
($\approx 100\, \rm km\, \rm s^{-1}$). The combination of high $v_{\star}$ and
high $v_{\rm w}$ induces a strong compression factor at the forward shock where
the gas cools rapidly and reduces the thickness of the shocked ISM into a thin,
unstable shell. These models best fit analytical approximations
of an infinitely thin bow shock~\citep{comeron_aa_338_1998}.

We conclude that for overlapping parameters, i.e. for similar $\dot{M}$ and
$v_{\star}$, our results agree well with existing models in terms of bow shock
morphology and stability. We extend the parameter space for stars with weak
winds, $\dot{M} \approx 10^{-9.5}$ in our $10\, M_{\odot}$ model
and use the typical particle density of the Galactic plane.

\subsubsection{Bow shocks around red supergiants}
\label{subsect:rsgphase}

We tested our numerical setup to reproduce the double bow shock around
Betelgeuse~\citep{mackey_apjlett_751_2012}. Including heat conduction did not
significantly change the results and we successfully reproduced the model using
the same cooling curve as in~\citet{mackey_apjlett_751_2012}. The 
simulations of red supergiant bow shocks of~\citet{mohamed_aa_541_2012} used a
more precise time-dependent cooling network~\citep{smith_MNRAS_339_2003} and,
because of their Lagragian nature, these models are intrinsically better in
terms of spatial resolution. To produce more detailed models which can predict
emission line ratio is beyond the scope of this
study but could be achieved using the native multi-ion non-equilibrium cooling
module of the code {\sc pluto}~\citep{tesileanu_aa_488_2008}.

Model RSG2020 shows a weak bow shock with a dense and cold shell expanding into the
former hot and smooth bow shock. Rayleigh-Taylor instabilities develop
at the discontinuity between the two colliding bow shocks as in the model of
Betelgeuse's multiple arched bow shock in~\citet{mackey_apjlett_751_2012}.

Our simulations with $v_{\star}=40\, \rm km\, \rm s^{-1}$ show radiative forward
shocks and unstable contact discontinuities. Model RSG1040 resembles the
simulations of~\citet{vanmarle_apjl_734_2011} and~\citet{decin_aa_548_2012}
which have a similar $\dot{M}\approx 3\times 10^{-6}\, \rm M_{\odot}\, \rm
yr^{-1}$ but a smaller $v_{\star}\approx28\, \rm km\, s^{-1}$ and denser ISM
($2\, \rm cm^{-3}$). We do not use the two-fluid approach of~\citet{vanmarle_apjl_734_2011}  
which allows the modelling of ISM dust grains
and explains the differences in terms of stability of the contact
discontinuity. Their simulation with type 1 grains is more unstable than our
model RSG1040, probably because they use a denser ISM.  Model RSG2040 has a
thinner region of shocked ISM compared to the region of shocked wind which
makes this model unstable. The instabilities of model RSG2040 are similar to the
clumpy forward shock of models A-C in~\citet{mohamed_aa_541_2012} which
have larger $\dot{M}$ and a denser medium.

Our simulations with $v_{\star}=70\, \rm km\, \rm s^{-1}$ show the largest
compression. Model RSG2070 has a strongly turbulent shell with dramatic
instabilities, consistent with the high $v_{\star}$ and high Mach number
model in~\citet{blondin_na_57_1998}. Our model RSG2070 illustrates 
the transverse acceleration instability where an isotropically expanding wind from the star
meets the collinear ISM flow and pushes the developing eddies sidewards. Model
RSG2070 is different from the model D with cooling of~\citet{mohamed_aa_541_2012} 
which has a similar $v_{\star}\approx 72.5\, \rm
km\, s^{-1}$ but a weaker wind $\dot{M}\approx 3.1\times 10^{-6}\, \rm
M_{\odot}\, \rm {yr}^{-1}$. Because of its particular initial conditions, i.e.
a hotter and diluted ISM with $n\approx 0.3\, \rm cm^{-3}$ and $T_{\rm
ISM}\approx 8000\, \rm K$, the gas does not cool efficiently at the forward
shock and the post-shock regions of the bow shock remain isothermal, see right
panel of fig. 10 of~\citet{mohamed_aa_541_2012}.

With similar model parameters, our results agree well with the existing models
and we conclude that heat conduction is not mandatory to model bow shocks from
cool stars. Because we neglect the effects of dust dynamics on the bow
shocks stability, our models differ slightly from existing models with
$v_{\star} \approx 30$$-$$40\, \rm km\, \rm s^{-1}$. However, this does not
concern the overall shape of the bow shocks but rather the (in)stability of
their contact discontinuities. We extended the parameter space by introducing
models with $v_{\star}=20\, \rm km\, \rm s^{-1}$.


\subsection{On the observability of bow shocks from massive runaway stars}
\label{subsect:observability}

\begin{figure*}
	\begin{minipage}[b]{0.325\textwidth}
		\includegraphics[width=1.0\textwidth,angle=0]{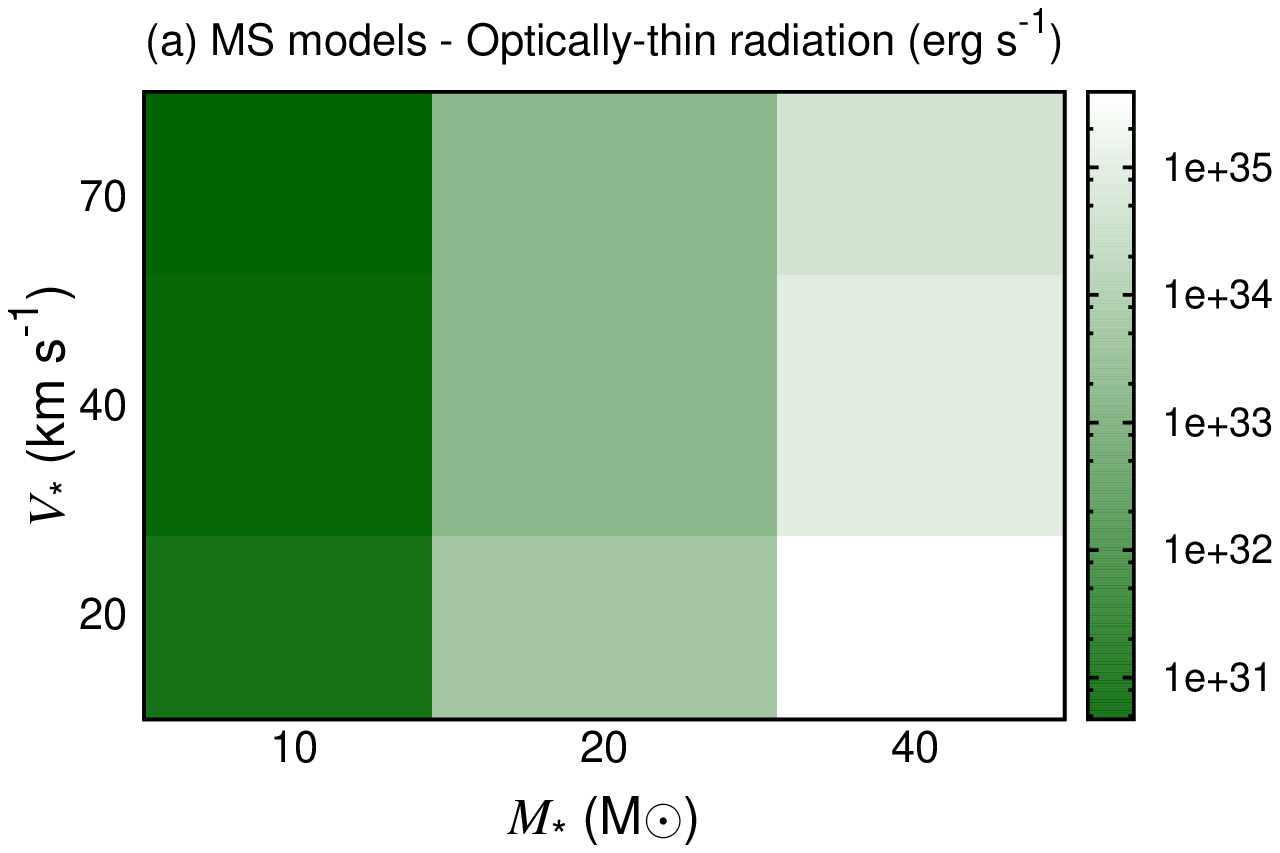}
	\end{minipage}
	\begin{minipage}[b]{0.325\textwidth}
		\includegraphics[width=1.0\textwidth,angle=0]{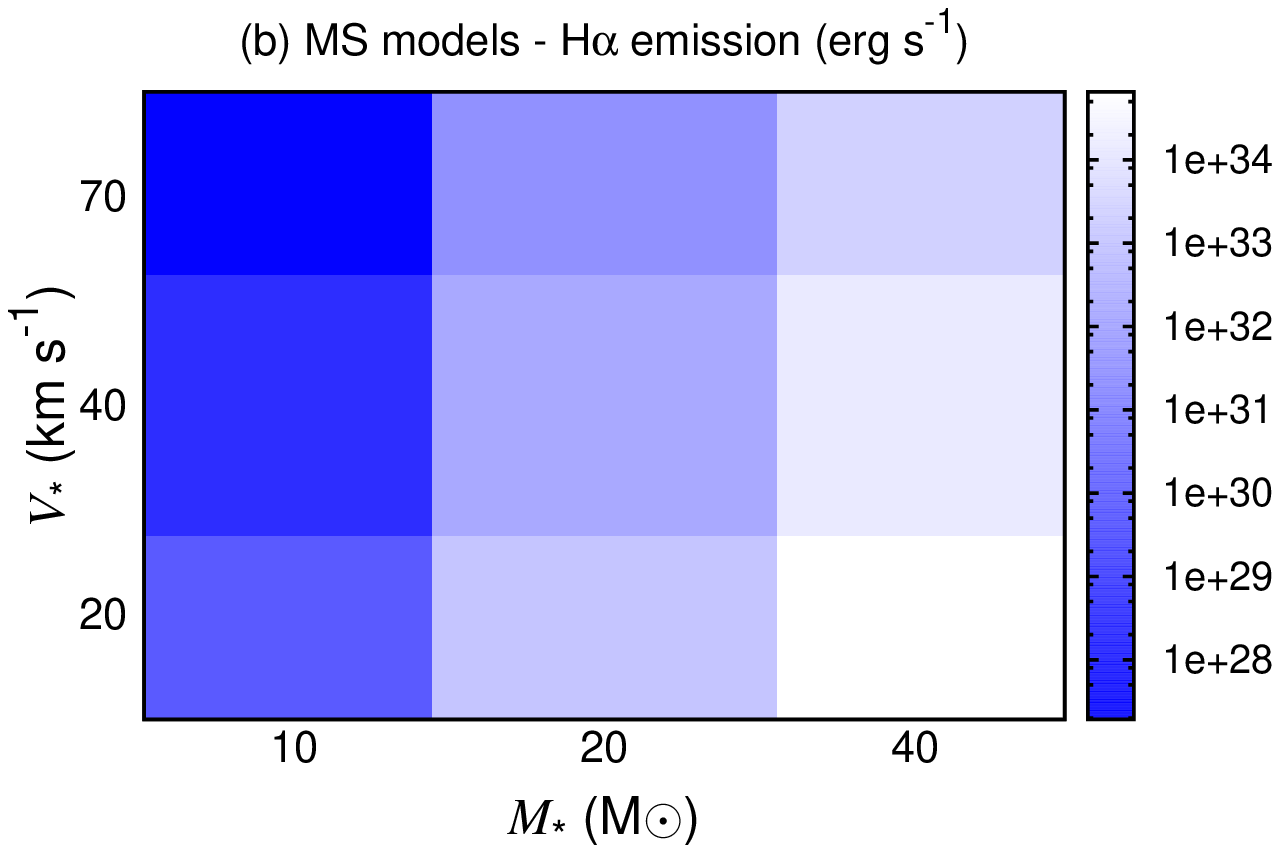}
	\end{minipage}
	\begin{minipage}[b]{0.325\textwidth}
		\includegraphics[width=1.0\textwidth,angle=0]{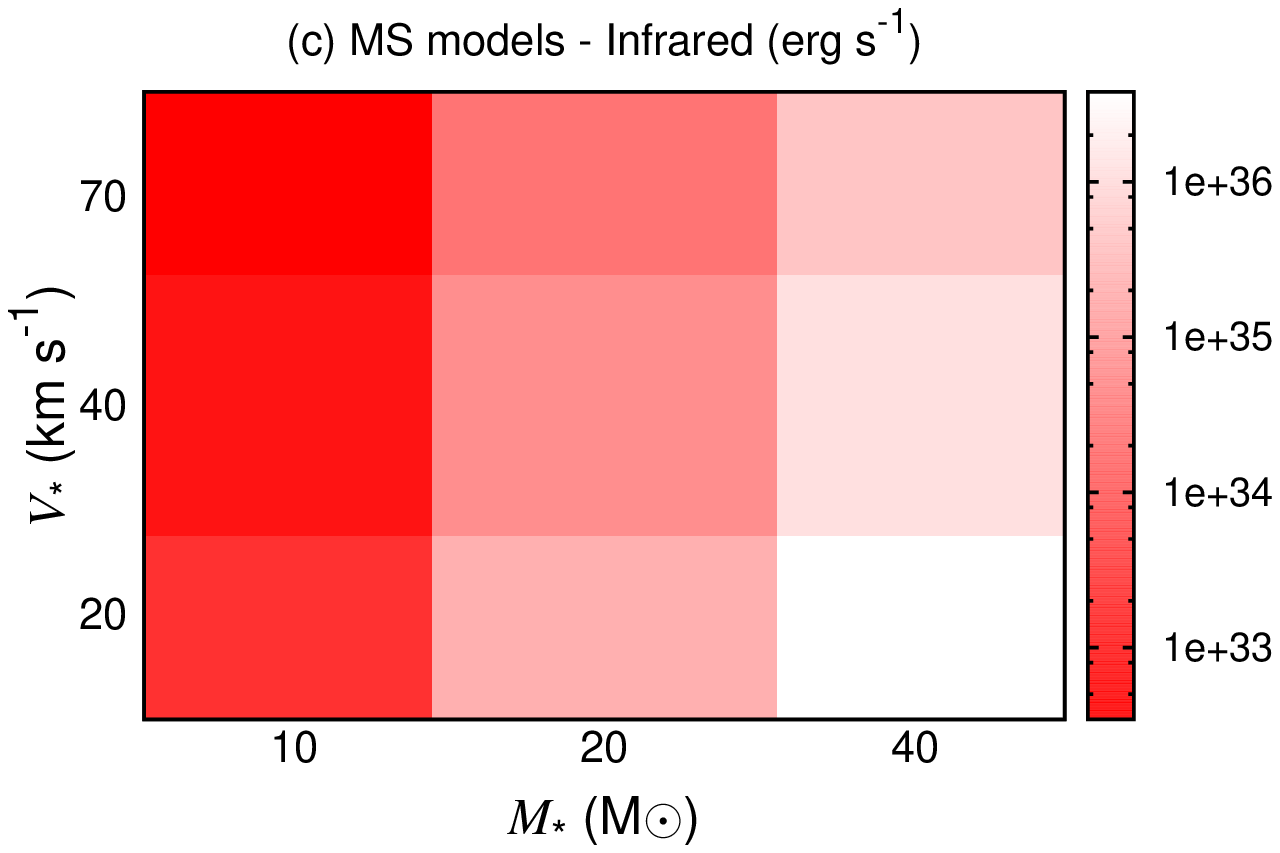}
	\end{minipage}  \\
	\begin{minipage}[b]{0.325\textwidth}
		\includegraphics[width=1.0\textwidth,angle=0]{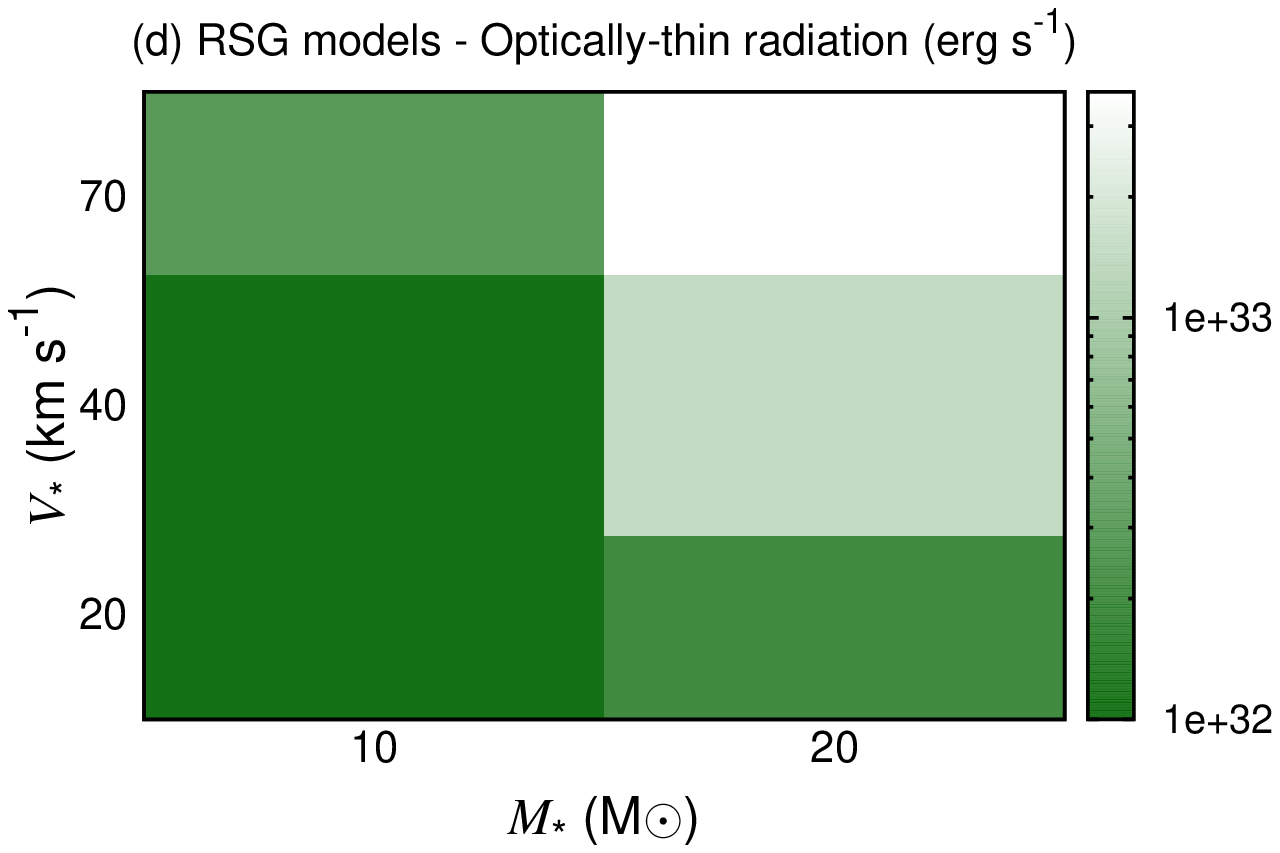}
	\end{minipage}
	\begin{minipage}[b]{0.325\textwidth}
		\includegraphics[width=1.0\textwidth,angle=0]{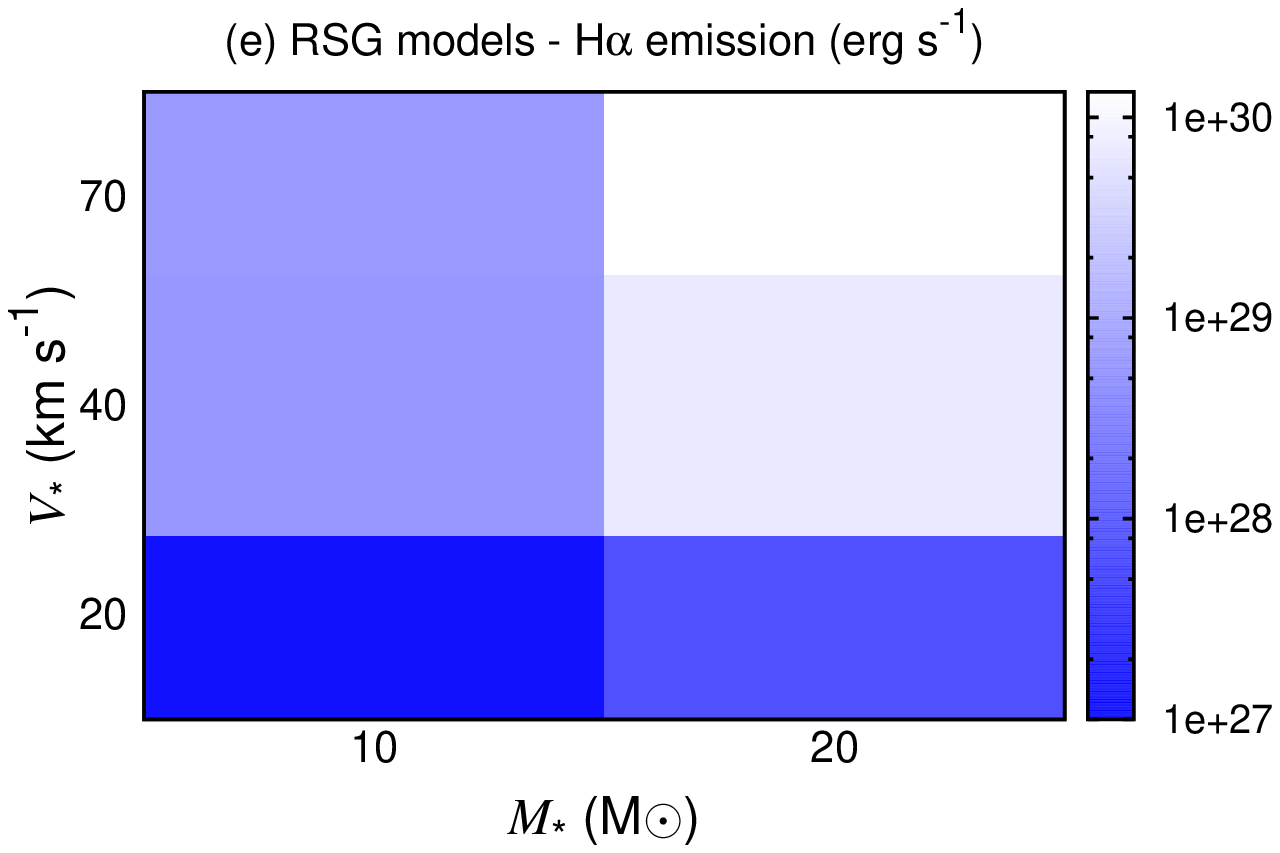}
	\end{minipage}
	\begin{minipage}[b]{0.325\textwidth}
		\includegraphics[width=1.0\textwidth,angle=0]{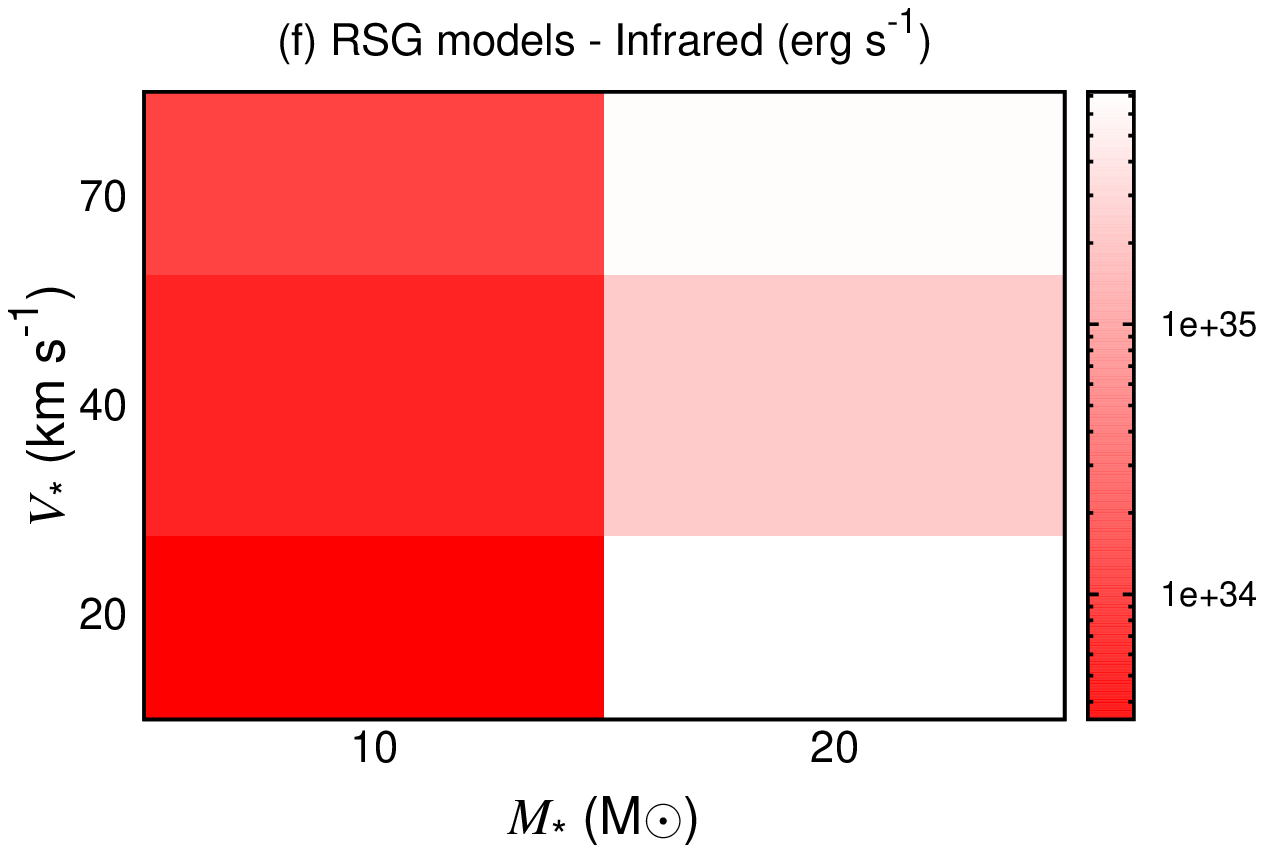}
	\end{minipage} 
	\caption{ Bow shocks luminosities (in $\rm erg\, \rm s^{-1}$) 
		  in our main sequence (top panels) and red supergiant (bottom panels) models. 
		  We show the luminosity of optically-thin cooling (left \textcolor{black}{green panels}), H$\alpha$ emission 
		  (middle \textcolor{black}{blue panels}) and reprocessed infrared starlight by dust 
		  grains (right \textcolor{black}{red panels}) of Table~\ref{tab:lum_val}.
		  On each plot the $x$-axis is the initial mass $M_{\star}$ 
		  (in $M_{\odot}$) and the $y$-axis is the space velocity $v_{\star}$ 
		  (in $\rm km\, \rm s^{-1}$) of our runaway stars.
		   }
	\label{fig:lum_models_ms_rsg}  
\end{figure*}

Fig.~\ref{fig:lum_models_ms_rsg} plots the luminosities of our bow shock
models for main sequence (top panels) and red supergiant (bottom panels) stars
as a function of $M_{\star}$ and $v_{\star}$. With respect to their optically-thin 
gas radiation, the brightest bow shocks produced by main-sequence stars are
generated by the more massive stars moving with a slow space velocity, e.g. the
$40\, M_{\odot}$ main sequence star moving with $v_{\star}=20\, \rm km\, \rm
s^{-1}$, and the brightest bow shocks produced by red supergiants are generated
by the more massive star of our sample, moving at high space velocity i.e. a $20\,
M_{\odot}$ red supergiant moving with $v_{\star}=70\, \rm km\, \rm s^{-1}$ (see
panels (a) and (d) of Fig.~\ref{fig:lum_models_ms_rsg}). The same points
arise from the luminosity of H$\alpha$ emission (see panels (b) and (e) of
Fig.~\ref{fig:lum_models_ms_rsg}). The infrared luminosity indicates that
the brightest bow shocks generated by a main sequence star are produced by high mass,
low velocity stars (see panel (c) in Fig.~\ref{fig:lum_models_ms_rsg}).
Concerning the bow shocks generated by red supergiants, their infrared
luminosities suggest that the brightest are produced by high-mass stars 
moving at either low or high space velocities (see panel (f) in
Fig.~\ref{fig:lum_models_ms_rsg}).

Because $L_{\rm IR}$ is larger than $L_{\rm H\alpha}$ or $L_{\rm gas}$,
the infrared \textcolor{black}{waveband is the most appropriate} to search for stellar-wind 
bow shocks around main sequence and red supergiant \textcolor{black}{stars}. According to our
study, bow shocks produced by high mass main sequence stars moving with
low space velocities should be the easiest ones to 
observe in the infrared. The most numerous runaway stars have a low space
velocity~\citep{eldridge_mnras_414_2011} and consequently bow shocks produced
by high-mass red supergiants moving with low space velocity are the most
numerous ones, and the probability to detect one of them is larger. 
\textcolor{black}{Many stellar wind bow shocks surrounding hot stars 
ejected from stellar cluster are detected by means of their 
$\le 24\, \mu m$ infrared signature~\citep[see][]{gvaramadze_519_aa_2010, gvaramadze_aa_535_2011}. 
Because our study focuses on the most probable bow shocks forming 
around stars exiled from their parent cluster, we expect them to be 
most prominent in that waveband.}

\begin{figure}
         \centering 
	 \includegraphics[width=0.49\textwidth,angle=0]{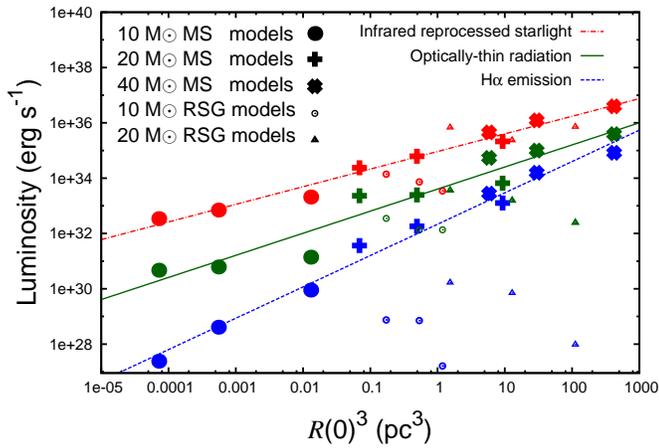} 
	 \caption{ Bow shocks luminosities (in $\rm erg\, \rm s^{-1}$) as a function 
		   of $R(0)^{3}$ (in $\rm {pc}^{3}$) for the main sequence (\textcolor{black}{large symbols})
		   and red supergiant models (\textcolor{black}{small symbols}).
	           The overplotted thin lines are least square fits of the luminosity of optically-thin gas 
	           radiation (\textcolor{black}{solid green line}), the infrared luminosity of reprocessed 
	           starlight (\textcolor{black}{dashed red line}) 
	           and the luminosity of H$\alpha$ emission (\textcolor{black}{dotted blue line}).
	            }
        \label{fig:scale_lum}
\end{figure}

Fig.~\ref{fig:scale_lum} plots the bow shock luminosities for our main sequence
models as a function of $R(0)^{3}$. It shows a strong scaling relation
between the luminosities and the volume of the bow shocks, i.e. the brightnesses 
of these bow shocks are governed by the wind momentum. The optical luminosities 
of our red supergiant models do not satisfy these fits because 
the gas is weakly ionized.  This behaviour concerns the overall luminosities 
of the bow shocks, not their surface brightnesses. Furthermore, this statement 
is only valid for the used ISM density, and some effects may 
make them dimmer, e.g. a lower density medium increasing their volume 
$\sim R(0)^{3} \sim 1/\sqrt{n_{\rm ISM}^{3}}$.

\section{Conclusion}
\label{section:cc}

We present a grid of hydrodynamical models of bow shocks around evolving
massive stars. The runaway stars initial masses range from $10$ to $40\, \mathrm{M}_{\odot}$
and their space velocities range from $20$ to $70\, \mathrm{{km}\, s^{-1}}$.
Their evolution is followed from the main sequence to the red supergiant
phase. Our simulations include thermal conduction and distinguish the treatment
of the optically-thin cooling and heating as a function of the evolutionary phase of
the star.

Our results are consistent with~\citet{comeron_aa_338_1998} in that our bow shocks
show a variety of shapes which usually do not fit a simple analytic
approximation~\citep{wilkin_459_apj_1996}. We stress the importance of heat
conduction to model the bow shocks around main sequence stars and find 
that this is not an important process to explain the morphology of bow 
shocks around red supergiants. We underline its
effects on their morphology and structure, especially concerning the transport
of ISM material to the hot region of the bow shocks generated by hot stars. The heat
transfer enlarges the bow shocks and considerably reduces the volume of shocked
wind so that optical emission mainly originates from shocked ISM material. 
We extend the analysis of our results by calculating the luminosities
of the bow shocks and detail how they depend on the star's mass loss and
space velocity.

Our bow shock models of hot stars indicate that the main coolants governing
their luminosities are the optical forbidden lines such as [O\,{\sc ii}] and
[O\,{\sc iii}]. The luminosity of optical forbidden lines is stronger than 
the luminosity of H$\alpha$ emission, which only
represents less than a tenth of the luminosity by optically-thin radiation. This agrees with
the observations of~\citet{gull_apj_230_1979} who noticed that [O\,{\sc iii}] is
the strongest optical emission line of the bow shock of $\zeta$ Oph. Our study
also shows that those forbidden emission lines are fainter than the infrared
emission of bow shocks produced by main sequence stars.

Our bow shock models with hot stars are brightest in H$\alpha$ in the
cold shocked ISM material near the contact discontinuity. Because their dust
surface mass density peaks at the same distance to the star as their H$\alpha$
emission, we suggest that their infrared emission is also maximum at the
contact discontinuity. The H$\alpha$ surface brightness is maximum upstream
from the star for small space velocities and are extended downstream from the
star for larger velocities. Our bow shock models can have
H$\alpha$ surface brightnesses above the detection threshold of the SuperCOSMOS
H-alpha Survey~\citep{parker_mnras_362_2005}.

Our bow shocks generated by red supergiants have a large infrared
luminosity. Their luminosity by optically-thin radiative cooling mainly
originates from shocked ISM material, whereas our models indicate that
their infrared luminosity principally comes from regions of shocked wind. The H$\alpha$
emission of our bow shocks around cool stars originates from their forward
shock. Its maximum is upstream from the star in the supersonic regime and is
lengthened in the wake of the bow shock in the hypersonic regime. 
Their H$\alpha$ emission is negligible compared to their luminosity of
optically-thin radiation because their gas is weakly ionized. In conclusion, 
these bow shocks are more likely to be observed in the infrared than in 
the optical or in H$\alpha$. This supports the hypothesis that the 
optically-detected bow shock of IRC$-$10414 is photoionized by an external 
source because the collisionally excited [N\,{\sc ii}] line in the shocked 
wind is brighter than the H$\alpha$ emission at the forward shock~\citep{meyer_mnras_2014}.

We also conclude that bow shocks produced by runaway main sequence and red
supergiants should be easier to detect in the infrared. The brightest and
most easily detectable bow shocks from main sequence stars are those of 
high mass stars ($\approx\, 40\, M_{\odot}$) with small space
velocity ($\approx\, 20\, \rm km\, \rm s^{-1}$). With the ISM density of the
Galactic plane, their luminosities are governed by their wind momentum and they
scale monotonically with their volume.  In the infrared, the most probable bow
shocks to detect around red supergiants are produced by high mass ($\approx\,
20\, M_{\odot}$) stars with small space velocity ($\approx\, 20\, \rm km\, \rm s^{-1}$).

The hereby presented grid of models will be enlarged in a wider study, and
forthcoming work will investigate the effects of an ISM background magnetic 
field. We also plan to focus on the latest stellar evolutionary stage in 
order to model the final explosion happening at the end of the massive star 
life, because the supernova ejecta interact with the shaped circumstellar 
medium.


\section*{Acknowledgements}

\textcolor{black}{We thank the anonymous reviewer for his valuable comments and suggestions which 
\textcolor{black}{greatly improved} the quality of the paper.}
DM is grateful to Fernando Comer\'on for his help and comments. 
DM also acknowledges Richard Stancliffe, Allard-Jan van Marle, Shazrene Mohamed 
and Hilding Neilson for very useful discussions. JM was partially supported by a fellowship from the Alexander 
von Humboldt Foundation. RGI thanks the Alexander von Humboldt Gesellschaft.
This work was supported by the Deutsche Forschungsgemeinschaft priority program 1573, 'Physics
of the Interstellar Medium'. Simulations were run thanks to a grant from John
von Neumann Institute of computing time on the JUROPA supercomputer at
J\"{u}lich Supercomputing Centre.




\addcontentsline{toc}{section}{\bibname}
{\footnotesize
\bibliography{refs.bib}
}


\appendix

\section{Emission maps and projected dust mass}
\label{maps_Ha}

The simulations are post-processed in order to obtain projected H$\alpha$
emission maps and ISM dust projected mass. The gas $T$ is calculated according
to Eq.~(\ref{eq:temperature}). For every cell of the computational domain and for
a given quantity $\xi(T)$ of units $[\xi]$ representing either rate of emission
(in $\rm erg\, s^{-1}\, cm^{-3}$) or a density (in $\rm g\, cm^{-3}$) we
calculate its projection $\mathcal{P}_{\xi}$. The integral of $\xi$ is performed
inside the bow shock along a path perpendicular to the plane ($O,R,z$), excluding 
the unperturbed ISM. Taking into account the projection factor, it is,
\begin{equation}
	\mathcal{P}_{\xi}(R,z) = 2 \int_{R^{'}=R}^{R^{'}=R_{max}} \xi(R^{'},z) \frac{R^{'}dR^{'}}{\sqrt{R^{'2}-R^{2}}}\, [\xi]\, \rm cm.
\label{eq:proj}
\end{equation}

For hot, photoionized medium we use the H$\alpha$ emissivity rate interpolated from the Table 4.4 
of~\citet{osterbrock_1989}, which is,
\begin{equation}
	\xi(T) \approx 1.21\times 10^{-22} T^{-0.9} n_{\rm e}n_{\rm p}\, \rm erg\, s^{-1}\, cm^{-3}\, sr^{-1},
\label{eq:Ha}
\end{equation}
where $n_{\rm e}$ and $n_{\rm p}$ are the number of electrons and protons per unit volume, respectively.
For cool, CIE medium we employ a similar formalism, taking into account the fact
that only the ions emit, i.e. the emission is proportional to $n_{\rm e}n_{\rm i}$ 
with $n_{\rm i}$ the number of ions per unit volume. The emission rate is, 
\begin{equation}
	\xi(T) \approx 1.21\times 10^{-22} T^{-0.9} n_{\rm i}n_{\rm e}\, \rm erg\, s^{-1}\, cm^{-3}\, sr^{-1}.
\label{eq:Haneut}
\end{equation}

The ISM projected dust mass is calculated integrating the number density. 
For a dust-to-gas ratio $X_{\rm d/\rm g}$ and for the total gas number density $n$, its expression is,
\begin{equation}
	\xi(T) = n X_{\rm d/\rm g} \mu m_{\rm p}\, \rm g\, cm^{-3}.
\label{eq:Ha}
\end{equation}
We use a dust-to-gas ratio $X_{\rm d/\rm g}=1/200$ by mass for the
ISM~\citep{neilson_apj_716_2010,neilson_aa_529_2011} and for the red supergiant
winds~\citep{lamer_cassinelli_1999}. The calculation of the dust density for bow
shocks around hot stars also requires us to exclude from the integral in
Eq.~\ref{eq:proj} the region which are only made of wind material, i.e. which
do not contain any dust.


\section{Estimation of the infrared emission of the bow shocks}
\label{maps_IR}

Learning from previous studies on the behaviour of dust in stellar bow
shock~\citep{vanmarle_apjl_734_2011,decin_aa_548_2012,decin_asr_50_2012}, the
infrared emission of a model is estimated as a part of the starlight absorbed by the dust
grains and reemitted at longer wavelengths, plus the gas collisional heating of
the dust particles.

We assumed that the shocked ISM material into the outer layer of the bow shock
is filled with spherical grains of radius $a=4.5\, \rm
nm$~\citep{vanmarle_apjl_734_2011} in a proportion of $X_{\rm d/\rm g}=1/200$ by
mass~\citep{neilson_apj_716_2010,neilson_aa_529_2011}. The interstellar grains
are assumed to be made of silicates whose density is $\rho_{g}=3.3\, \rm g\, \rm
cm^{-3}$~\citep{draine_apj_285_1984}. The dust in the red supergiant wind is
treated as in~\citet{mackey_apjlett_751_2012}, but considering grains of radius
$a=5.0\, \rm nm$ only. \textcolor{black}{ Such an approach is in accordance with 
interpretation of $24\, \mu m$ infrared emission suggesting that small-sized dust 
grains are not destroyed in ionized regions in the vicinity of young massive stars~\citep{pavlyuchenkov_ARep_57_2013}.
} We assume that no dust crosses the material
discontinuity, i.e. the shocked wind is dust-free in bow shocks around main
sequence stars.  

The flux from the starlight $L_{\star}$ is intercepted at a distance $d$ from
the star by the dust, which geometrical cross section is $\sigma_{\rm d}=\pi
a^{2}\, \rm cm^{2}$. A part of the flux from the star is absorbed by the dust 
to be instantaneously re-radiated as, 
\begin{equation}
	\itl{ \Gamma}_{\star}^{\rm dust} = 
	\frac{L_{\star}}{ 4\pi d^{2} }n_{\rm d}\sigma_{\rm d}(1-A)\, \rm erg\, s^{-1}\, cm^{-3},
\label{eq:radlum}
\end{equation}
where $A=1/2$ is the dust grain albedo~\citep{vanburen_apj_329_1988}.
\textcolor{black}{ This assumes that the dust is not decoupled from the gas, 
which is realistic for ionized bow shocks, whereas it may not be true 
in bow shocks around cool stars~\citep{vanmarle_apjl_734_2011}. 
This would influence both our estimation of $n_{\rm d}$ and $\itl \Gamma_{\star}^{\rm dust}$. 
}

Besides, we estimate the dust collisional heating rate $\it \Gamma_{\rm coll}^{\rm
dust}(T)$. On the one hand, the collisional heating for a photoionized medium is
computed following~\citet{ostriker_apj_184_1973}, 
\begin{equation}
	\itl{\Gamma}_{\rm coll,photo}^{\rm dust}(T) = \frac{ 2^{5/2} }{ \sqrt{\pi m_{\rm p}} } f Q 
	n n_{\rm d} \sigma_{\rm d} \Big( k_{\rm B}T \Big)^{3/2} \, \rm erg\, s^{-1}\, cm^{-3},
\label{eq:dustheathot}
\end{equation}
where $n_{\rm d}$ is the dust number density, $m_{\rm p}$ is the mass of the
proton and $Q\, \simeq 1$ is a correction due to the electrical properties of
the grains. On the other hand, it is calculated for the CIE medium
following~\citet{hollenbach_apj_41_1979}, with,
\begin{equation}
	\itl{\Gamma}_{\rm coll,CIE}^{\rm dust}(T) = 2 k_{\rm B} n n_{\rm d} \sigma_{\rm d}
 f v_{\rm p} \times (T-T_{\rm d})\, \rm erg\, s^{-1}\, cm^{-3},
\label{eq:dustheatcool}
\end{equation}
where $k_{\rm B}$ is the Boltzman constant, $v_{\rm p}=\sqrt{k_{\rm B}T/m_{\rm
p}}$ is the proton thermal velocity, $f \approx 10$ is a parameter representing
the effects of the \textcolor{black}{species other than} the protons and $T_{\rm d}$ is the dust
temperature,
\begin{equation}
	T_{\rm d} = 2.3\left(  \frac{ fn_{\rm d} }{ a(\mu m) } \Big(\frac{T}{10^{4}\, \rm K}\Big)^{3/2}  \right)^{1/5}\, \rm K.
\label{eq:dusttemp}
\end{equation}
In Eq.~\ref{eq:dusttemp}, $a(\mu m)$ is the dust radius expressed in $\mu m$.

This method to calculate the infrared emission from a stellar wind bow shock is \textcolor{black}{rather simple}. 
It assumes that the starlight is reemitted by the smallest possible grains and
therefore constitute an upper limit of the corresponding luminosity. 
For all models it was found that radiative heating is dominant 
over collisional heating for all regions within the bow shock.


\end{document}